\documentclass[12pt,a4paper]{article}
\usepackage{amsmath,caption,amsfonts,setspace,mathrsfs,mathtools,bbm}
\usepackage[top=1.25in, bottom=1.25in, left=1.2in, right=1.2in]{geometry}

\usepackage[round]{natbib}
\usepackage{color,soul}
\renewcommand{\baselinestretch}{1.4} 
\DeclareCaptionStyle{italic}[justification=centering]{labelfont={bf},textfont={it},labelsep=colon}
\usepackage{graphicx,psfrag,epsf,textcomp}
\usepackage{enumerate, dsfont}
\usepackage{url,xcolor} 
\usepackage{booktabs, subfig, bm, paralist,mathpazo,tikz,todonotes,longtable,microtype}
\usepackage{mathrsfs}
\usepackage{calligra,amssymb}
\usepackage[pdftex,colorlinks=true]{hyperref}
\definecolor{darkblue}{rgb}{0,0,.6}
\hypersetup{citecolor=darkblue,linkcolor=darkblue,urlcolor=darkblue}

\newcommand{\blind}{1}

\graphicspath{{plots/}}

\newsavebox\CBox

\definecolor{a0}{rgb}{0.0, 0.5, 0.0}
\definecolor{bistre}{rgb}{0.24, 0.17, 0.12}
\definecolor{amethyst}{rgb}{0.6, 0.4, 0.8}
\definecolor{blue-violet}{rgb}{0.54, 0.17, 0.89}
\definecolor{Rcolor}{RGB}{150,160,190}
\definecolor{blush}{rgb}{0.87, 0.36, 0.51}
\definecolor{brightturquoise}{rgb}{0.03, 0.91, 0.87}
\definecolor{burntorange}{rgb}{0.8, 0.33, 0.0}
\usepackage{orcidlink}

\newcommand\X{\mathcal{X}}
\newcommand\Y{\mathcal{Y}}
\newcommand{\Rlogo}{\protect\includegraphics[height=1.8ex,keepaspectratio]{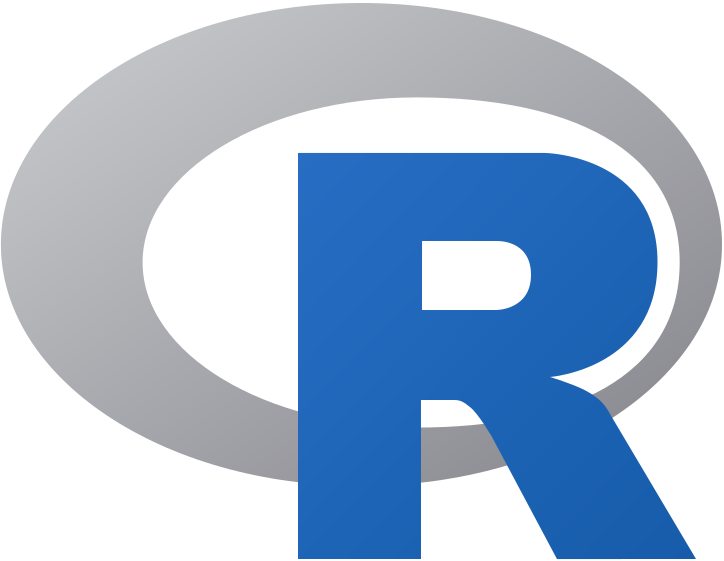}}
\date{}

\begin{document}

\def\spacingset#1{\renewcommand{\baselinestretch}{#1}\small\normalsize} \spacingset{1}

\if0\blind
{
  \title{\bf Function-on-function linear quantile regression}
  \maketitle
} \fi

\if1\blind
{
    \title{\bf Function-on-function linear quantile regression}
    \author{Ufuk Beyaztas \orcidlink{0000-0002-5208-4950} \footnote{Postal address: Department~of~Statistics, Marmara~University, Goztepe Campus, 34722, Istanbul, Turkey; Email: ufuk.beyaztas@marmara.edu.tr} \\ 
    Department~of~Statistics \\
    Marmara~University \\
    \\
    Han Lin Shang \orcidlink{0000-0003-1769-6430}  \\
     Department of Actuarial Studies and Business Analytics \\
     Macquarie University}
  \maketitle
} \fi

\begin{abstract}
In this study, we propose a function-on-function linear quantile regression model that allows for more than one functional predictor to establish a more flexible and robust approach. The proposed model is first transformed into a finite-dimensional space via the functional principal component analysis paradigm in the estimation phase. It is then approximated using the estimated functional principal component functions, and the estimated parameter of the quantile regression model is constructed based on the principal component scores. In addition, we propose a Bayesian information criterion to determine the optimum number of truncation constants used in the functional principal component decomposition. Moreover, a stepwise forward procedure and the Bayesian information criterion are used to determine the significant predictors for including in the model. We employ a nonparametric bootstrap procedure to construct prediction intervals for the response functions. The finite sample performance of the proposed method is evaluated via several Monte Carlo experiments and an empirical data example, and the results produced by the proposed method are compared with the ones from existing models.
 \end{abstract}

\newpage
\doublespacing

\section{Introduction} \label{sec:intro}

With advancements in technology and data storage, it is progressively common to obtain data whose sample elements are collected and saved over a continuum, such as space grids, time, depth, and wavelength. By interpolation and smoothing, these data are represented in curves, images, shapes, or more general objects called functional data. The availability of functional data has regularly increased in almost all scientific branches. The need for developing statistical approaches for this kind of data is also increasing. Consult \cite{ramsay1991}, \cite{ramsay2002, ramsay2006}, \cite{ferraty2006}, \cite{horvath2012}, \cite{cuevas2014}, \cite{Hsing}, and \cite{KoRe} for the theoretical results and case studies of functional data analysis methods. Among many others, function-on-function regression (FFR) models have received considerable attention among researchers from the statistics community to investigate the relationship between a functional response and one or more functional explanatory variables. See, e.g., \cite{yao2005}, \cite{harezlak2007}, \cite{senturk2008}, \cite{matsui2009}, \cite{ivanescu2015}, \cite{chiou2016}, and \cite{BeyaztasComp}.

Let $\left\lbrace \Y_i, \X_i: i = 1, \ldots, n \right\rbrace $ denote a random sample from a pair $\left( \Y, \X \right)$, where both the response $\Y$ and predictor $\X$ variables are square-integrable random functions, $\Y = \left( \Y(t) \right)_{t \in \mathcal{I}}$ and $\X = \left( \X(s) \right)_{s \in S}$, on the bounded and closed intervals $\mathcal{I}, S \in \mathbb{R}$. Without loss of generality, we assume that both the response and predictor are mean-zero processes so that $\text{E}[\Y(t)] = \text{E}[\X(s)] = 0$. Then, the FFR model of $\Y$ on $\X$ is defined as follows:
\begin{equation}\label{ffrm}
\text{E}[\Y(t) | \X(s)] = \int_{s \in S} \X(s) \beta(s,t) ds + \epsilon(t),
\end{equation}
where $\beta(s,t)$ is the smooth bivariate coefficient function and $\epsilon(t)$ is the error function with mean-zero. In the FFR model, the conditional mean of the functional response $\Y$ given a functional predictor $\X$, $\text{E}[\Y | \X]$, is recovered via optimizing least squares (LS) loss, i.e., $\left[ \Y(t) - \int_{s \in S} \X(s) \beta(s,t) ds \right]^2 $. However, the LS loss is sensitive to outliers, which are observations far from the bulk of the data. Outliers are very common in empirical applications. In such cases, biased regression estimates are obtained under the LS loss, leading to unreliable inferences. Also, optimal regression estimates for the FFR model in~\eqref{ffrm} may not be obtained when the error term follows a non-normal heavy-tailed distribution. To remedy these problems, the quantile regression may be used as an alternative to model~\eqref{ffrm}.

Quantile regression (QR), which was introduced by \cite{Koenker1978}, has become a general framework to investigate the effects of predictors at different quantile levels of the response. QR renders it possible to characterize the entire conditional distribution of the response variable. Thus, it provides a more general picture of the response variable's conditional distribution function that might not be reflected by mean regression. There are several advantages of QR over mean regression:
\begin{inparaenum}
\item[1)] QR is more robust to outliers than mean regression when outlying points are present in the response variable since it belongs to a robust model family \cite{Koenker2005};
\item[2)] Compared with mean regression, QR produces more efficient results when the errors follow a non-normal heavy-tailed distribution; and
\item[3)] QR provides more efficient results than mean regression in the presence of conditional heteroscedasticity, the case when the variance depends on one or more predictors.
\end{inparaenum}
Besides, QR allows obtaining prediction intervals based on several quantiles of the response. Consult \cite{Koenker2005} for the theoretical materials and empirical applications of QR.

In FRMs, several QR models have been proposed when one of the observed response or predictor variables consists of random functions rather than scalar observations. See, e.g., \cite{cardot2005}, \cite{ferraty2005}, \cite{chenmuller2012}, \cite{Tang2014}, \cite{Yu2016}, \cite{yao2017}, \cite{Ma2019}, and \cite{Liu2020}. The numerical results obtained from the studies mentioned above have shown that the QR produces better prediction and estimation accuracy than standard linear FRMs when the errors follow a non-normal heavy-tailed distribution and in the presence of outliers.

To the best of our knowledge, there has been no comprehensive work for QR in the case of function-valued response and predictors (function-on-function regression). In this paper, we propose a function-on-function linear quantile regression (FFLQR) model, which allows for more than one functional predictor by extending the traditional FFR model into the QR setting to establish a more flexible and robust approach. Our proposal also aims to present a more comprehensive description of the relationship between the functional response and predictors by focusing on conditional quantiles at different quantile levels. In traditional QR, directly modeling conditional quantiles and estimating the regression coefficients is performed by minimizing the check loss function. Although they are practically observed in discrete-time points in the functional data framework, the functional random variables intrinsically belong to an infinite-dimensional space. Thus, as in FRM, the direct estimation of the proposed method is an ill-posed problem. The common practical approach to remedy this problem is based on projecting infinite-dimensional functional objects onto a finite-dimensional space using dimension reduction methods. For this purpose, several general basis expansion methods, such as B-spline, Fourier, and wavelet basis have been proposed \cite{ramsay1991,ramsay2006, ivanescu2015}. However, general basis expansion methods may require a large number of basis functions to project functional objects onto a finite-dimensional space, leading to poor estimation and prediction accuracy. On the other hand, the dimension reduction techniques including the functional principal component (FPC) analysis provide more informative approximation since they are data-driven and uses the information of functional predictors gathered from their covariance functions \cite{Hall2007}. Therefore in this paper, the FPC analysis is first used to transform the functional random variables into a finite-dimensional space. This approximation allows converting infinite-dimensional FFLQR into a multivariate regression model of principal component scores. In what follows, we estimate the proposed FFLQR model using the FPC approximations of the functional response and predictor variables. The prediction performance of the proposed method is based on the truncation constants used in the decompositions of the functional objects. In this paper, the optimum values of the truncation constants are determined via a Bayesian information criterion (BIC). In practice, the exact form of the actual model is unspecified since the model's significant predictors are unknown. To this end, we employ a stepwise forward procedure along with the BIC to determine significant predictors. We apply a nonparametric bootstrap method to construct pointwise prediction intervals for the response function at different quantile levels.

The remainder of this paper is organized as follows. Section~\ref{sec:methodology} presents the methodology of the proposed method. The results of Monte Carlo experiments and empirical data analysis are given in Section~\ref{sec:results}. Section~\ref{sec:conc} concludes the paper.

\section{Methodology} \label{sec:methodology}

We consider a random sample $\left\lbrace \Y_i, \bm{\X}_i: i = 1, \ldots, n \right\rbrace $ from a pair $\left( \Y, \bm{\X} \right)$, where $\Y = \left( \Y(t) \right)_{t \in \mathcal{I}}$ is a functional response and $\bm{\X} = \left[ \X_{i1}, \ldots, \X_{iM} \right]^\top$ with $\X_m = \left( \X_m(s) \right)_{s \in S}$ is an $M$ dimensional vector of functional predictors, defined on the bounded and closed intervals $\mathcal{I}, S \in \mathbb{R}$. Without loss of generality, we assume that $\mathcal{I}, S = [0,1]$ and both the response and predictors are mean-zero processes so that $\text{E}[\Y(t)] = \text{E}[\X_m(s)] = 0$ for $m = 1, \ldots, M$.

Given any $\tau \in (0,1)$, let $Q_{\tau}\left[ \Y(t) | \bm{\X} \right]$ denote the conditional quantile of the response function $\Y$ given the vector of functional predictors $\bm{\X}$. Let us assume that $Q_{\tau}\left[ \Y(t) | \bm{\X} \right] $ can be written as a linear functional of $\bm{\X}$ along with smooth bivariate functions $\left[ \beta_{1 \tau}(s,t), \ldots, \beta_{M \tau}(s,t) \right]$. Then, we define the proposed FFLQR model as follows:
\begin{equation} \label{eq:cq}
Q_{\tau}\left[ \Y_i(t) | \bm{\X}_i \right] = \sum_{m=1}^M \int_0^1 \X_{im}(s) \beta_{m \tau}(s,t) ds.
\end{equation}
We consider minimizing the check loss function, $\rho_{\tau}(u) = u \left\lbrace \tau - \mathbb{1} (u < 0) \right\rbrace$ where $\mathbb{1}\{\cdot\}$ denotes an indicator function \cite{Koenker1978, Koenker2005}, to solve the $\tau$\textsuperscript{th} conditional quantile $Q_{\tau}\left[ \Y_i(t) | \bm{\X}_i \right]$ as follows:
\begin{equation}\label{eq:fcheck}
\underset{\begin{subarray}{c}
  \beta_{1 \tau}(s,t), \ldots, \beta_{M \tau}(s,t)
  \end{subarray}}{\arg\min}~ \sum_{i=1}^N \rho_{\tau} \left[ \Y_i(t) - \sum_{m=1}^M \int_0^1 \X_{im} (s) \beta_{m \tau}(s,t) ds \right].
\end{equation}

As stated in Section~\ref{sec:intro}, the functional objects intrinsically belong to an infinite-dimensional space. Consequently, the minimizing problem stated in~\eqref{eq:fcheck} is an ill-posed problem due to the infinite-dimensional nature of $\Y$ and $\bm{\X}$. To overcome this problem, we consider FPC decomposition for each functional object in~\eqref{eq:cq}. Let $\mathcal{C}_{\Y}(t_1,t_2) = \text{Cov}[\Y(t_1), \Y(t_2)]$ and $\mathcal{C}_{\X_m}(s_1,s_2) = \text{Cov}[\X(s_1), \X(s_2)]$, respectively, denote the covariance functions of $\Y(t)$ and $\X_m(s)$ satisfying $\int_0^1 \int_0^1 \mathcal{C}_{\Y}^2(t_1,t_2) dt_1 dt_2 < \infty$ and $\int_0^1 \int_0^1 \mathcal{C}_{\X_m}^2(s_1,s_2) ds_1 ds_2 < \infty$. Then, by Mercer's theorem, the covariance functions can be represented as follows:
\begin{align*}
\mathcal{C}_{\Y} &= \sum_{k \geq 1} w_k \phi_k(t_1) \phi_k(t_2), \qquad \forall t_1, t_2 \in [0,1], \\
\mathcal{C}_{\X_m} &= \sum_{l \geq 1} \kappa_{ml} \psi_{ml}(s_1) \psi_{ml}(s_2), \qquad \forall s_1, s_2 \in [0,1],
\end{align*}
where $\left\lbrace \phi_k(t): k = 1, 2, \ldots \right\rbrace$ and $\left\lbrace \psi_{ml}(s): l = 1, 2, \ldots \right\rbrace$ denote the orthonormal eigenfunctions corresponding to the non-negative eigenvalues $\left\lbrace w_k: k = 1, 2, \ldots \right\rbrace$ and $\left\lbrace \kappa_{ml}: l = 1, 2, \ldots \right\rbrace$ with $w_k \geq w_{k+1}$ and $\kappa_{m1} \geq \kappa_{ml+1}$, respectively. To obtain FPC decompositions of functional variables, we use predetermined truncation constants $K_{\Y}$ (for $\Y(t)$) and $K_{\X_m}$ (for $\X_m(s)$). By Karhunen-Lo\`{e}ve expansion, the functional response and functional predictors can be represented as follows:
\begin{align}
\Y_i(t) &\approx \sum_{k=1}^{K_{\Y}} \xi_{ik} \phi_k(t) = \bm{\xi}_i^\top \bm{\phi}(t), \qquad \forall t \in [0,1], \label{eq:by} \\
\X_{im}(s) &\approx \sum_{l=1}^{K_{\X_m}} \zeta_{iml} \psi_{ml}(s) = \bm{\zeta}_{im}^\top \bm{\psi}_m(s), \qquad \forall s \in [0,1], \label{eq:bx}
\end{align}
where the random variables $\xi_{ik} = \int_0^1 \Y_i(t) \phi_k(t) dt$ and $\zeta_{iml} = \int_0^1 \X_{im}(s) \psi_{ml}(s) ds$ are the projections of $\Y_i(t)$ and $\X_{im}(s)$ onto their corresponding orthonormal bases, respectively. Similarly, the smooth bivariate coefficient functions $\beta_{m \tau}(s,t)$ can be expressed in terms of eigenfunctions as follows:
\begin{equation}\label{eq:bb} 
\beta_{m \tau}(s,t) \approx \sum_{l=1}^{K_{\X_m}} \sum_{k=1}^{K_{\Y}} \beta_{mlk}^{(\tau)} \psi_{ml}(s) \phi_k(t) = \bm{\psi}_m^\top(s) \bm{\beta}_m^{(\tau)} \bm{\phi}(t), \qquad \forall t,s \in [0,1], 
\end{equation}
where $\beta_{mlk}^{(\tau)} = \int_0^1 \int_0^1 \beta_m(s,t) \psi_{ml}(s) \phi_k(t) ds dt$ links the smooth dependence of $\beta_{m \tau}(s,t)$ on the quantile $\tau$. 

Next, using a similar parameterization as in \cite{chiou2016, Harjit} and by orthonormalities of $\bm{\phi}(t)$ and $\bm{\psi}_m(s)$, the $\tau$\textsuperscript{th} conditional quantile of the response function in \eqref{eq:cq} can be expressed as follows:
\begin{align}
Q_{\tau}\left[ \Y_i(t) | \bm{\X}_i \right] &= \sum_{m=1}^M \int_0^1 \bm{\zeta}_{im}^\top \bm{\psi}_m(s) \bm{\psi}_m^\top(s) \bm{\beta}_m^{(\tau)} \bm{\phi}(t) \nonumber \\
\bm{\xi}_i^\top \bm{\phi}(t) &= \sum_{m=1}^M \bm{\zeta}_{im}^\top \bm{\beta}_m^{(\tau)} \bm{\phi}(t), \nonumber \\
\bm{\xi}_i^\top \int_0^1 \bm{\phi}(t) \bm{\phi}^\top(t) dt &= \sum_{m=1}^M \bm{\zeta}_{im}^\top \bm{\beta}_m^{(\tau)} \int_0^1 \bm{\phi}(t) \bm{\phi}^\top(t) dt, \nonumber \\
\bm{\xi}_i^\top &= \sum_{m=1}^M \bm{\zeta}_{im}^\top \bm{\beta}_m^{(\tau)}\label{eq:redf}.
\end{align}
Let $\bm{\Xi} = \left[\bm{\xi}_1^\top, \ldots, \bm{\xi}_n^\top \right]^\top$, $\bm{\Pi} = \left[\bm{\zeta}_1, \ldots, \bm{\zeta}_n \right]^\top$ where $\bm{\zeta}_i = \left[\bm{\zeta}_{i1}^\top, \ldots, \bm{\zeta}_{iM}^\top \right]^\top$, and $\bm{\mathcal{B}} = \left[\bm{\beta}_1^{(\tau)}, \ldots, \bm{\beta}_M^{(\tau)} \right]^\top$ denote the matrices of the FPC scores and coefficient vectors. Then, we have the following multivariate model in matrix form:
\begin{equation}
\bm{\Xi} = \bm{\Pi} \bm{\mathcal{B}}.
\end{equation}
In what follows, the parameter matrix $\bm{\mathcal{B}}$ can be estimated by minimizing the check loss function as follows:
\begin{equation*}
\widehat{\bm{\mathcal{B}}} = \underset{\bm{\mathcal{B}}}{\arg\min} \left[ \sum_{i=1}^N \rho_{\tau} \left( \bm{\Xi}_i - \bm{\Pi}_i \bm{\mathcal{B}} \right) \right],
\end{equation*}
where $\widehat{\bm{\mathcal{B}}} = \left[\widehat{\beta}_1^{(\tau)}, \ldots, \widehat{\beta}_M^{(\tau)} \right]^\top$. Note that the input of $\rho_{\tau}$ in the minimization problem given above is a vector, and thus, minimization is done according to each coordinate of the vector separately to obtain the estimates of each element of $\bm{\mathcal{B}}$. Accordingly, the $m$\textsuperscript{th} estimated coefficient function can be reconstructed as
\begin{equation*}
\widehat{\beta}_{m \tau}(s,t) = \sum_{l=1}^{K_{\X_m}} \sum_{k=1}^{K_{\Y}} \psi_{ml}(s) \widehat{\beta}_{mlk}^{(\tau)} \phi_k(t) = \bm{\psi}_m^\top(s) \widehat{\bm{\beta}}_m^{(\tau)} \bm{\phi}(t),
\end{equation*}
and for a given $\tau \in (0,1)$, the estimated conditional quantile of response in~\eqref{eq:cq} is obtained
\begin{equation}\label{eq:est_qn}
\widehat{Q}_{\tau}\left[ \Y_i(t) | \bm{\X}_i \right] = \sum_{m=1}^M \int_0^1 \X_{im}(s) \widehat{\beta}_{m \tau}(s,t) ds.
\end{equation}
Note that, if the functional response and/or one or more functional predictors are not mean-zero processes, i.e., $\text{E}[\Y(t)] \neq 0$ and/or $\text{E}[\X_m(s)] \neq 0$ ($m \in [1, \ldots, M]$), then the conditional quantile of $\Y(t)$ given $\bm{\X}$ includes an intercept function, say $\beta_{0 \tau}(t)$, as follows:
\begin{equation*}
Q_{\tau}\left[ \Y_i(t) | \bm{\X}_i\right] = \beta_{0 \tau}(t) + \sum_{m=1}^M \int_0^1 \X_{im}(s) \beta_{m \tau}(s,t) ds.
\end{equation*}
In such a case, the estimate of the intercept function can be obtained by including the FPC decomposition of $\beta_{0 \tau}(t)$, i.e., $\beta_{0 \tau}(t) = \sum_{k=1}^{K_{\Y}} g_{k} \phi_k(t)$, into the check loss function.

The results given above demonstrate that:
\begin{inparaenum}
\item[1)] the infinite-dimensional minimization problem~\eqref{eq:fcheck} can be reduced to a simple finite-dimensional QR problem using the metrics $\lbrace \bm{\psi}_m(s): m = 1, \ldots, M \rbrace$ and $\bm{\phi}(t)$ in the spaces of FPC scores $\left\lbrace \bm{\zeta}_{im}^\top: m = 1, \ldots, M \right\rbrace $ and $\bm{\xi}_i^\top$; and
\item[2)] the coefficient functions in the proposed FFLQR model can be approximated using the FPC basis functions and the estimated parameter of the QR model constructed based on the FPC scores. This finite-dimensional QR model can be easily estimated using the available \Rlogo \ package ``quantreg'' \cite{quantreg}.
\end{inparaenum}

The performance of the proposed FFLQR depends on the truncation parameters $K_{\Y}$ and $K_{\X_m}$ (for $m = 1, \ldots, M$). In this paper, the optimum numbers of truncation constants are determined based on the BIC. Let us denote all the possible models by $\mathcal{J} = \left\lbrace K_{\Y}, K_{\X_m} | K_{\Y} = 1, \ldots, K_{\Y_{\max}},~K_{\X_m} = 1, \ldots, K_{\X_{m, \max}} \right\rbrace$. Then, we assume that there exists an optimal estimate for the FFLQR associated with $K_{\Y_0},~K_{\X_{m_0}} \in \mathcal{J}$ as follows:
\begin{equation*}
\widehat{Q}_{\tau}\left[ \Y_i(t) | \bm{\X}_i\right] = \sum_{m=1}^M \int_0^1 \X_{im}(s) \widehat{\beta}_{m \tau}^{(K_{\Y_0}, K_{\X_0})}(s,t) ds,
\end{equation*}
where $\widehat{\beta}_{m \tau}^{(K_{\Y_0}, K_{\X_0})}(s,t)$ is the estimated coefficient function with $K_{\Y_0}$ and $K_{\X_{m_0}}$ and $K_{\X_0} = \left\lbrace K_{\X_{1_0}}, \ldots, K_{\X_{M_0}} \right\rbrace$. For computational simplicity, we apply the same truncation constant $K_{\X}$ to all the predictors. For each combination of $K_{\Y}$ and $K_{\X}$, the estimates of coefficient functions are obtained by
\begin{small}
\begin{equation*}
\widehat{\bm{\beta}}_{\tau}^{(K_{\Y}, K_{\X})}(s,t) = \underset{\begin{subarray}{c}
  \beta_{1 \tau}^{(K_{\Y}, K_{\X})}(s,t), \ldots, \beta_{M \tau}^{(K_{\Y}, K_{\X})}(s,t)
  \end{subarray}}{\arg\min}~ \sum_{i=1}^N \rho_{\tau} \left[ \Y_i(t) - \sum_{m=1}^M \int_0^1 \X_{im} (s) \beta_{m \tau}^{(K_{\Y}, K_{\X})}(s,t) ds \right],
\end{equation*}
\end{small}
where $\widehat{\bm{\beta}}_{\tau}^{(K_{\Y}, K_{\X})}(s,t) = \left[ \widehat{\beta}_{1 \tau}^{(K_{\Y}, K_{\X})}(s,t), \ldots, \widehat{\beta}_{M \tau}^{(K_{\Y}, K_{\X})}(s,t) \right]^\top$. Then, according to the definition of BIC given by \cite{Schwarz}, we obtain the following BIC for the FFLQR:
\begin{small}
\begin{equation*}
\text{BIC}(K_{\Y}, K_{\X}) = \bigg \Vert \ln \left[ \sum_{i=1}^n \rho_{\tau} \left( \Y_i(t) - \sum_{m=1}^M \int_0^1 \X_{im}(s) \widehat{\beta}_{m \tau}^{(K_{\Y}, K_{\X})}(s,t) ds \right) \right] \bigg \Vert_{\mathcal{L}_2} + \omega \ln(n),
\end{equation*}
\end{small}
where $\omega = K_{\Y} + K_{\X}$ and $\Vert \cdot \Vert_{\mathcal{L}_2}$ denotes the $\mathcal{L}_2$ norm. In short, $K_{\Y_{\max}} \times K_{\X_{\max}}$ different models (i.e., $K_{\Y} = 1, \ldots, K_{\Y_{\max}},~K_{\X} = 1, \ldots, K_{\X_{\max}})$ are built and the optimum $K_{\Y}$ and $K_{\X}$ ($\widetilde{K}_{\Y}$ and $\widetilde{K}_{\X}$, respectively) are determined by
\begin{equation*}
\left( \widetilde{K}_{\Y}, \widetilde{K}_{\X} \right) = \underset{\begin{subarray}{c} K_{\Y}, K_{\X} \end{subarray}}{\arg\min}~ \text{BIC}(K_{\Y}, K_{\X}).
\end{equation*}

\subsection{Variable selection procedure}\label{sec:vsp}

When considering the proposed FFLQR model~\eqref{eq:cq}, the vector of functional predictors $\bm{\X}$ may include too many variables, and not all of them may have a significant effect on the model. Consequently, there may need a variable selection procedure to determine only the significant functional predictors. To this end, we consider a forward variable selection procedure together with the extension of BIC introduced by \cite{Lee2014} to the proposed method. Denote by $\mathcal{D} = \left\lbrace d_1, \ldots, d_a \right\rbrace \subset \left\lbrace 1, \ldots, M \right\rbrace$ the candidate model having the functional predictors $\left\lbrace \X_{d_1}, \ldots, \X_{d_a} \right\rbrace$ in the model. Using the check loss function, the estimates of coefficient function for this model are given by
\begin{equation*}
\widehat{\bm{\beta}}_{\tau}^{\mathcal{D}}(s,t) = \underset{\begin{subarray}{c}
  \beta_{d_1 \tau}^{\mathcal{D}}(s,t), \ldots,  \beta_{d_a \tau}^{\mathcal{D}}(s,t)
  \end{subarray}}{\arg\min}~ \sum_{i=1}^N \rho_{\tau} \left[ \Y_i(t) - \sum_{m \in \mathcal{D}} \int_0^1 \X_{im}(s) \beta_{m \tau}^{\mathcal{D}}(s,t) ds \right],
\end{equation*}
where $\widehat{\bm{\beta}}_{\tau}^{\mathcal{D}}(s,t) = \left[ \widehat{\beta}_{d_1 \tau}^{\mathcal{D}}(s,t), \ldots, \widehat{\beta}_{d_a \tau}^{\mathcal{D}}(s,t) \right]^\top$. Let $\vert \mathcal{D} \vert$ denote the cardinality $a$ of $\mathcal{D}$. Then, following the definition of BIC presented by \cite{Lee2014}, the BIC for the candidate model is defined as follows:
\begin{equation*}
\text{BIC} \left( \mathcal{D} \right) = \bigg \Vert \ln \left[ \sum_{i=1}^n \rho_{\tau} \left( \Y_i(t) - \sum_{m \in \mathcal{D}} \int_0^1 \X_{im}(s) \widehat{\beta}_{m \tau}^{\mathcal{D}}(s,t) ds \right) \right] \bigg \Vert_{\mathcal{L}_2} + \vert \mathcal{D} \vert \frac{\ln (n)}{2 n}.
\end{equation*}
Accordingly, we present the considered variable selection procedure as follows.

\begin{itemize}
\item[Step 1.] Construct $M$ - FFLQR models based on the typical functional response and a functional predictor variable:
\begin{equation*}
Q_{\tau}\left[ \Y_i(t) | \X_{im} \right] = \int_0^1 \X_{im}(s) \beta_{m \tau}(s,t) ds, \quad m = 1, \ldots, M
\end{equation*}
and calculate the $\text{BIC} \left( \mathcal{D} \right)$ for each of these models. Then, determine the initial model to having smallest $\text{BIC} \left( \mathcal{D} \right)$. Let $\X_i^{(1)}(s)$ and $\text{BIC}^{(1)} \left( \mathcal{D} \right)$ denote the functional predictor in the initial model and its $\text{BIC} \left( \mathcal{D} \right)$ value, respectively.
\item[Step 2.] Construct ($M-1$) - FFLQR model based on the common functional response and a vector of functional predictors:
\begin{equation*}
Q_{\tau}\left[ \Y_i(t) | \bm{\X}_{im}\right] = \sum_{m=1}^2 \int_0^1 \X_{im}(s) \beta_{m \tau}(s,t) ds,
\end{equation*}
where $\bm{\X}_{im} = \left[ \X_i^{(1)}(s), \X_{im}(s) \right]$ and $\X_{im}(s) \neq \X_i^{(1)}(s)$, and calculate the $\text{BIC} \left( \mathcal{D} \right)$ for each of these models. The vector of functional predictors having the smallest $\text{BIC} \left( \mathcal{D} \right)$, say $\bm{\X}^{(2)}$, is chosen as the predictor vector for the current model if $\text{BIC}^{(2)} \left( \mathcal{D} \right) / \text{BIC}^{(1)} \left( \mathcal{D} \right) < 0.95$, where $\text{BIC}^{(2)} \left( \mathcal{D} \right)$ is the computed $\text{BIC} \left( \mathcal{D} \right)$ when $\bm{\X}^{(2)}$ is used in the current model. In other words, the second predictor enters the model if it contributes at least 5\% to the model (the threshold value 5\% is determined via an extensive Monte Carlo experiment). Repeat this procedure until all the significant functional predictors are included in the model.
\end{itemize}

Note that applying the above variable selection procedure, together with the procedure of determination of optimum truncation constants discussed in Section~\ref{sec:methodology}, may not be computationally efficient when there are a large number of predictors to be considered. In our numerical analyses, we first determine the significant variables using fixed truncation constants, say $K_{\Y} = K_{\X} = 2$. Then, the optimum values of $K_{\Y}$ and $K_{\X}$ are determined based on the significant functional predictors. In this context, we perform several simulations (but not reported in the paper for the sake of space), and the results have shown that the choices of $K_{\Y}$ and $K_{\X_m}$ do not have a significant effect on the determination of significant predictors. In high-dimensional settings, a parallel implementation of the determination of optimum truncation constants and significant variables (such as using ``doParallel'' \cite{doParallel} and ``foreach'' \cite{foreach} \Rlogo \ packages) can be used to reduce computational costs.

\subsection{Computational framework}

In this section, we summarize the computational framework of the proposed method. We note that all the numerical calculations for the proposed method are performed using \Rlogo \ version 4.1.1 on an Intel Core i7 6700HQ 2.6 GHz PC. In the proposed method, first, the FPC decomposition of the functional variables are obtained using some available functions such as \texttt{create.bspline.basis} and \texttt{pca.fd} in the \Rlogo \ package ``fda'' \cite{fda}. Then, the \texttt{rqs.fit} function in the \Rlogo \ package ``quantreg'' \cite{quantreg} along with the computed principal component scores is used to approximate the coefficient functions in the finite-dimensional space. Finally, the functional forms of the estimated coefficient functions are obtained using the estimated FPC basis functions. We build some functions for other calculations such as BIC and variable selection procedure. An example \Rlogo \ code (with definitions) for the proposed method can be found at \url{https://github.com/UfukBeyaztas/FFLQR}.

\section{Numerical results} \label{sec:results}

In this section, several Mote Carlo experiments under different scenarios and empirical data analysis are performed to investigate the finite-sample predictive performance of our proposed FFLQR model. We compare our results with those calculated via the LS \cite{ramsay2006}, functional partial least squares (FPLS) \cite{BeyaztasShang2020}, and classical FPC based FFR models. The LS method uses a general basis function expansion method (B-spline basis function is used in our calculations) to project the functional variables into the finite-dimensional space. Then, the LS loss function is used to estimate the model parameters. In FPLS, an iterative procedure similar to FPC is used to obtain the FPLS basis functions and corresponding coefficients. Then, the LS estimator is used to approximate the functional coefficients.

\subsection{Monte Carlo experiments}

\begin{figure}[!htb]
  \centering
  \includegraphics[width=5cm]{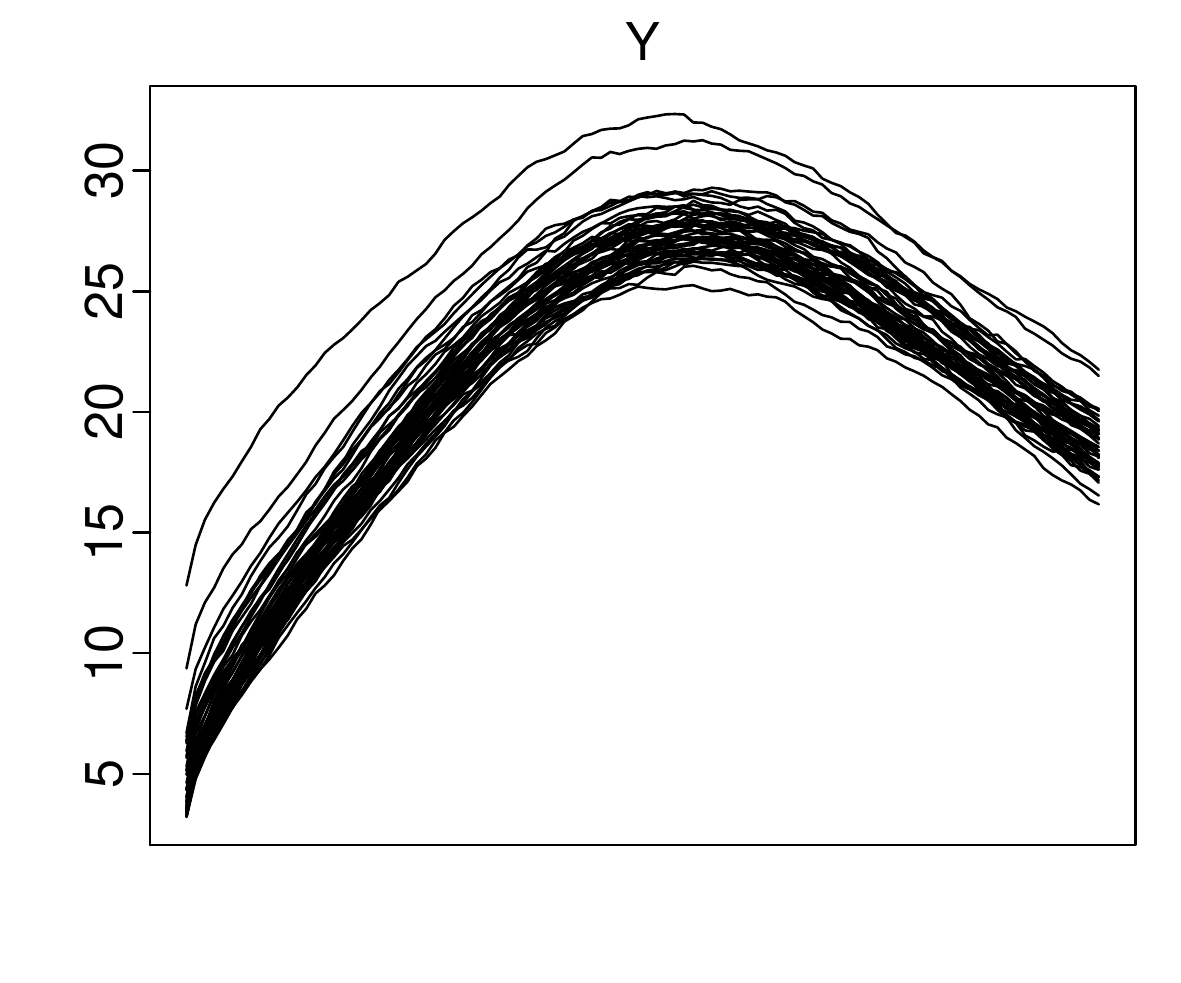}
  \includegraphics[width=5cm]{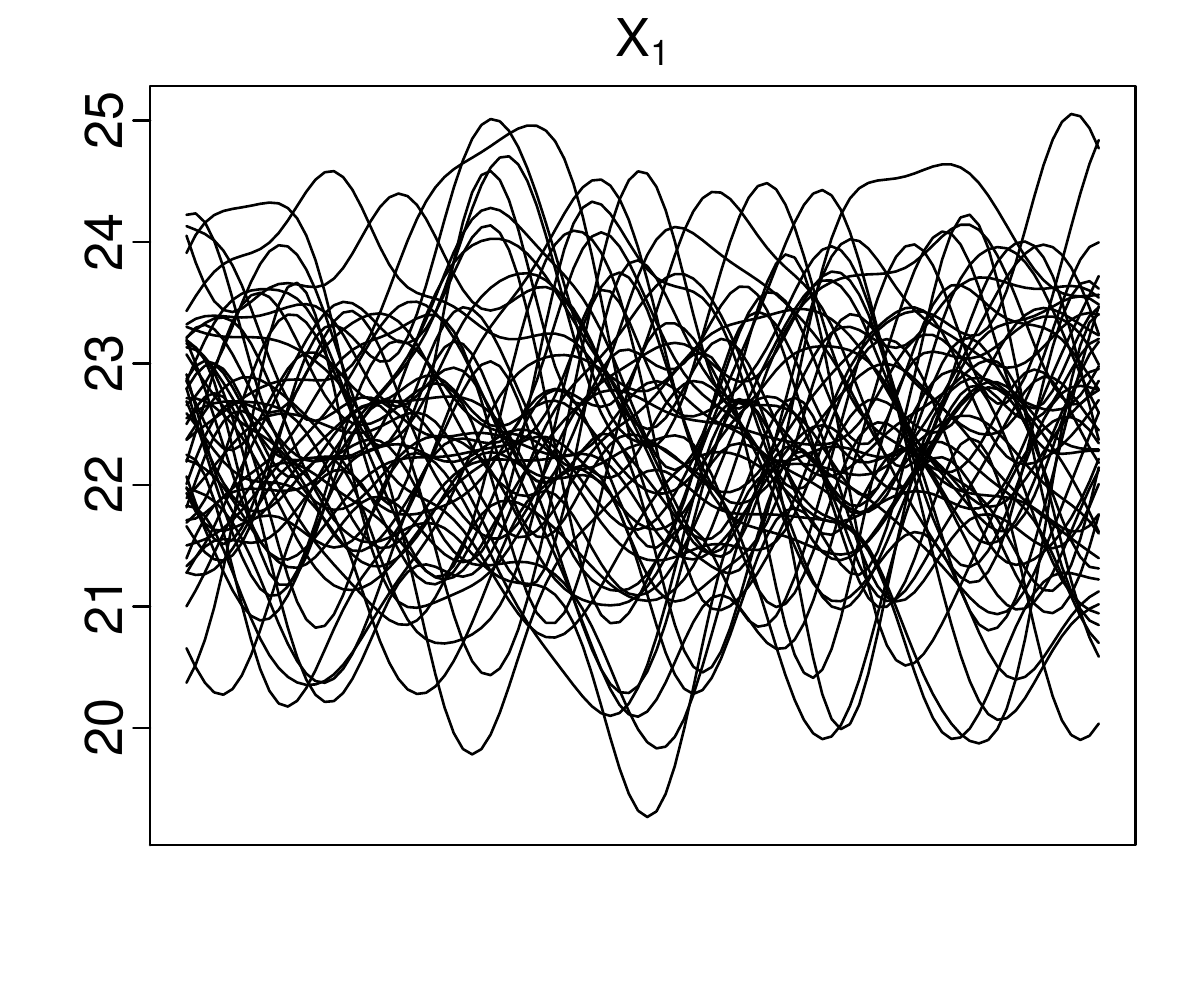}
  \includegraphics[width=5cm]{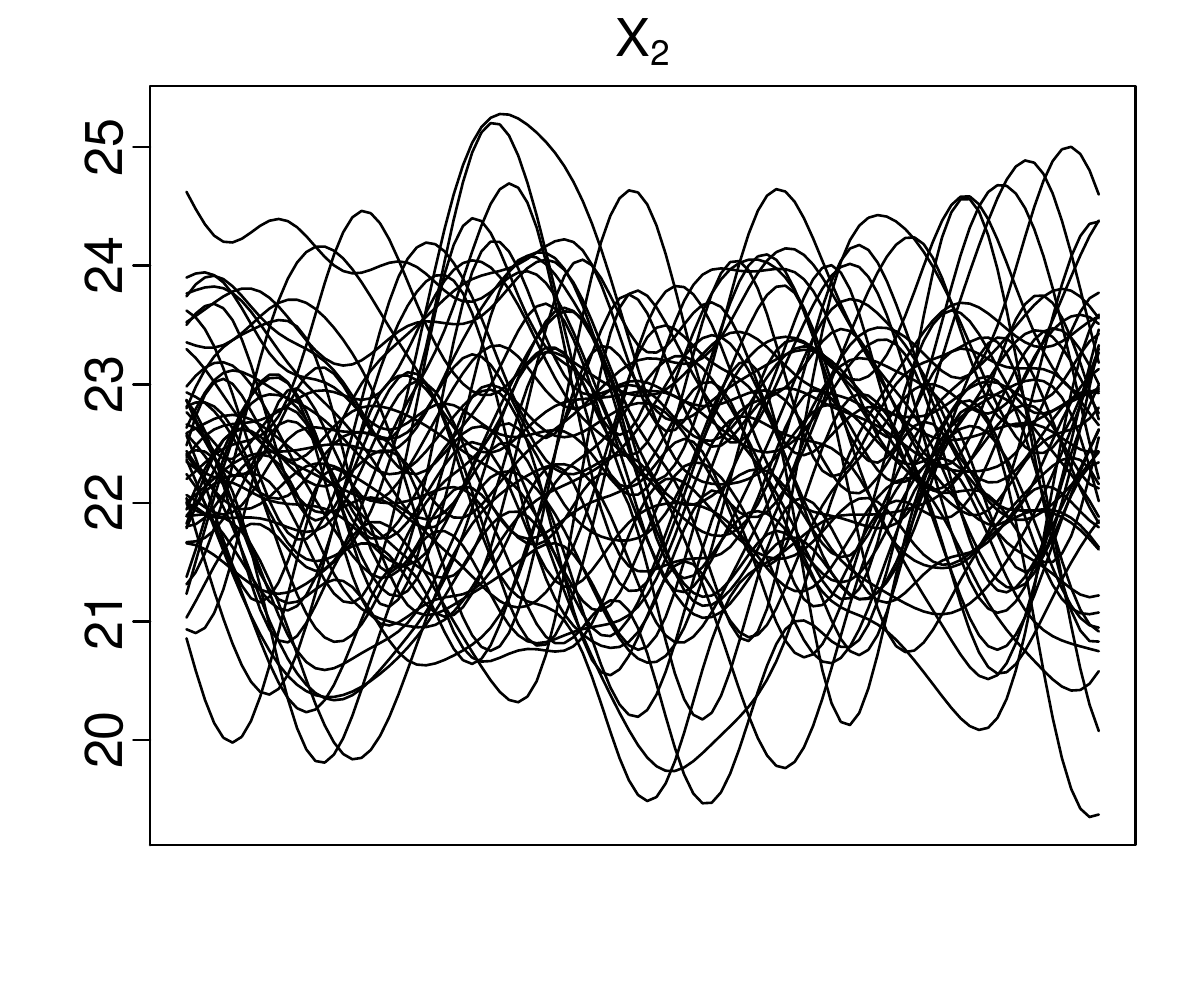}
\\ 
  \includegraphics[width=5cm]{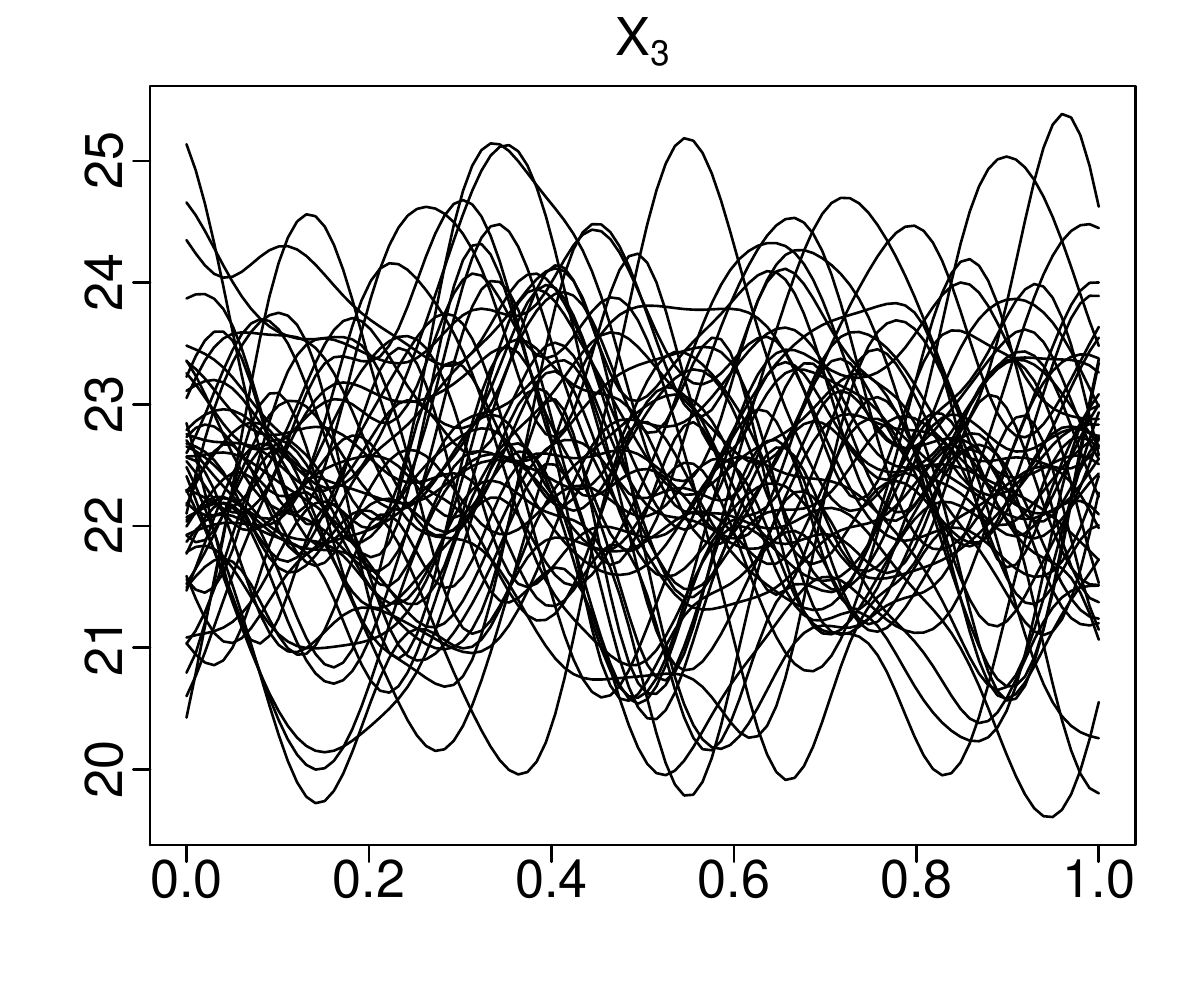}
  \includegraphics[width=5cm]{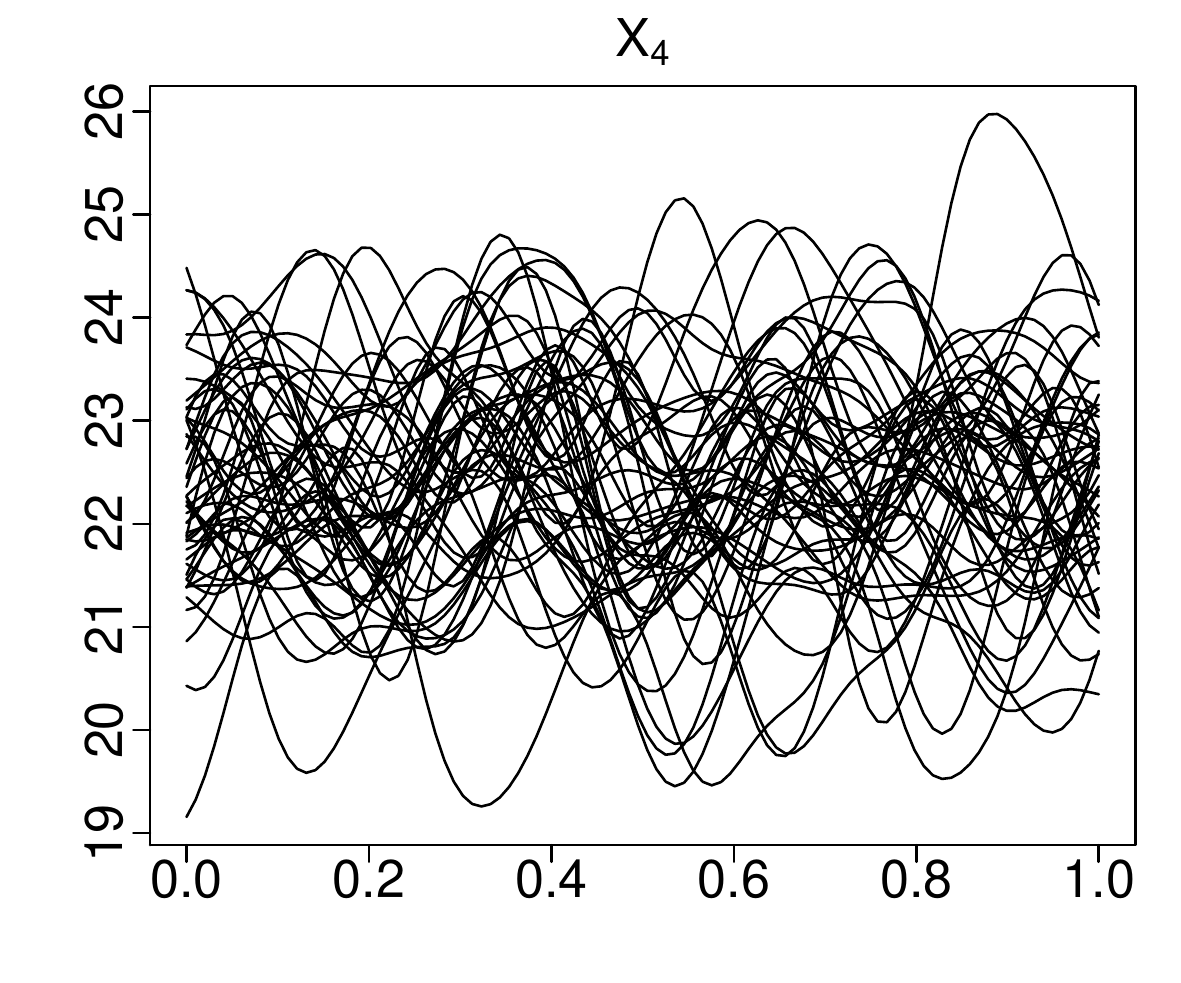}
  \includegraphics[width=5cm]{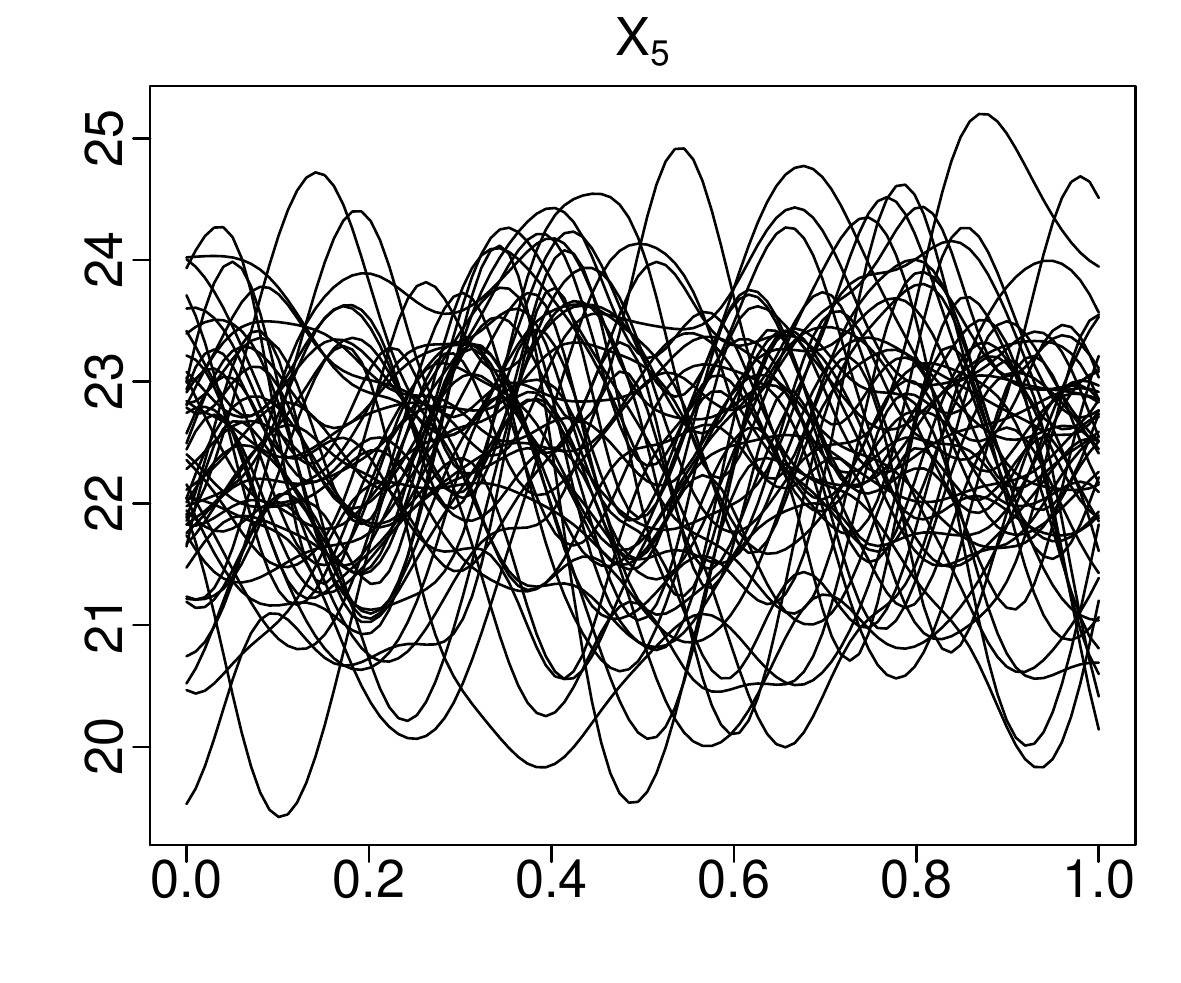}
  \caption{Plots of the generated 50 functions of the functional variables when $\sigma = 1$ and error terms follow $\chi^2_{(1)}$ distribution. The functions of the response variable are generated as $\Y_i(t) = \sum_{m \in \mathcal{D}} \int_0^1 \X_{im}(s) \beta_m(s,t) ds + \epsilon_i(t)$ where $\mathcal{D} = \left\lbrace 2, 4, 5 \right\rbrace$.}
  \label{fig:Fig_1}
\end{figure}

In the Monte Carlo experiments, we perform $\text{MC}$ = $200$ Monte Carlo simulation runs, each of which consists of $M = 5$ functional predictors $\bm{\X}_i = \left\lbrace \X_{i1}(s), \ldots, \X_{i5}(s) \right\rbrace$ with $i = 1, 2, \dots, 500$ observed at 100 equally spaced points in the unit interval $s\in[0, 1]$. We consider the following process, which is a modified version of those considered by \cite{LuoQi}, to generate the functional predictors:
\begin{equation}\label{eq:fpreds}
\X_{im}(s) = 10 + \sum_{j=0}^{\ell} \frac{V_{i,m+j}(s)}{\sqrt{\ell+1}},
\end{equation}
where $[V_{i,1}(s), V_{i,2}(s),\dots, V_{i,9}(s)]$ are generated from a Gaussian process with mean-zero and a positive definite covariance function $\pmb{\Sigma}_V(s,s^{\prime}) = e^{-100 (s - s^{\prime})^2}$. In~\eqref{eq:fpreds}, the parameter $\ell$ ($\ell > 0$) denotes the lag parameter that controls the correlation level between predictor functions. Herein, the large $\ell$ corresponds to high correlations between the predictors. In our analyses, we consider $\ell = 4$ to generate functional predictors. In addition, we consider the following processes to generate the smooth bivariate regression coefficients:
\begin{align*}
\beta_1(s,t) &= (1-s)^2 (t-0.5)^2, \\ 
\beta_2(s,t) &= e^{-3 (s-1)^2 -5 (t - 0.5)^2}, \\ 
\beta_3(s,t) &= e^{-5 (s-0.5)^2 - 5(t-0.5)^2} + 8 e^{-5(s-1.5)^2 - 5(t-0.5)^2}, \\
\beta_4(s,t) &= \sin(1.5 \pi s) \sin(\pi t), \\ 
\beta_5(s,t) &= \sqrt{st}.
\end{align*}
Then, the functions of the response variable are generated as follows:
\begin{equation*}
\Y_i(t) = \sum_{m \in \mathcal{D}} \int_0^1 \X_{im}(s) \beta_m(s,t) ds + \epsilon_i(t),
\end{equation*}
where $\mathcal{D} = \left\lbrace 2, 4, 5 \right\rbrace$ denotes the index set of significant functional predictors. The error functions $\epsilon_i(t)$ are generated from the Ornstein-Uhlenbeck process:
\begin{equation*}
\epsilon_i(t) = \gamma + \left[\epsilon_0(t) - \gamma\right] e^{-\theta t} + \sigma \int_0^t e^{-\theta (t-u)} dW_u,
\end{equation*}
where $\gamma$, $\theta$ and $\sigma > 0$ are real constants, and $W_u$ denotes the Wiener process. Given $\epsilon_0(t)$, which is the initial value of $\epsilon_i(t)$ and independently taken from $W_u$, the functions $\epsilon_i(t)$ are generated from the joint distribution of $\lbrace \epsilon_i(t) \rbrace_{i=1}^n$. To investigate the robustness of the QR over the mean regression, we consider two distributions for the error terms:
\begin{inparaenum}
\item[(i)] $N(0,1)$ (symmetric) and
\item[(iii)] $\chi^2_{(1)}$ (skewed).
\end{inparaenum}
In addition, we consider two signal-to-noise ratio levels: $\sigma = 0.1$ and $\sigma = 1$. Examples of the generated functions for the response and predictor variables are presented in Figure~\ref{fig:Fig_1}.

To further investigate the robust nature of the proposed method, we consider a second scenario for the response variable, where the observations associated with the response variable are contaminated by outliers at $[5\%, 10\%, 20\%]$ contamination levels. Since the QR characterizes the conditional distribution of the response variable, only the generated functional response is contaminated by the outliers. The outlier-contaminated response functions are generated by contaminating $n \times [5\%, 10\%, 20\%]$ randomly selected functions by a random function; $\widetilde{\Y}_i = \Y_i + \vert N(10, 0.04) \vert$.

The mean squared prediction error (MSPE) metric is considered to compare the methods' predictive performance. In doing so, the generated data are first divided into training and test samples with sizes $n_{\text{train}} = 200$ and $n_{\text{test}} = 300$. Then, the training samples are used to construct models, and the functions in the test sample are used for validation. The MSPE is computed as follows:
\begin{equation*}
\text{MSPE} = \frac{1}{n_{\text{test}}} \sum_{i=1}^{n_{\text{test}}} \left\Vert \Y_i(t) - \widehat{\Y}_i(t) \right\Vert^2_{\mathcal{L}_2},
\end{equation*}
where $\widehat{\Y}_i(t)$ is the prediction of the $i^\textsuperscript{th}$ observation in the test sample and is calculated using~\eqref{eq:est_qn}. To construct pointwise prediction intervals for the response function in the test sample, we employ a case-sampling-based bootstrap. In doing so, $R = 100$ bootstrap pseudo-samples, $\left( \Y^*, \bm{\X}^* \right)$, with sizes $n_{\text{train}}$ are drawn with replacement from $\left( \Y, \bm{\X} \right)$. Then, based on the bootstrap pseudo-samples, the smooth bivariate functional coefficients are estimated to obtain $R = 100$ sets of bootstrap replicates of the predicted response functions $\left\lbrace \widehat{\Y}_i^{*,1}(t), \ldots, \widehat{\Y}_i^{*,R}(t) \right\rbrace$. Finally, the $100(1-\alpha)\%$ bootstrap prediction interval for $\Y_i(t)$ is obtained taking the $\alpha/2$\textsuperscript{th} and $1 - \alpha/2$\textsuperscript{th} quantiles of the generated $R$ sets of bootstrap replicates of the $i$\textsuperscript{th} predicted response:
\begin{equation*}
\left[ Q_{\alpha/2}^i(t), Q_{1 - \alpha/2}^i(t)\right],
\end{equation*}
where $Q_{\alpha}^i(t)$ denote the $\alpha$\textsuperscript{th} quantile of the bootstrap replicates. We consider the coverage probability deviance (CPD) and interval score (score) metrics to evaluate the performance of the methods:
\begin{align*}
\text{CPD} &= (1 - \alpha) - \frac{1}{n_{\text{test}}} \sum_{i=1}^{n_{\text{test}}} \mathbb{1} \left\{  Q_{\alpha/2}^i(t) \leq \Y_i(t) \leq Q_{1 - \alpha/2}^i(t) \right\}, \\
\text{score} &= \frac{1}{n_{\text{test}}} \sum_{i=1}^{n_{\text{test}}} \bigg\vert \bigg\vert \left[ \left\{ Q_{1 - \alpha/2}^i(t) - Q_{\alpha/2}^i(t) \right\} \right. \\
&+ \frac{2}{\alpha} \left( Q_{\alpha/2}^i(t) - \Y_i(t) \right) \mathbb{1} \left\{ \Y_i(t) < Q_{\alpha/2}^i(t) \right\} \\
&+ \left. \frac{2}{\alpha} \left( \Y_i(t) - Q_{1 - \alpha/2}^i(t) \right) \mathbb{1} \left\{ \Y_i(t) > Q_{1 - \alpha/2}^i(t) \right\} \right] \bigg\vert \bigg\vert_{\mathcal{L}_2}.
\end{align*}
The CPD is the absolute difference between the nominal and empirical coverage probabilities. The small CPD value indicates that the prediction interval constructed by the method covers most of the observations in the test sample. In contrast, its large values indicate the prediction interval fails to cover observations. The interval score simultaneously evaluates the coverage probability and width of the constructed prediction interval, and a smaller interval score corresponds to a sharp prediction interval (more accurate and narrower). Note that we also calculate pointwise prediction intervals for the response variable using the proposed FFLQR by fitting the same model on the generated data for two quantile levels $\tau_1 = \alpha/2$ and $\tau_2 = (1-\alpha/2)$, and compare its performance with the metrics obtained from the bootstrap method.

Before presenting our results, we note that we compare our proposed FFLQR method with the LS, FPLS, and FPC under three models: 
\begin{inparaenum}
\item[1)] the \textit{full model}, where all five predictors are included in the model;
\item[2)] the \textit{true model}, where only the significant predictors in the index set $\mathcal{D} = \lbrace 2, 4, 5 \rbrace$ are included to the model; and 
\item[(3)] the \textit{selected model}, where the variables used in the model are determined using the variable selection procedure discussed in Section~\ref{sec:vsp}.
\end{inparaenum}
To make a logical comparison, throughout the simulations, we set $\tau = 0.5$ under our FFLQR model, and we focus on evaluating function-on-function median regression versus mean regression.

\begin{figure}[!htb]
  \centering
  \includegraphics[width=4.9cm]{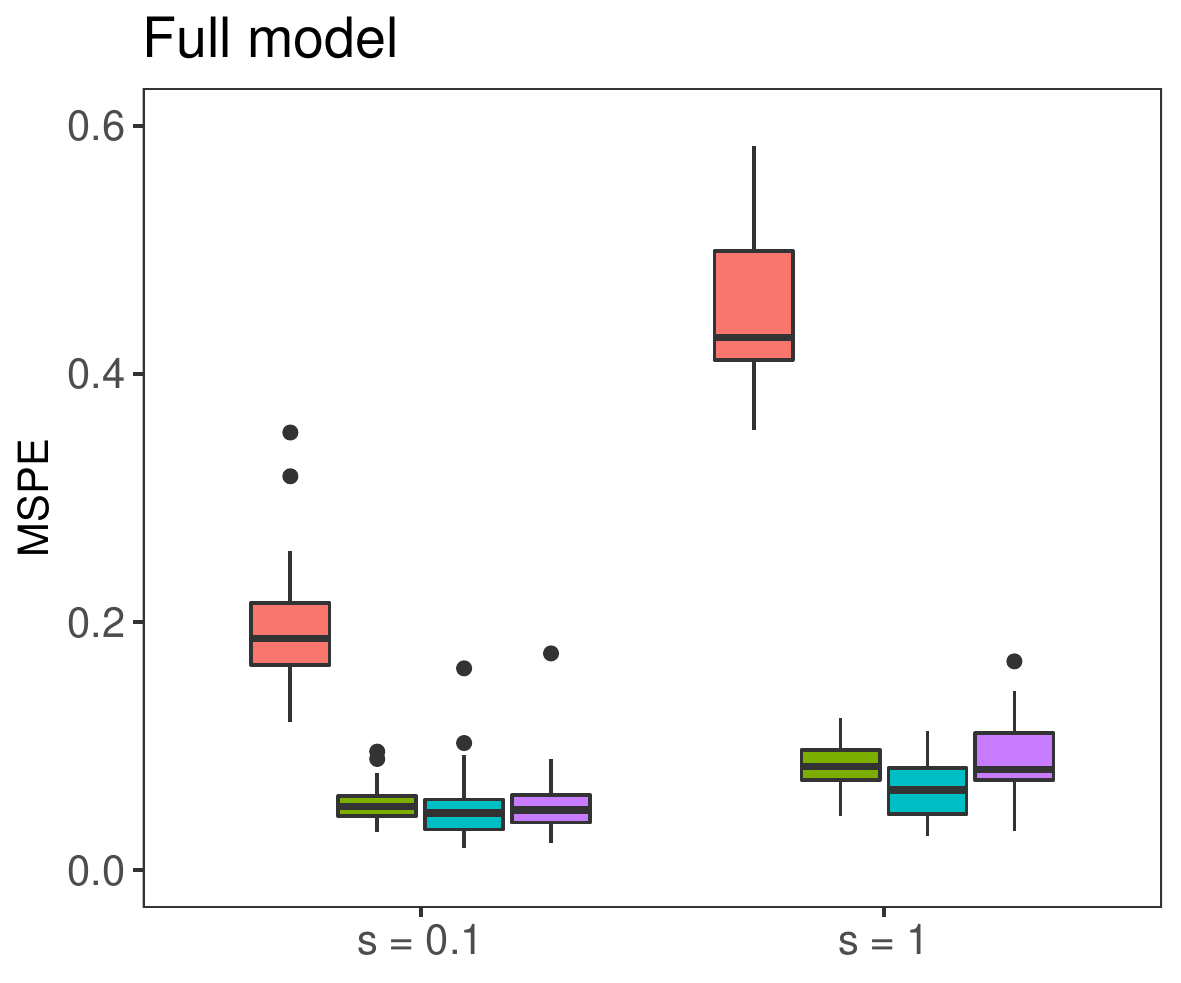}
  \includegraphics[width=4.9cm]{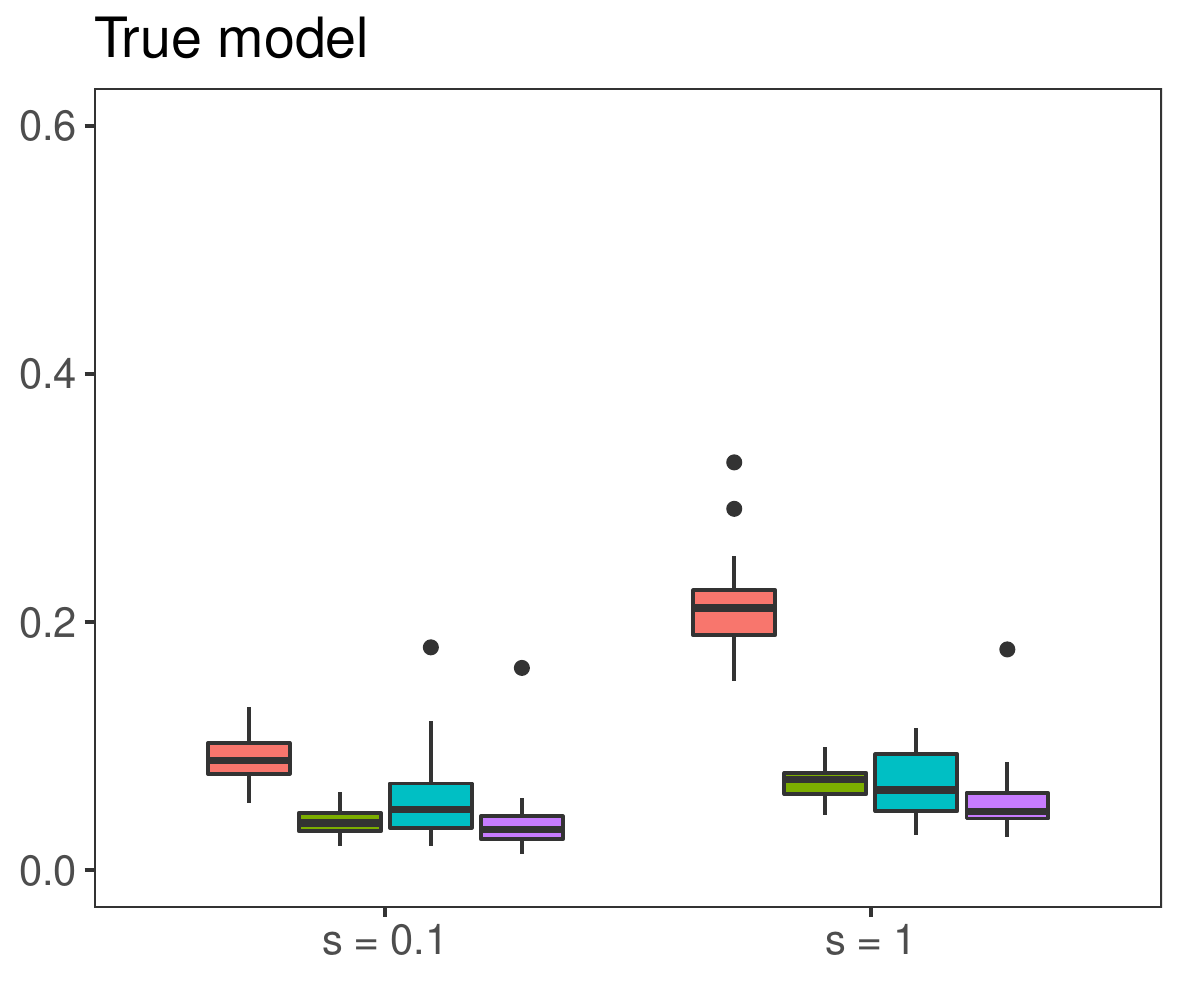}
  \includegraphics[width=4.9cm]{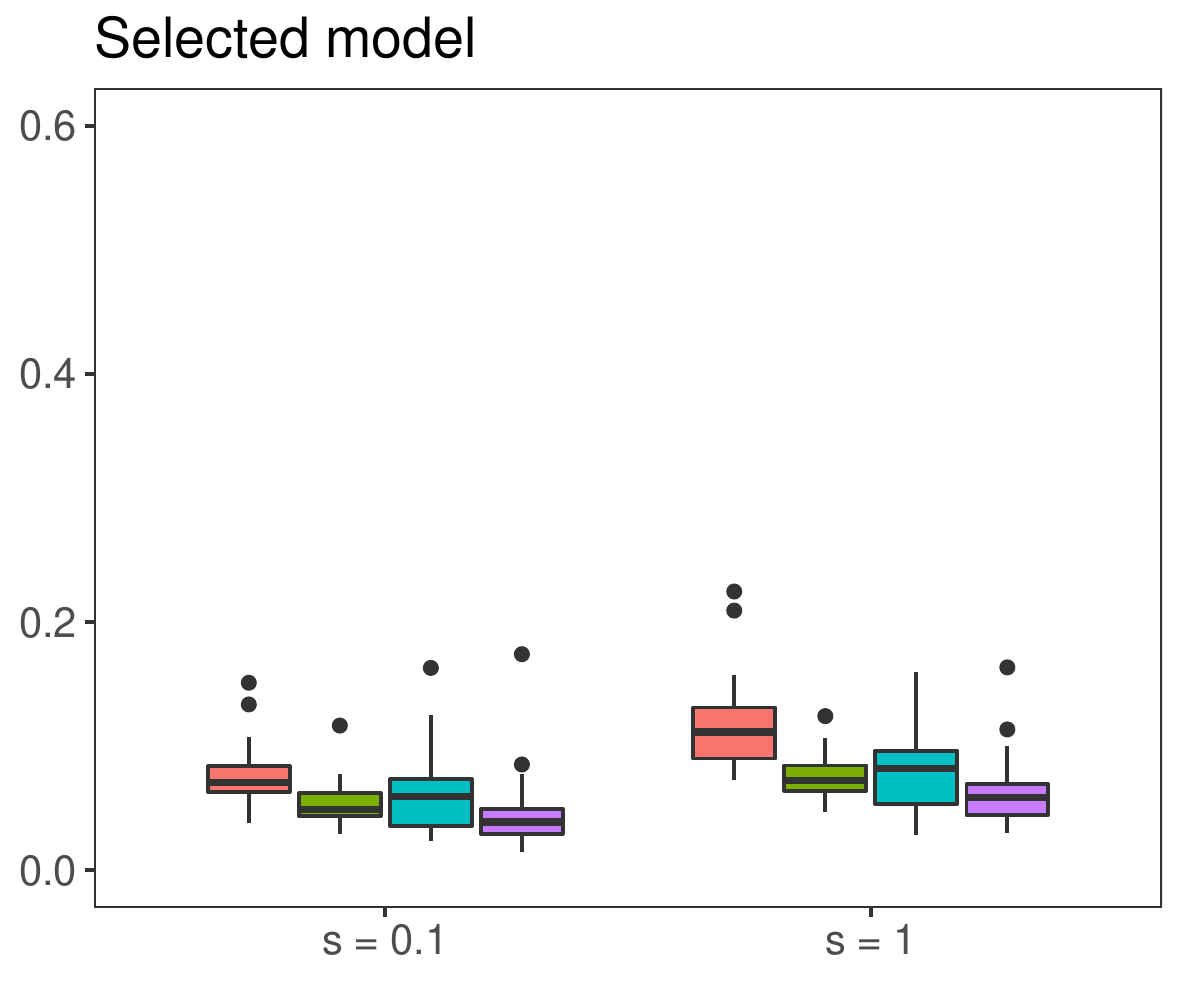}
\\  
  \includegraphics[width=4.9cm]{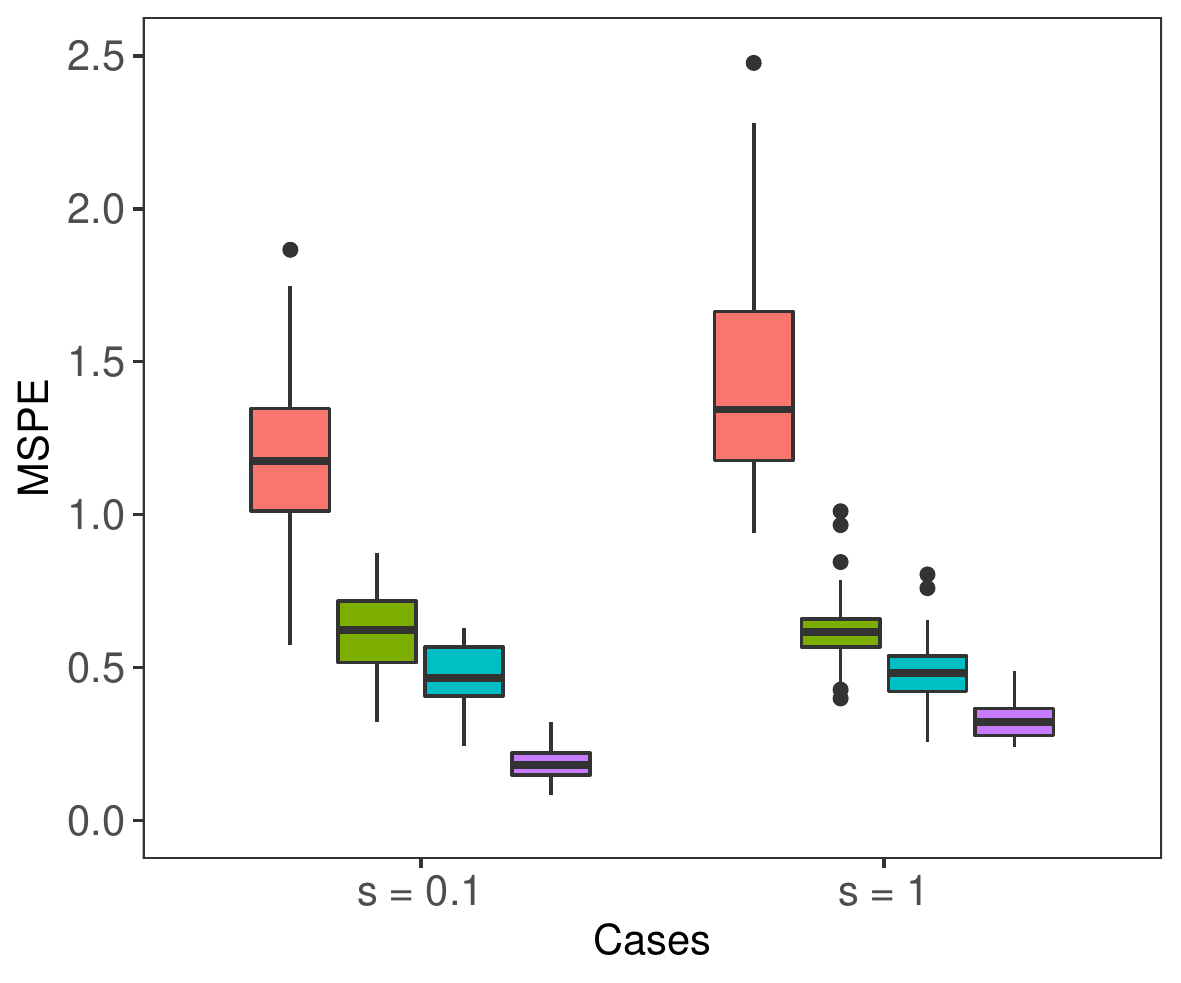}
  \includegraphics[width=4.9cm]{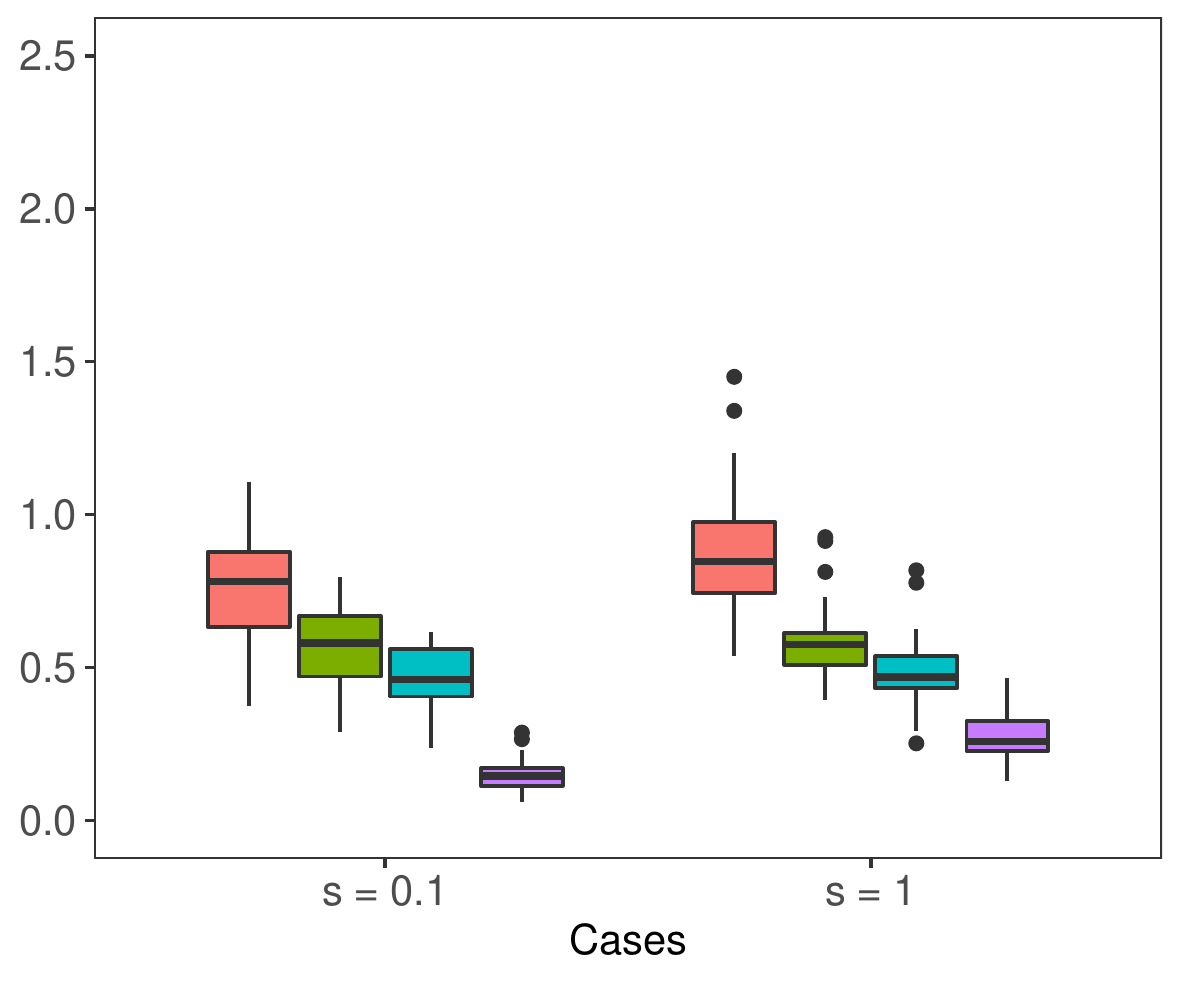}
  \includegraphics[width=4.9cm]{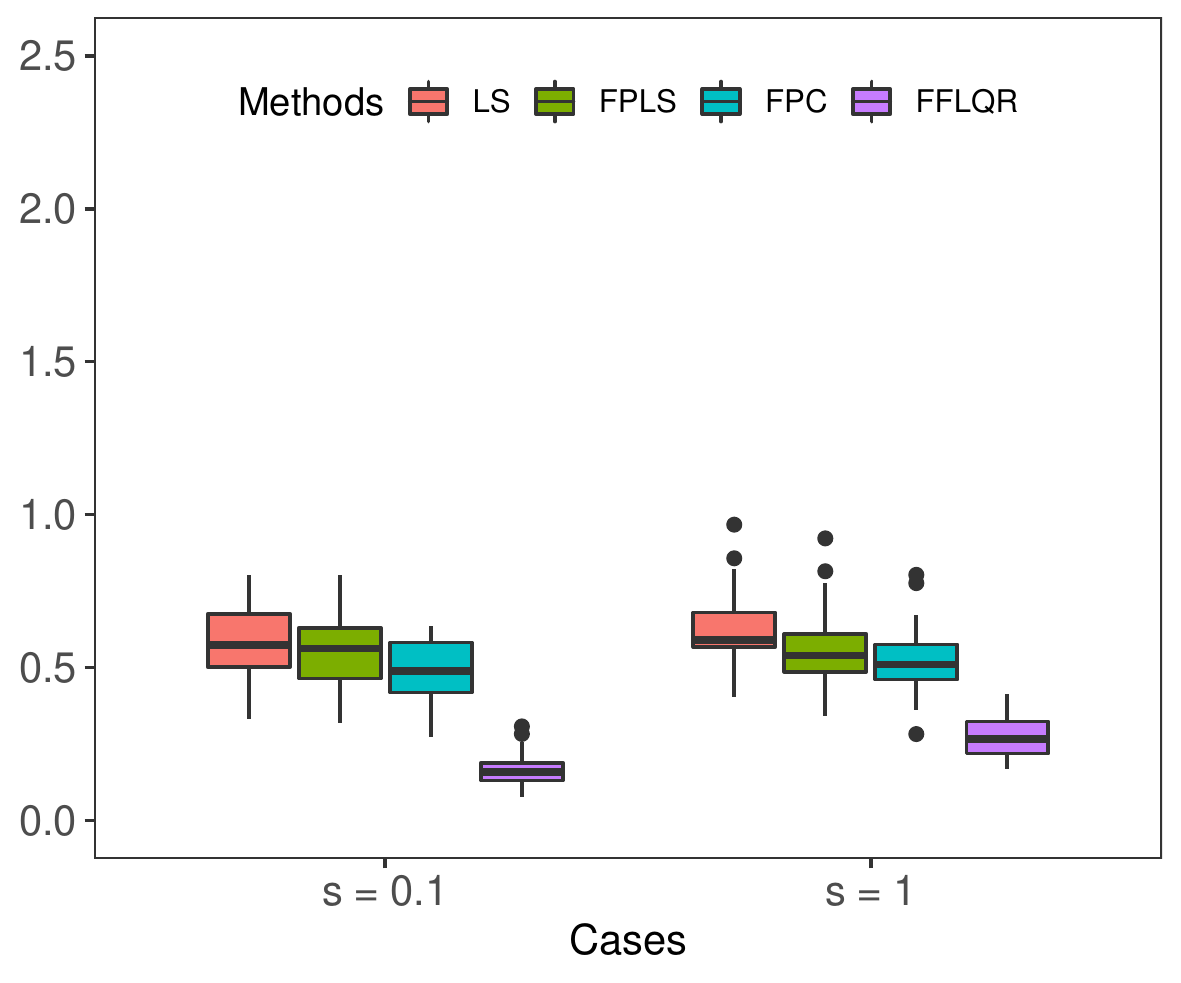}
  \caption{Predictive model performances: Calculated MSPE values of the LS, FPLS, FPC, and FFLQR methods when no outliers are present in the data; full model (first column), true model (second column), and selected model (third column). Data are generated based on two cases where the signal-to-noise ratio levels are $s = 0.1$ and $s = 1$ and two error distributions; $N(0,1)$ (first row) and $\chi^2_{(1)}$ (second row).}
  \label{fig:Fig_2}
\end{figure}

The calculated MSPE values when no outliers are present in the data are given in Figure~\ref{fig:Fig_2}. The results demonstrate that:
\begin{inparaenum}
\item[1)] Both true and selected models produce significantly smaller MSPE values than those of full models for all methods. An interesting but not too surprising result produced by our simulations is that, compared with true models, selected models tend to produce smaller MSPEs. This result could be because some true coefficient functions have a smaller effect on the response function than others. In such a case, to estimate the true model, the functional predictors corresponding to these coefficient functions have to be estimated, which increases the number of coefficients to be estimated and reduces the prediction accuracy. See, for example \cite{LuoQi} for similar results.
\item[2)] For all settings, all the methods produce larger MSPE values with increasing signal-to-noise ratio levels.
\item[(3)] When the error terms follow symmetric distribution (i.e., $N(0,1)$), the proposed FFLQR model produces competitive performance compared with the existing methods. However, when the errors follow the non-normal $\chi^2_{(1)}$ distribution, the proposed method significantly outperforms the LS, FPLS, and FPC methods.
\end{inparaenum}

\begin{figure}[!htbp]
  \centering
  \includegraphics[width=4.4cm]{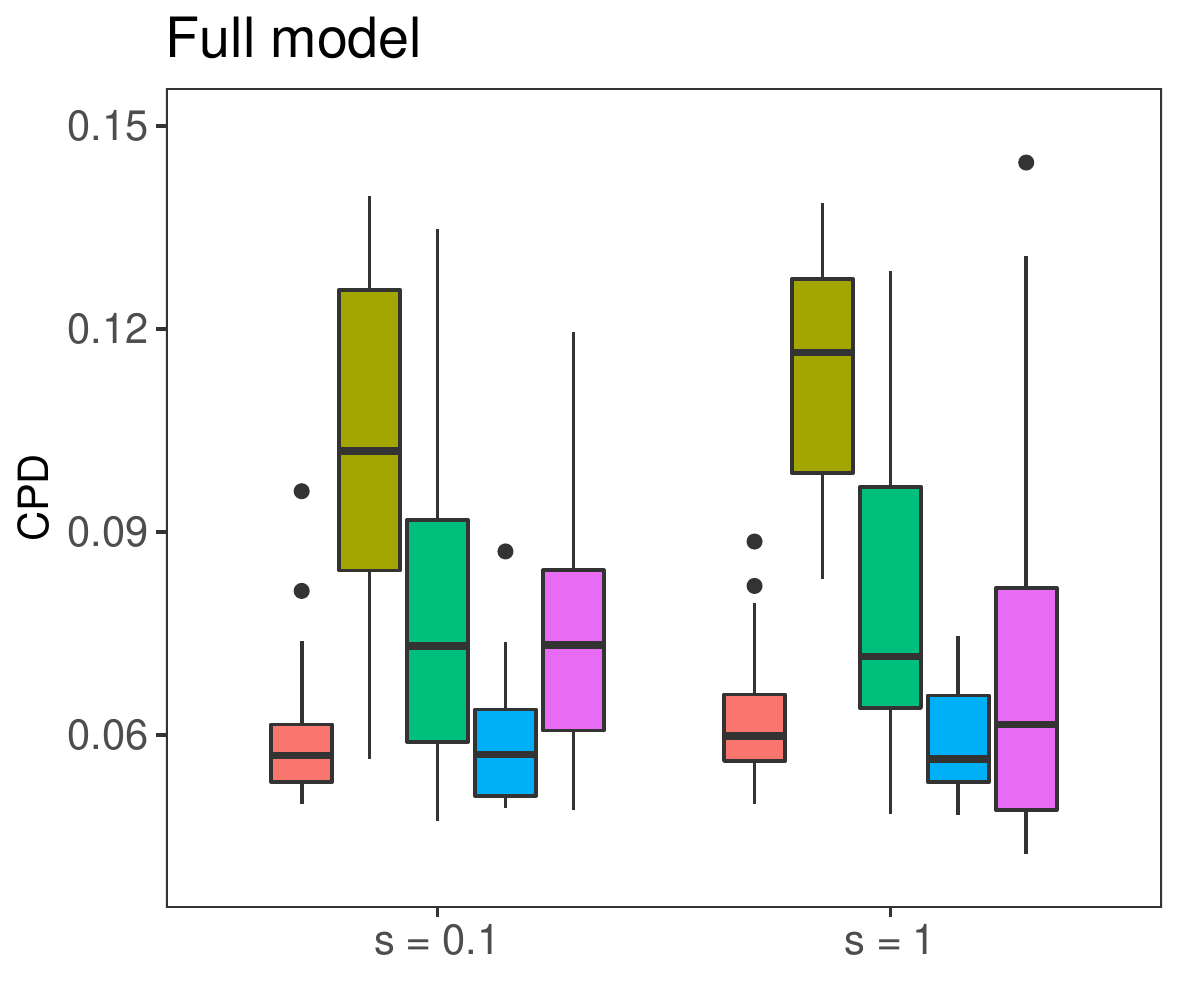}
  \includegraphics[width=4.4cm]{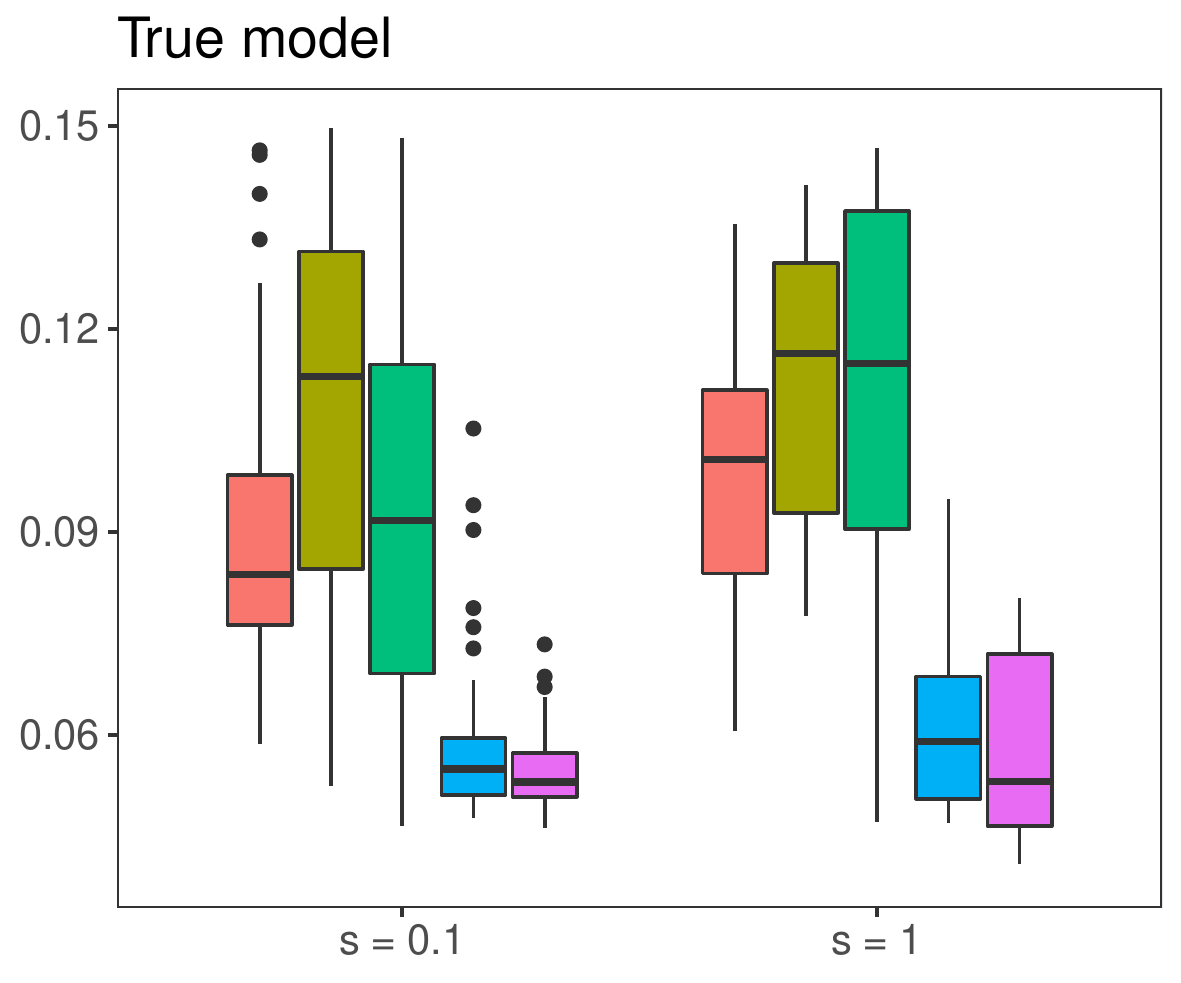}
  \includegraphics[width=4.4cm]{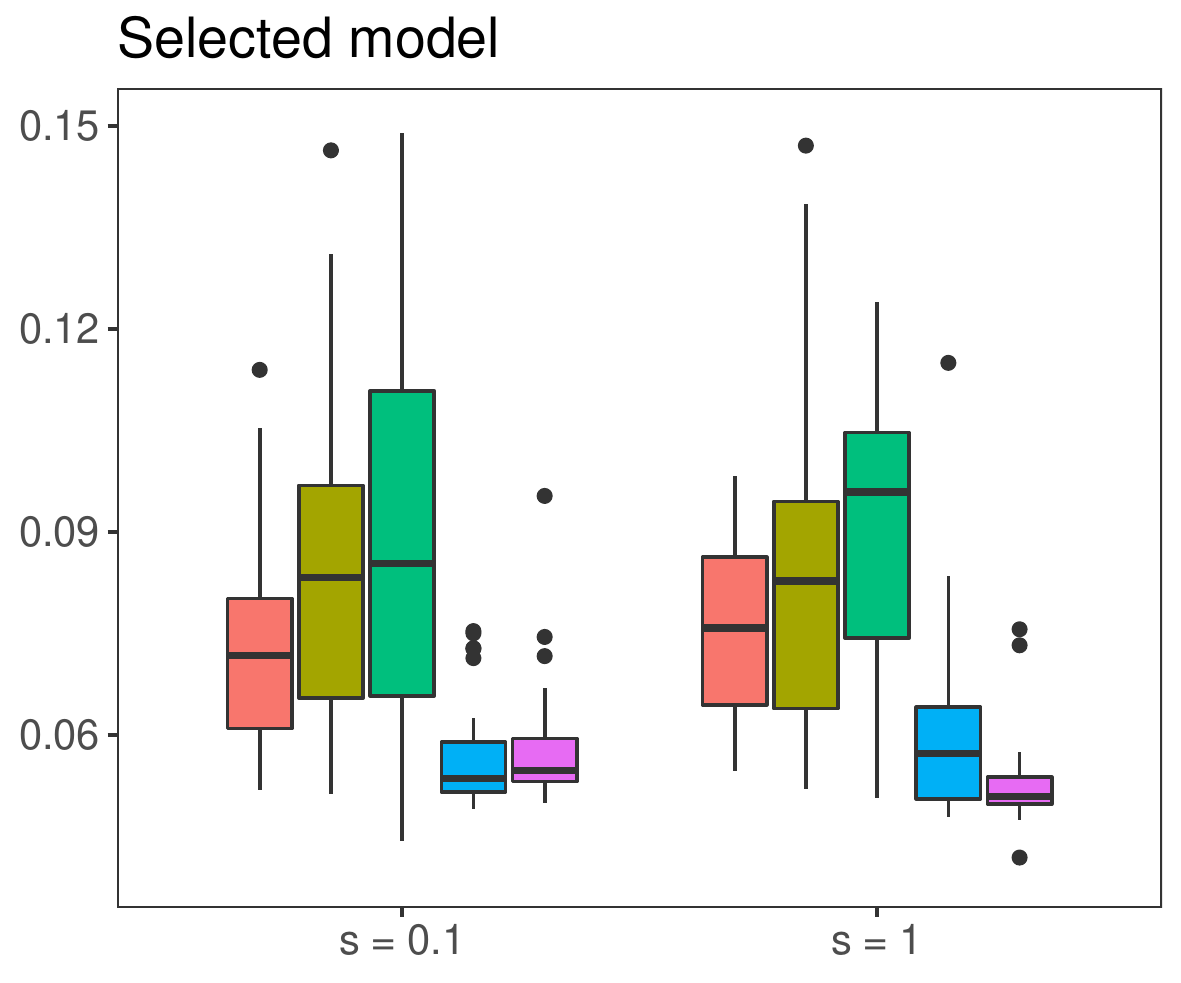}
\\  
  \includegraphics[width=4.4cm]{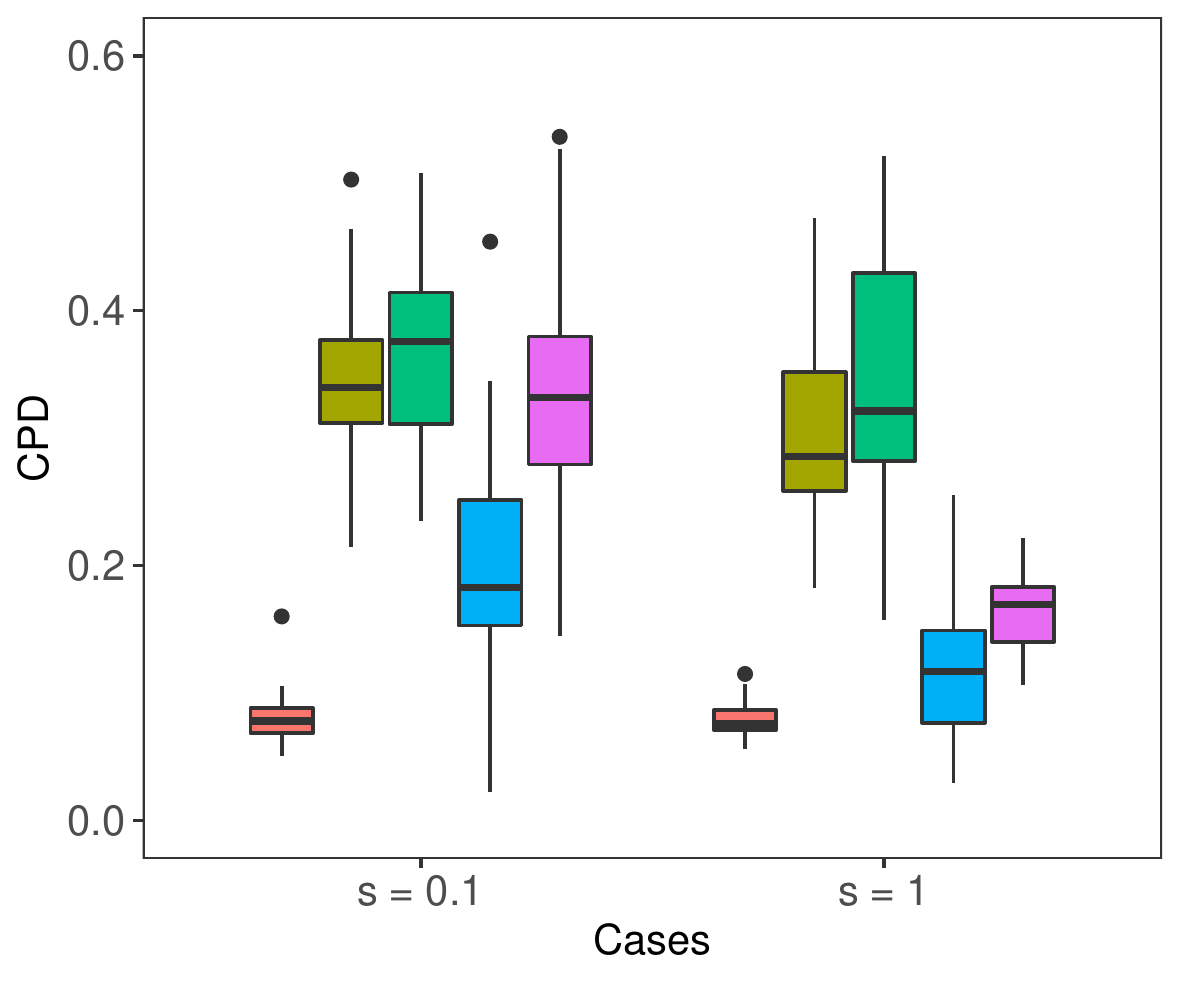}
  \includegraphics[width=4.4cm]{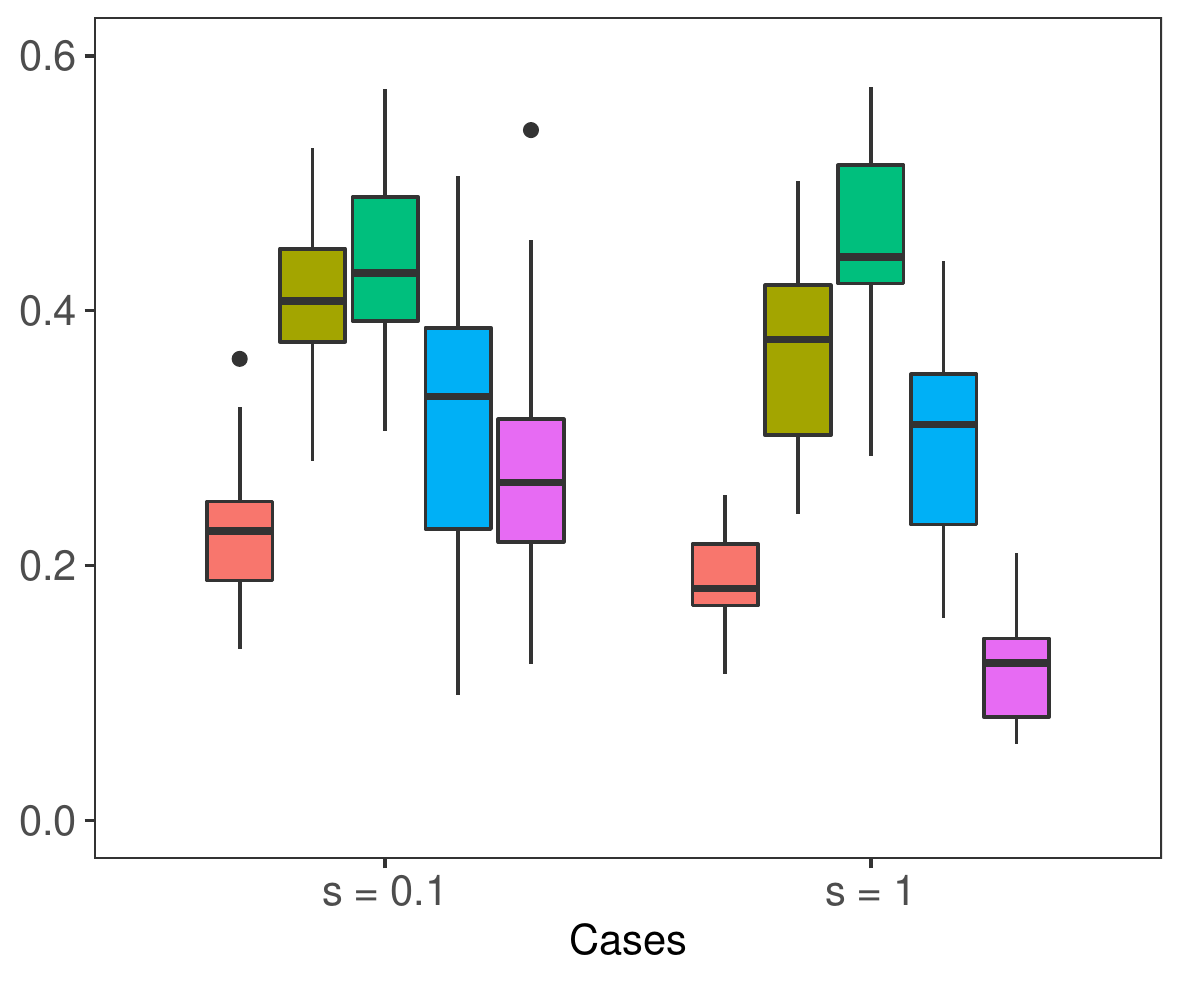}
  \includegraphics[width=4.4cm]{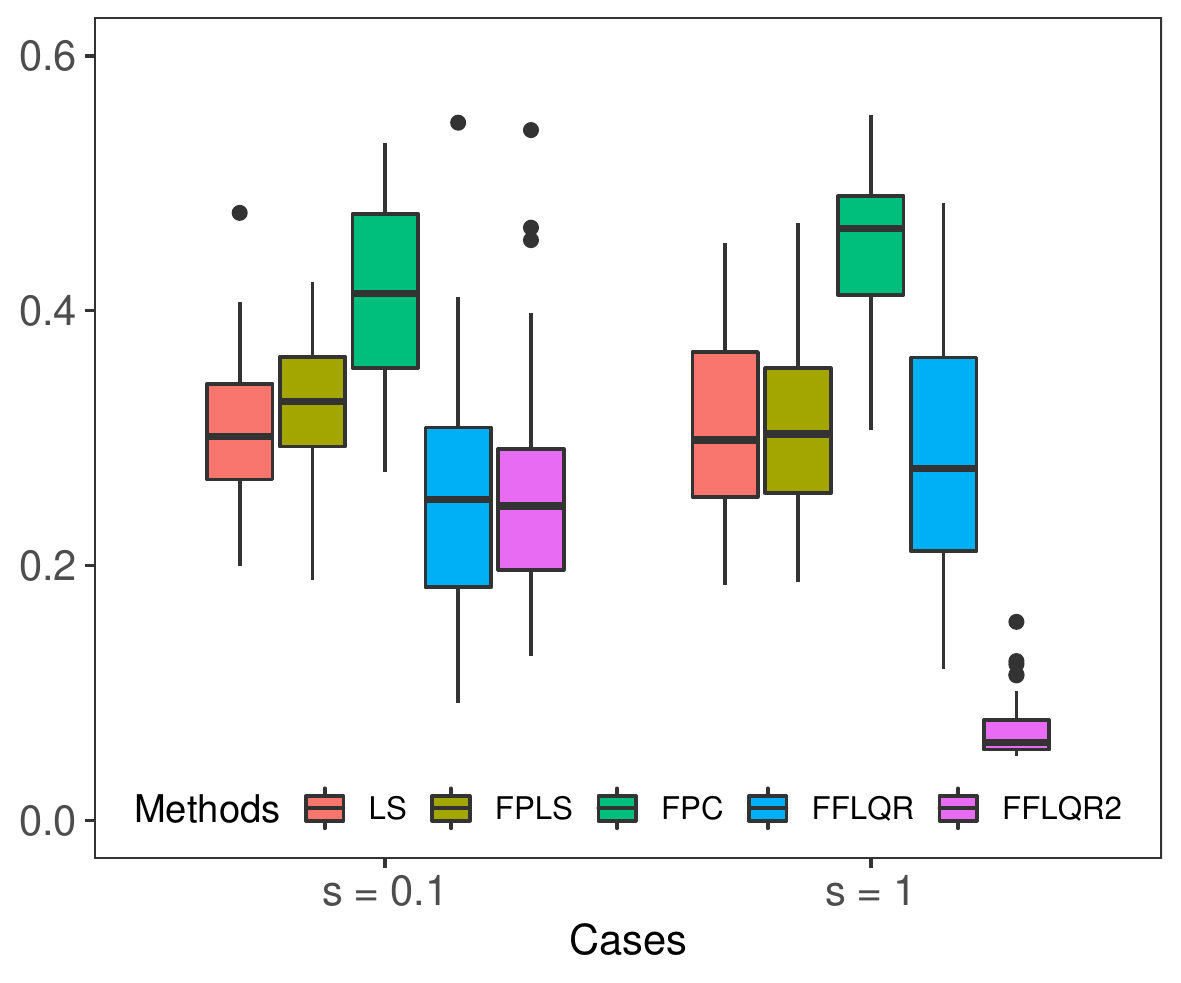}
  \caption{\textbf{Predictive model performances}: Calculated CPD values of the LS, FPLS, FPC, and FFLQR methods when no outliers are present in the data; full model (first column), true model (second column), and selected model (third column). Data are generated based on two cases where the signal-to-noise ratio levels are $s = 0.1$ and $s = 1$ and two error distributions; $N(0,1)$ (first row) and $\chi^2_{(1)}$ (second row). ``FFLQR2'' denotes the performance metrics obtained directly from the proposed FFLQR.}
  \label{fig:Fig_3}
\end{figure}

\begin{figure}[!htb]
  \centering
  \includegraphics[width=5.3cm]{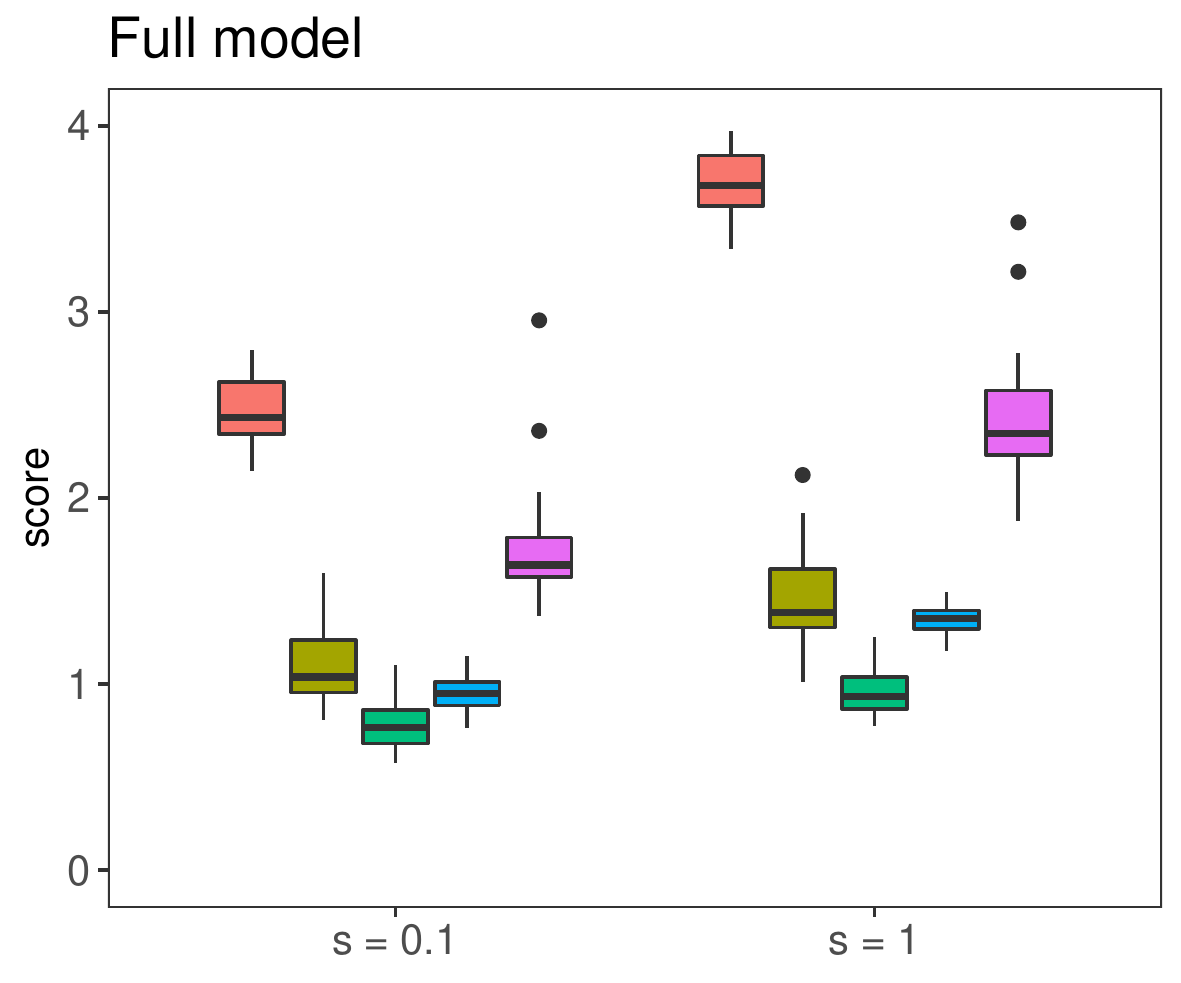}
  \includegraphics[width=5.3cm]{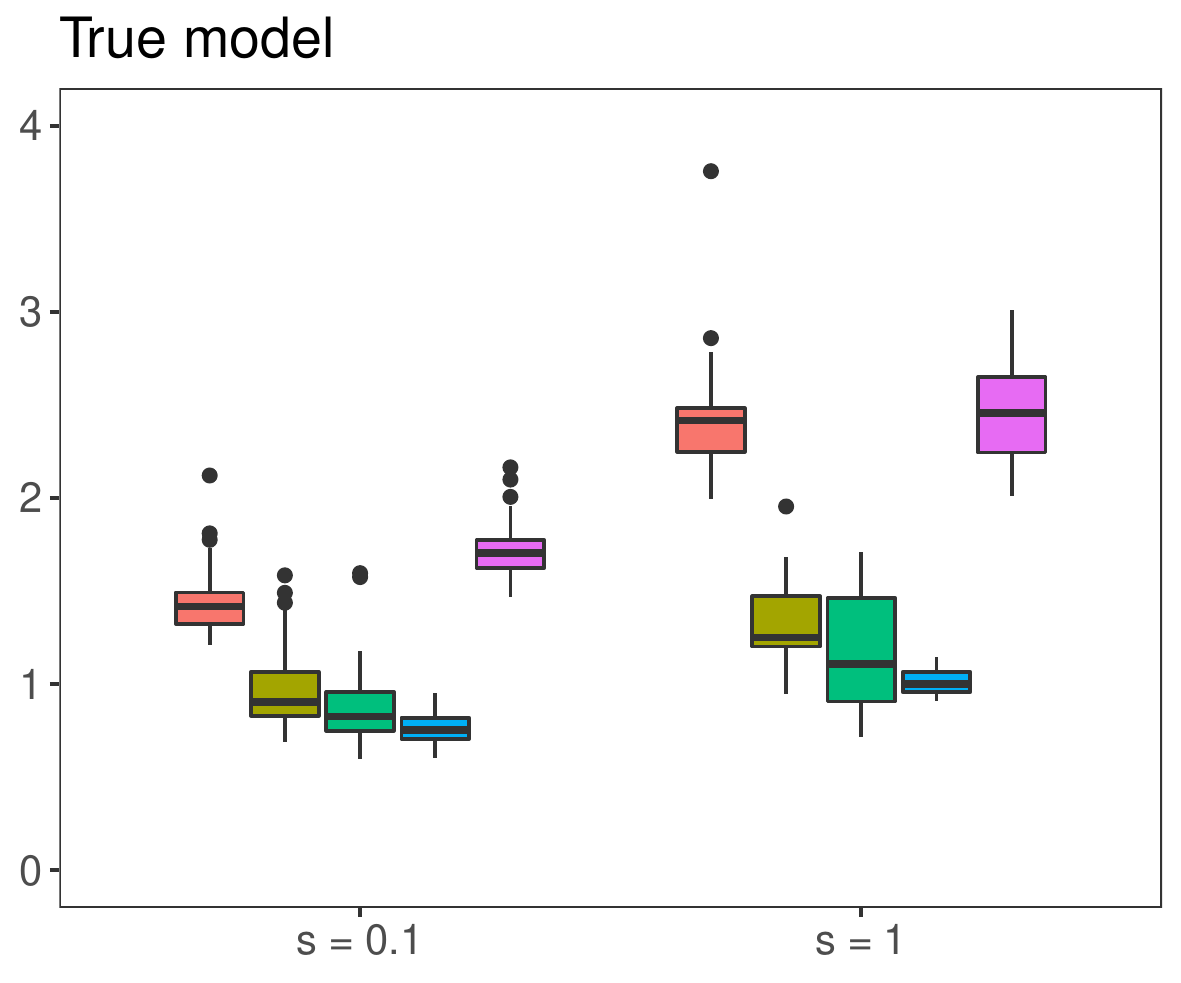}
  \includegraphics[width=5.3cm]{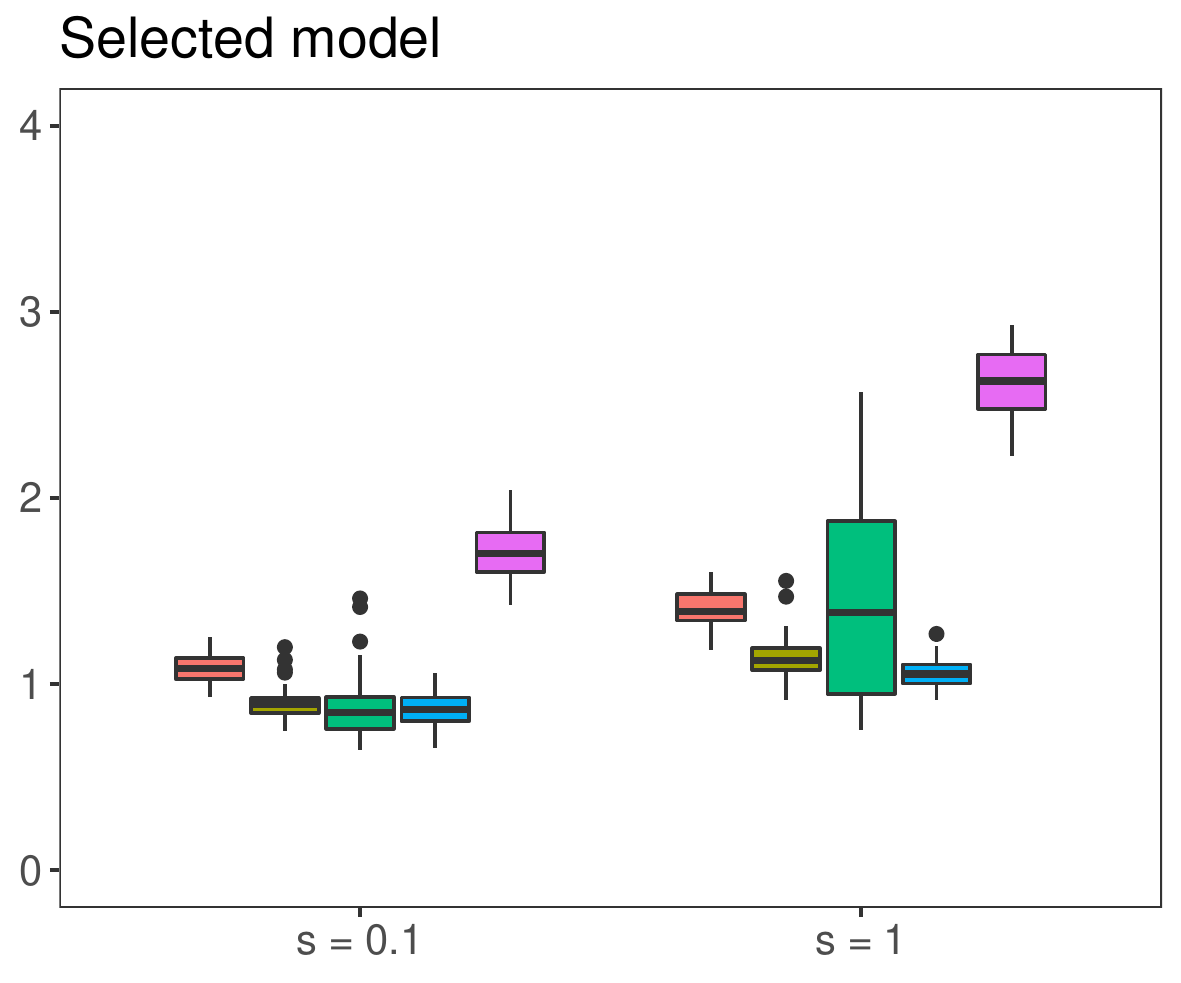}
\\  
  \includegraphics[width=5.3cm]{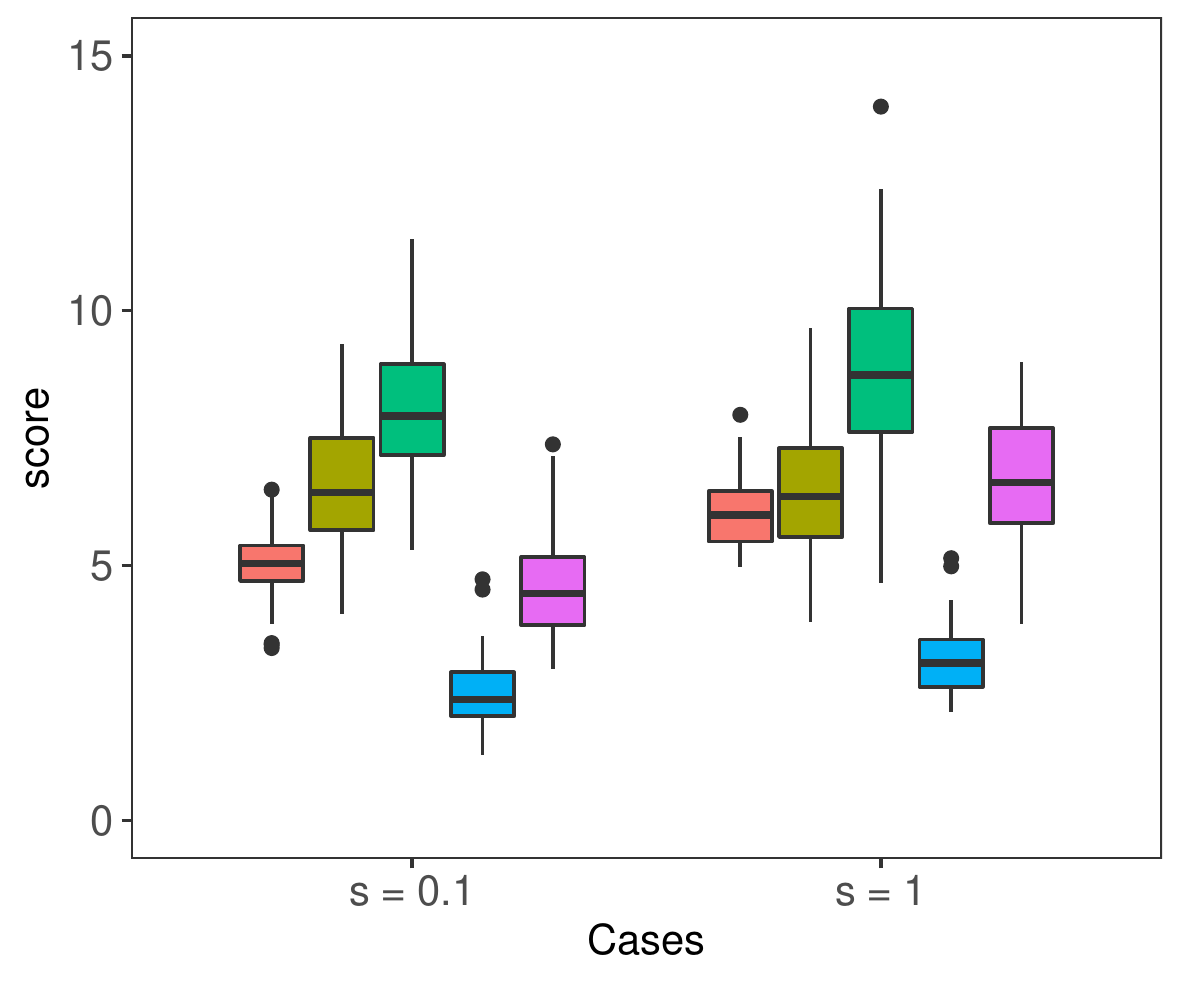}
  \includegraphics[width=5.3cm]{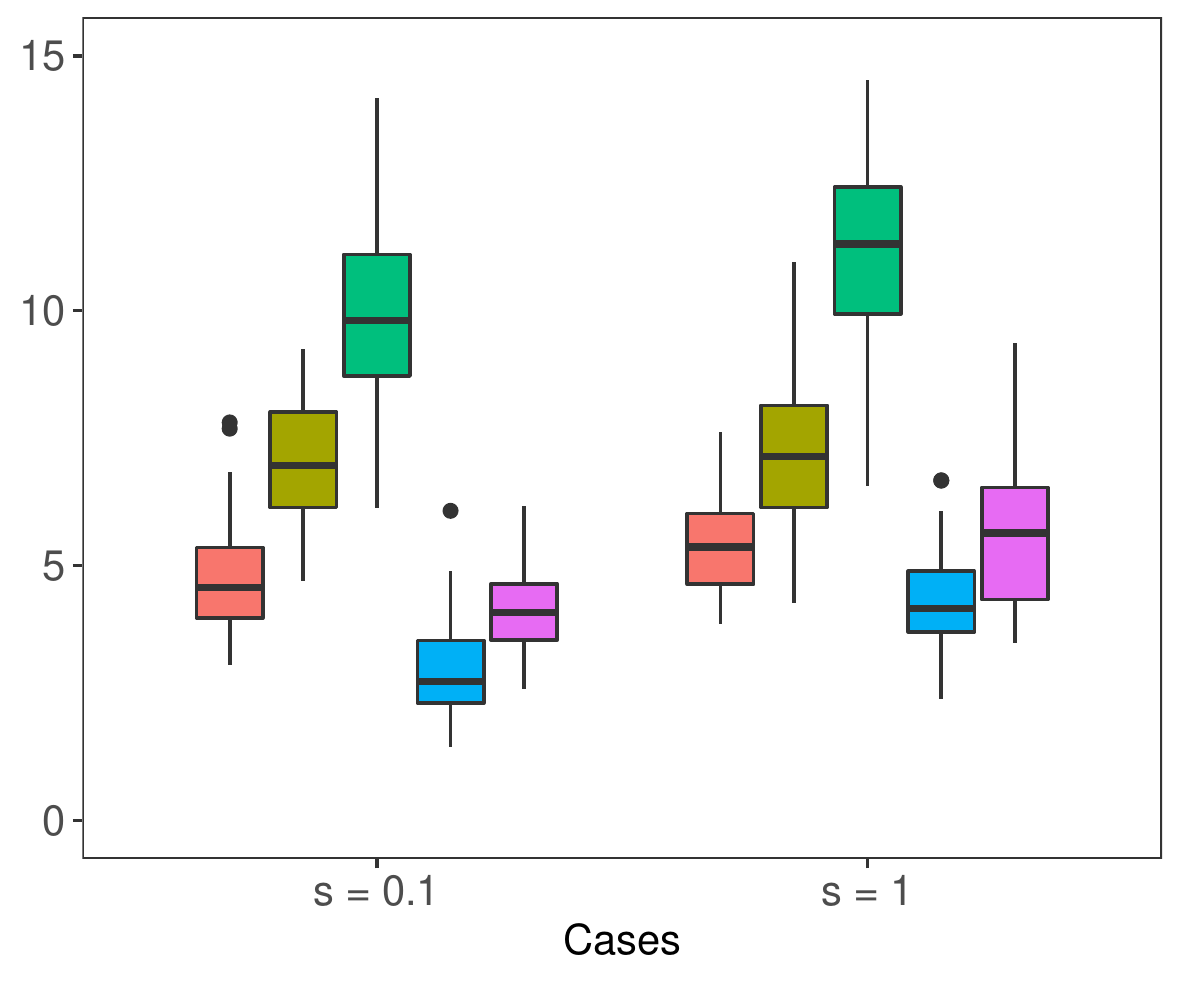}
  \includegraphics[width=5.3cm]{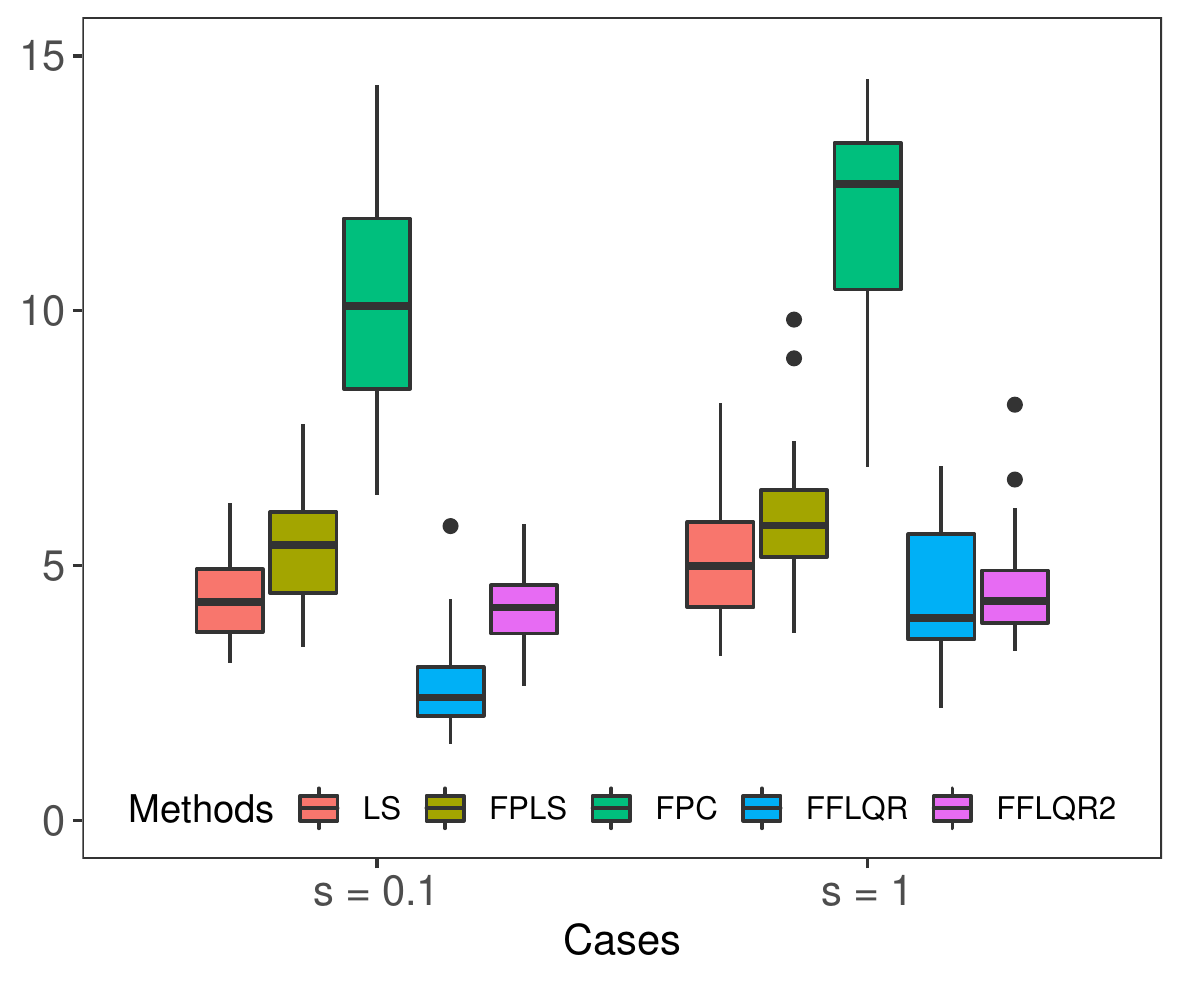}
  \caption{\textbf{Predictive model performances}: Calculated score values of the LS, FPLS, FPC, and FFLQR methods when no outliers are present in the data; full model (first column), true model (second column), and selected model (third column). Data are generated based on two cases where the signal-to-noise ratio levels are $s = 0.1$ and $s = 1$ and two error distributions; $N(0,1)$ (first row) and $\chi^2_{(1)}$ (second row). ``FFLQR2'' denotes the performance metrics obtained directly from the proposed FFLQR.}
  \label{fig:Fig_4}
\end{figure}

When no outliers are present in the data, the calculated CPD and interval score values are shown in Figures~\ref{fig:Fig_3} and~\ref{fig:Fig_4}, respectively. The proposed method produces improved CPD and score values, among others, under the symmetric errors from these figures. This result indicates that the proposed method produces more accurate bootstrap prediction intervals with narrower prediction interval lengths compared with other methods. The prediction intervals obtained directly from the proposed FFLQR model (not bootstrap-based) produce competitive or even better CPD values than the bootstrap-based prediction intervals constructed by the proposed method. However, they have generally larger score values than the bootstrap prediction intervals. In other words, the non-bootstrap-based prediction intervals are wider than those of bootstrap-based prediction intervals.

\begin{figure}[!htb]
  \centering
  \includegraphics[width=5.7cm]{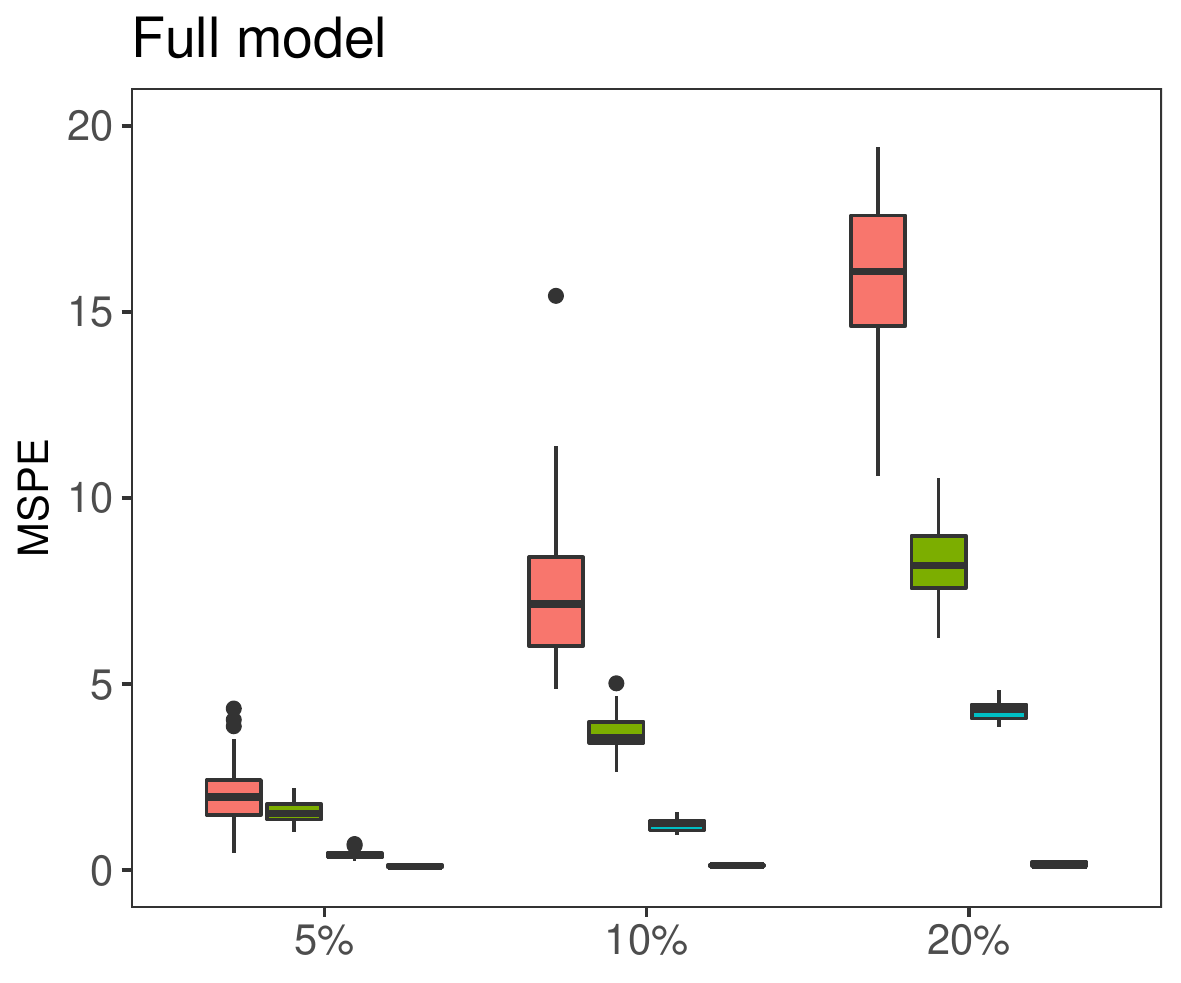}
  \includegraphics[width=5.7cm]{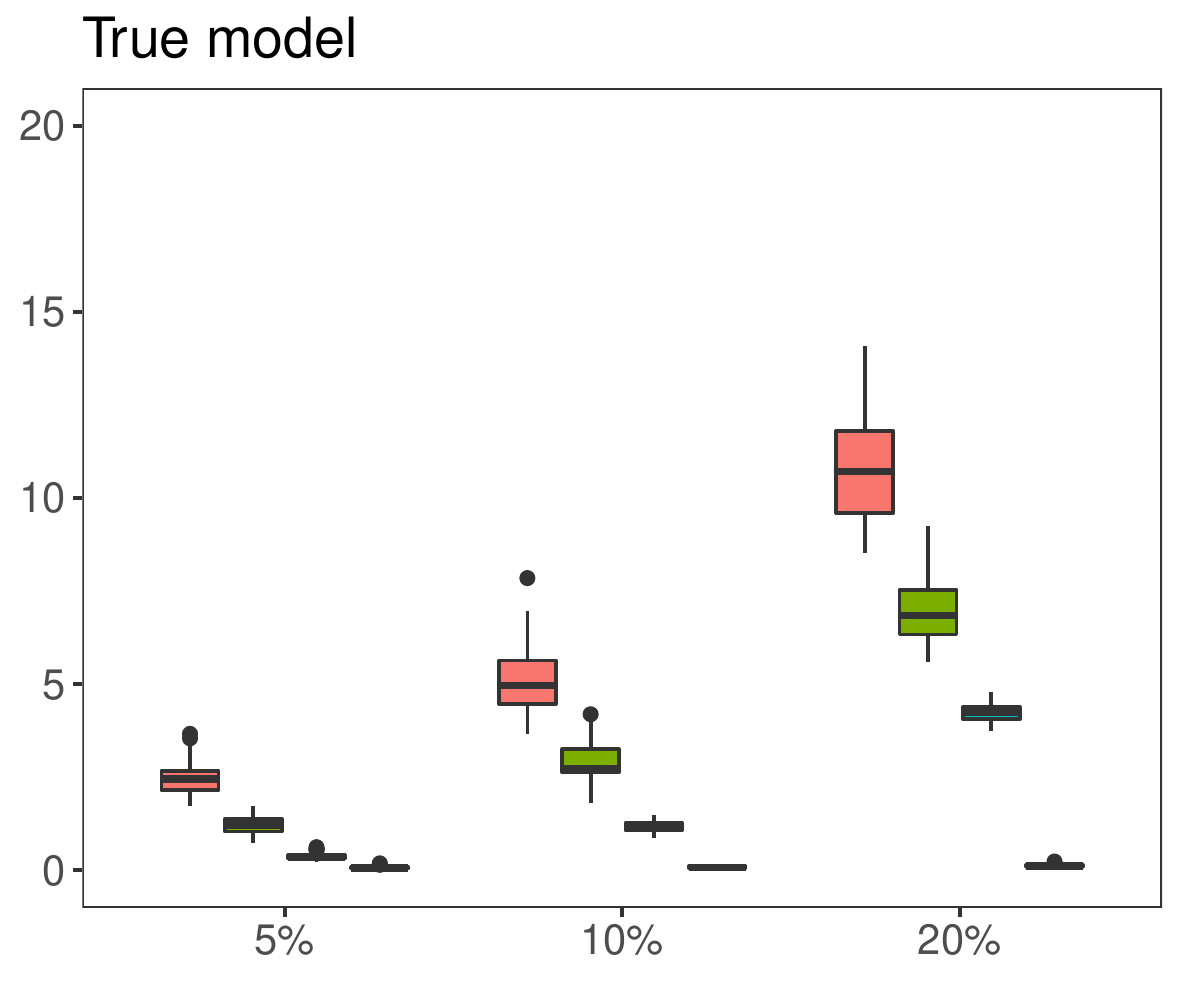}
  \includegraphics[width=5.7cm]{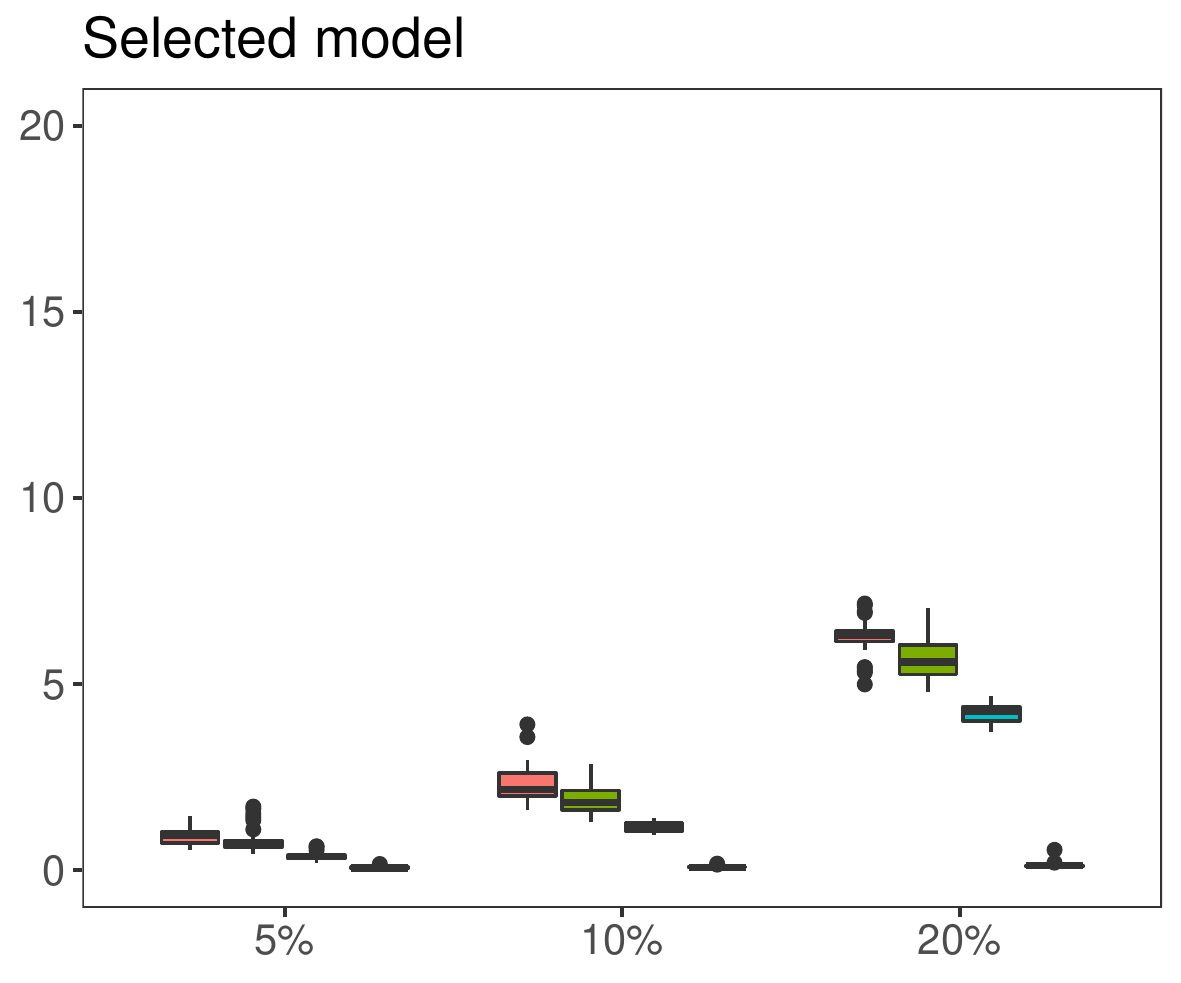}
\\  
  \includegraphics[width=5.7cm]{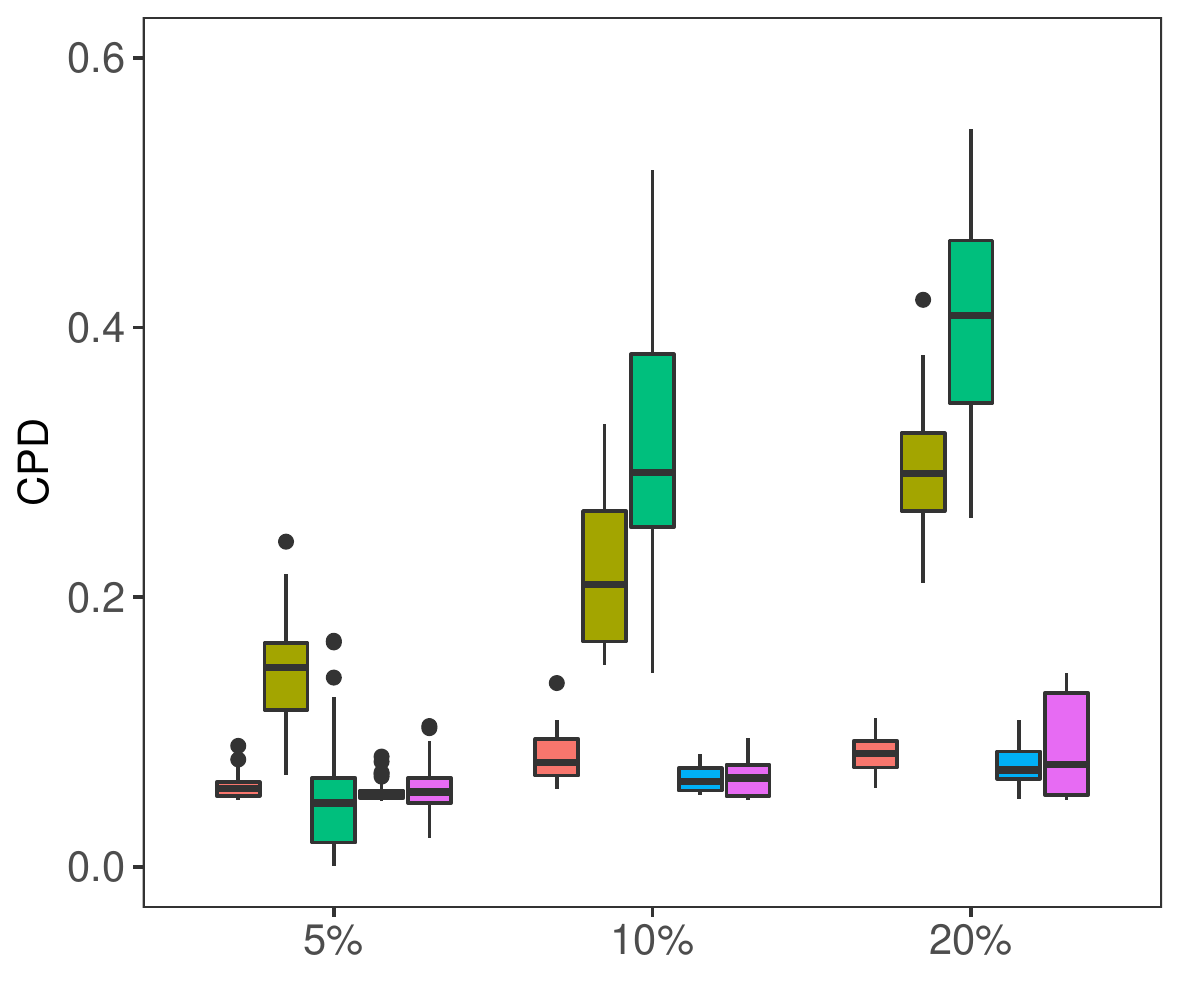}
  \includegraphics[width=5.7cm]{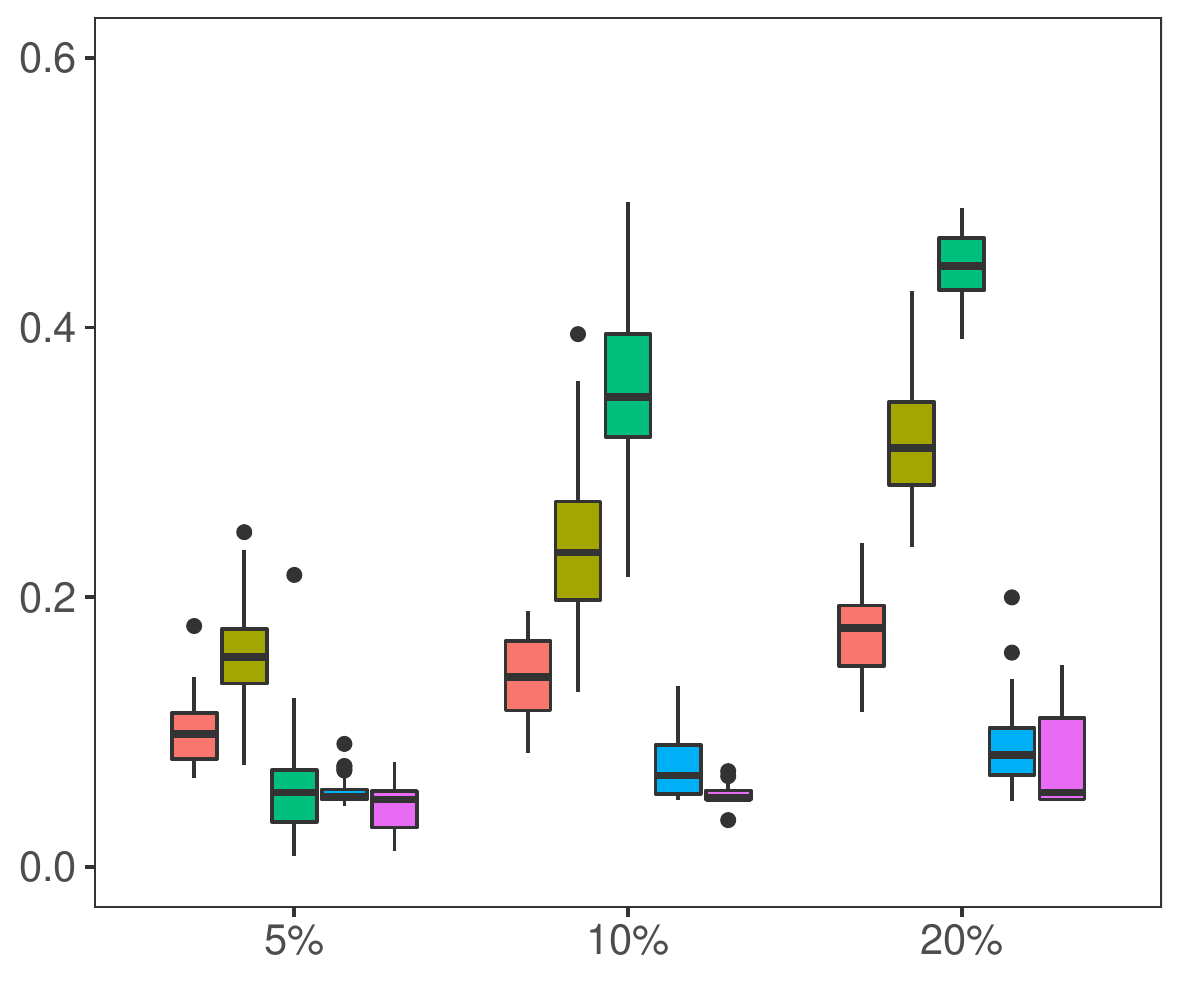}
  \includegraphics[width=5.7cm]{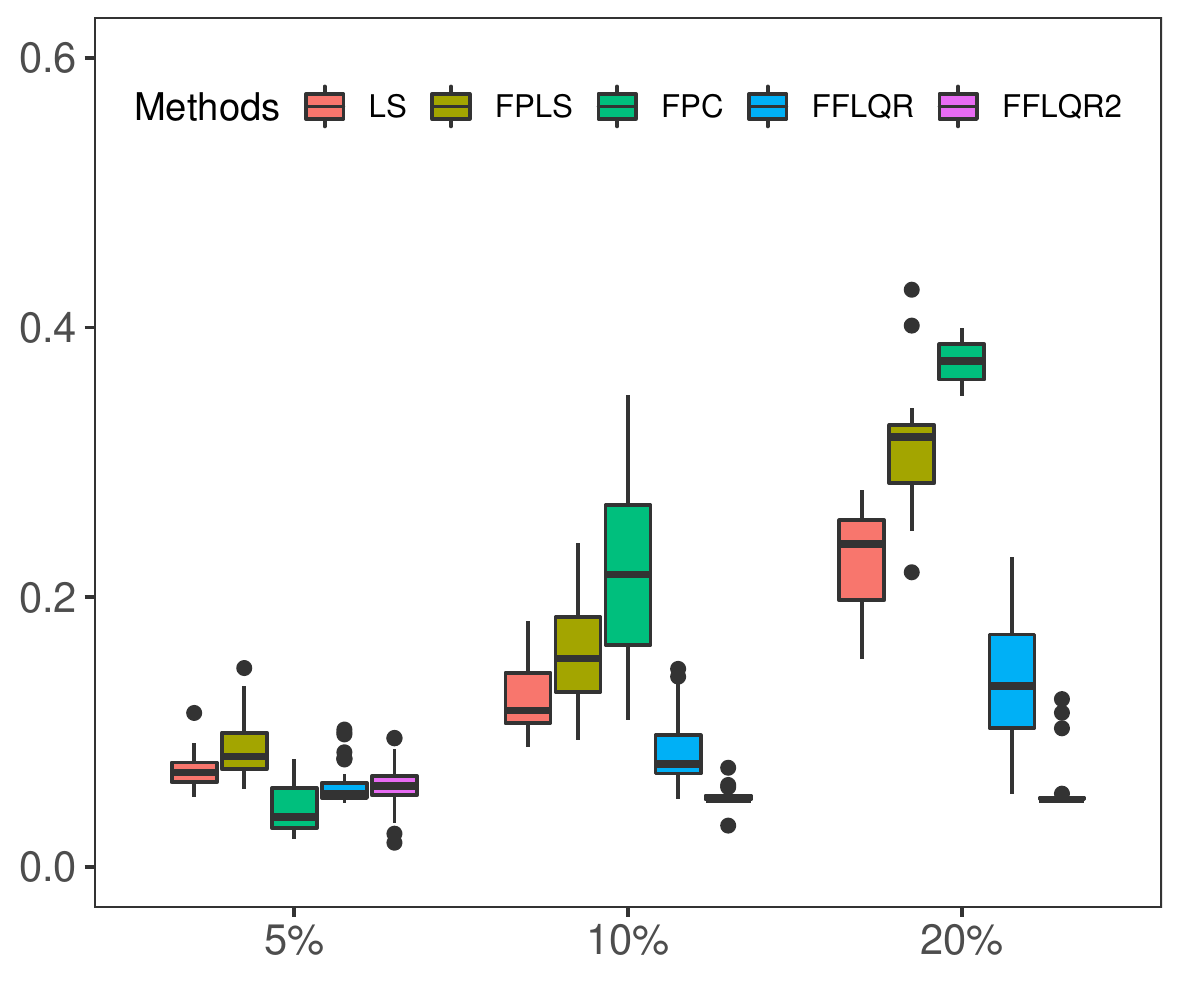}
\\  
  \includegraphics[width=5.7cm]{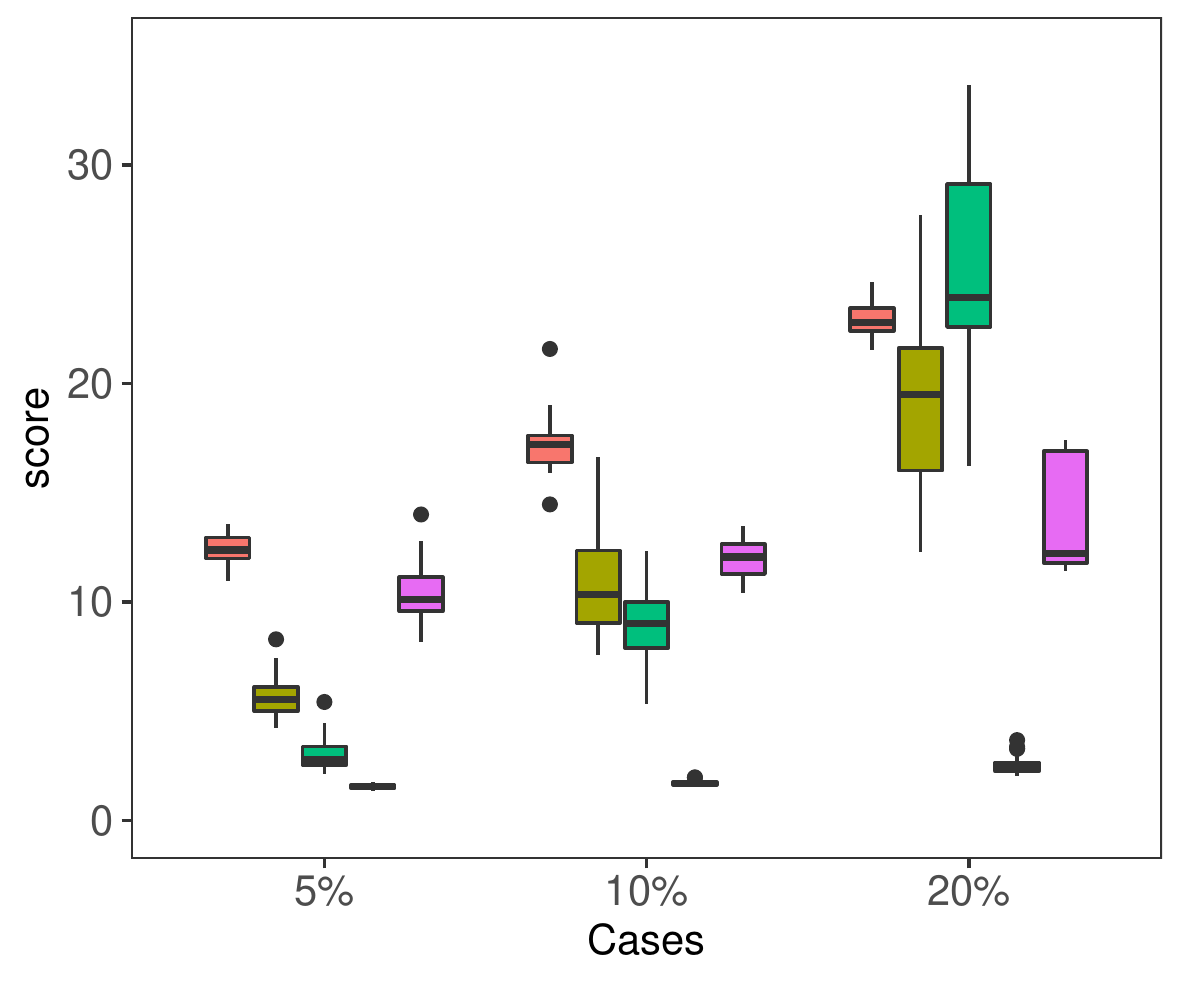}
  \includegraphics[width=5.7cm]{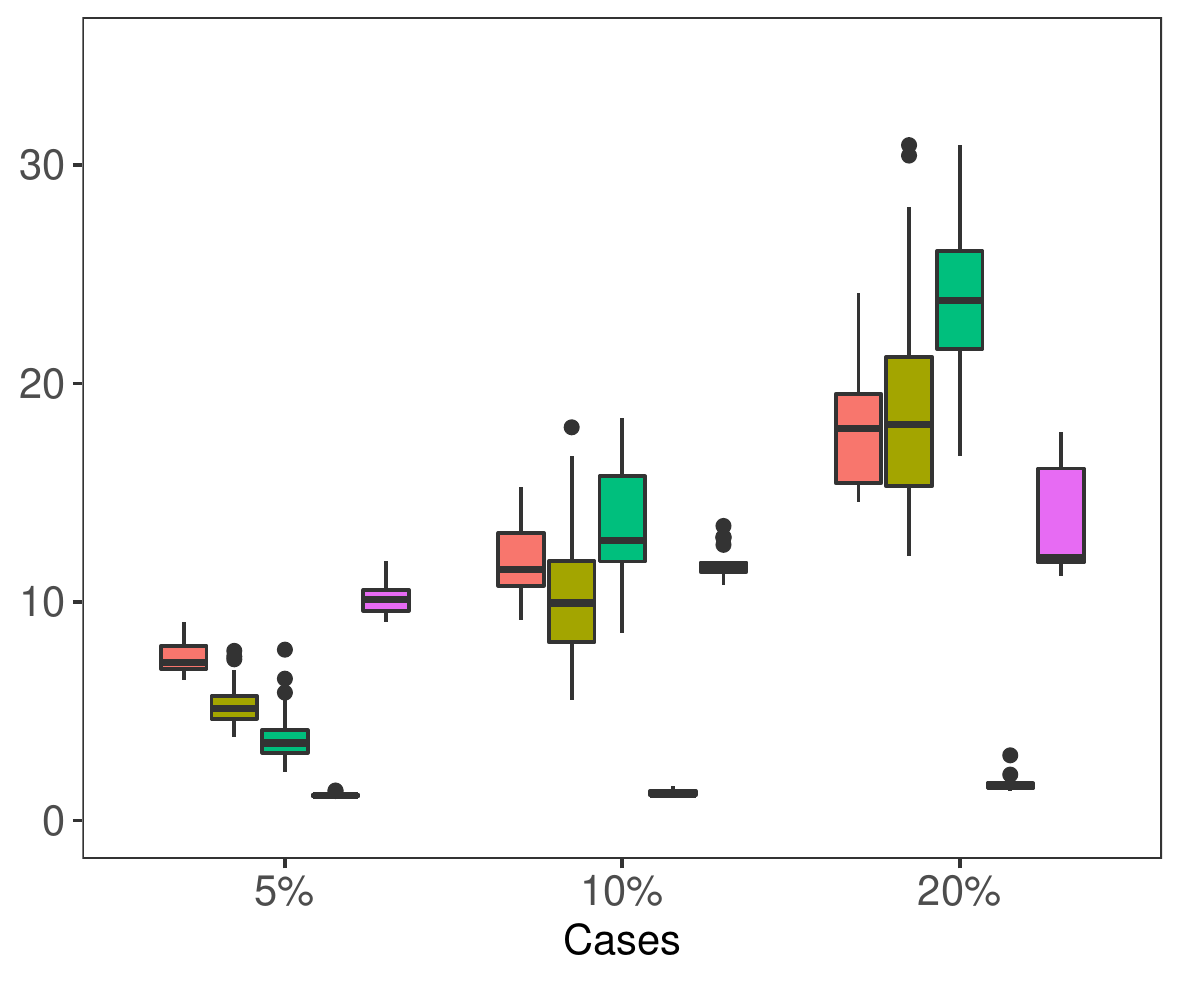}
  \includegraphics[width=5.7cm]{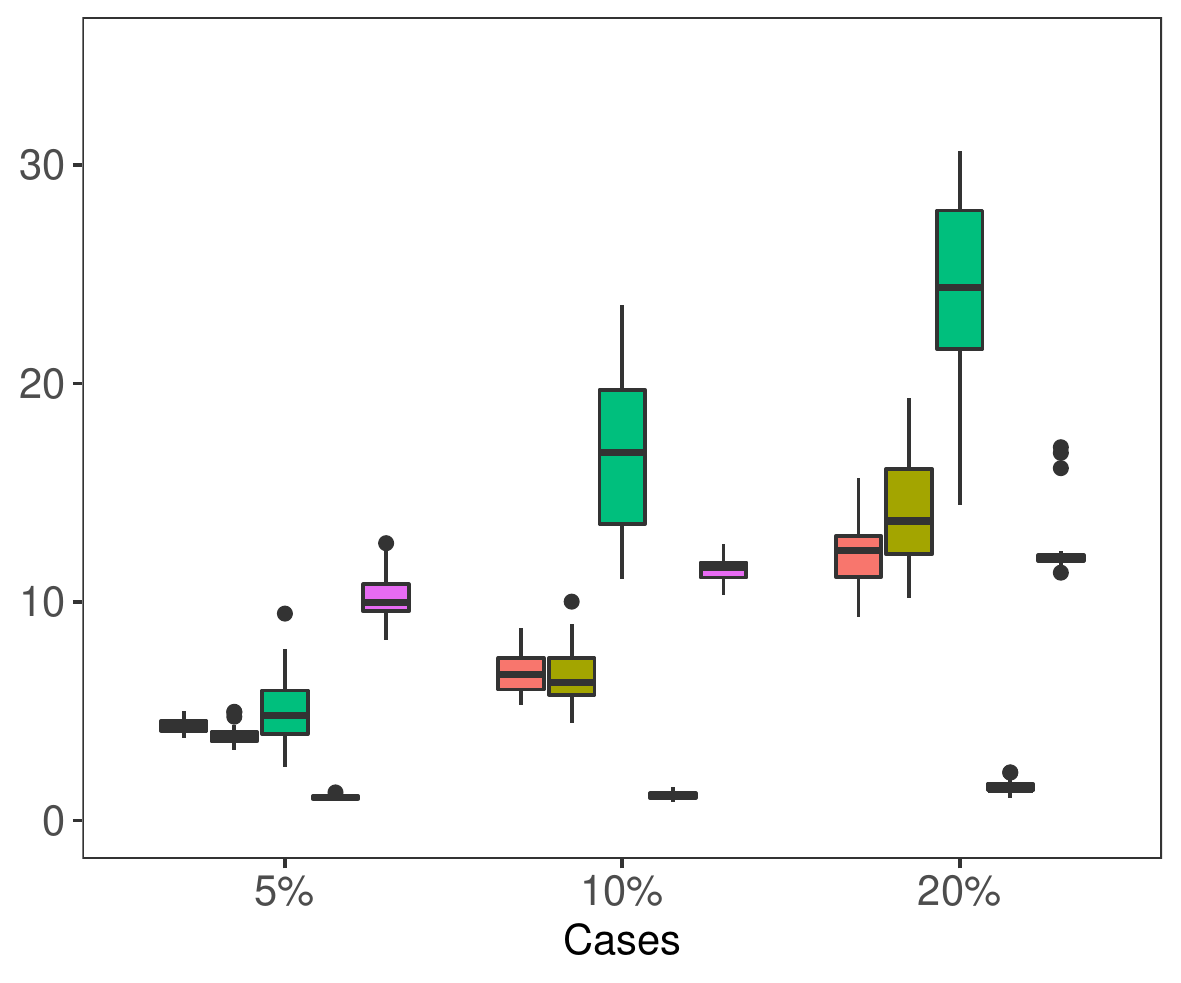}
  \caption{\textbf{Predictive model performances}: Calculated MSPE (first row), CPD (second row), and score (third row) values of the LS, FPLS, FPC, and FFLQR methods when $[ 5\%, 10\%, 20\% ]$ of the data are contaminated by magnitude outliers; full model (first column), true model (second column), and selected model (third column). Data are generated when the signal-to-noise ratio level is $[s = 1$ and the error terms follow $N(0,1)$ distribution. ``FFLQR2'' denotes the interval scores obtained directly from the proposed FFLQR.}
  \label{fig:Fig_5}
\end{figure}

The calculated MSPE, CPD, and interval score values, when the data are contaminated by outliers, are presented in Figure~\ref{fig:Fig_5}. The results indicate that the proposed method produces smaller prediction errors than the existing methods for all models and contamination levels. Also, the proposed method produces significantly smaller bootstrap-based interval score values than those obtained by LS, FPLS, and FPC methods. The non-bootstrap-based prediction intervals produce competitive CPD values with larger score values (i.e., wider prediction interval lengths) to the bootstrap-based prediction intervals. Therefore, the bootstrap procedure seems a better approach to construct prediction intervals.

We also compare the proposed method with LS, FPLS, and FPC methods regarding their computing times. In doing so, the data are generated under $N(0,1)$ errors with signal-to-ratio level $\sigma = 1$. Then, all the methods under the selected model are performed, and the computing times of the methods are recorded. Our results indicate that the proposed method requires more computing time compared with the other three methods. For one Monte Carlo experiment, our proposed method requires approximately nine seconds, while other methods require approximately four seconds.

\subsection{Data analysis: Drosophila life cycle gene expression time-series data}

\begin{figure}[!htbp]
  \centering
  \includegraphics[width=5.7cm]{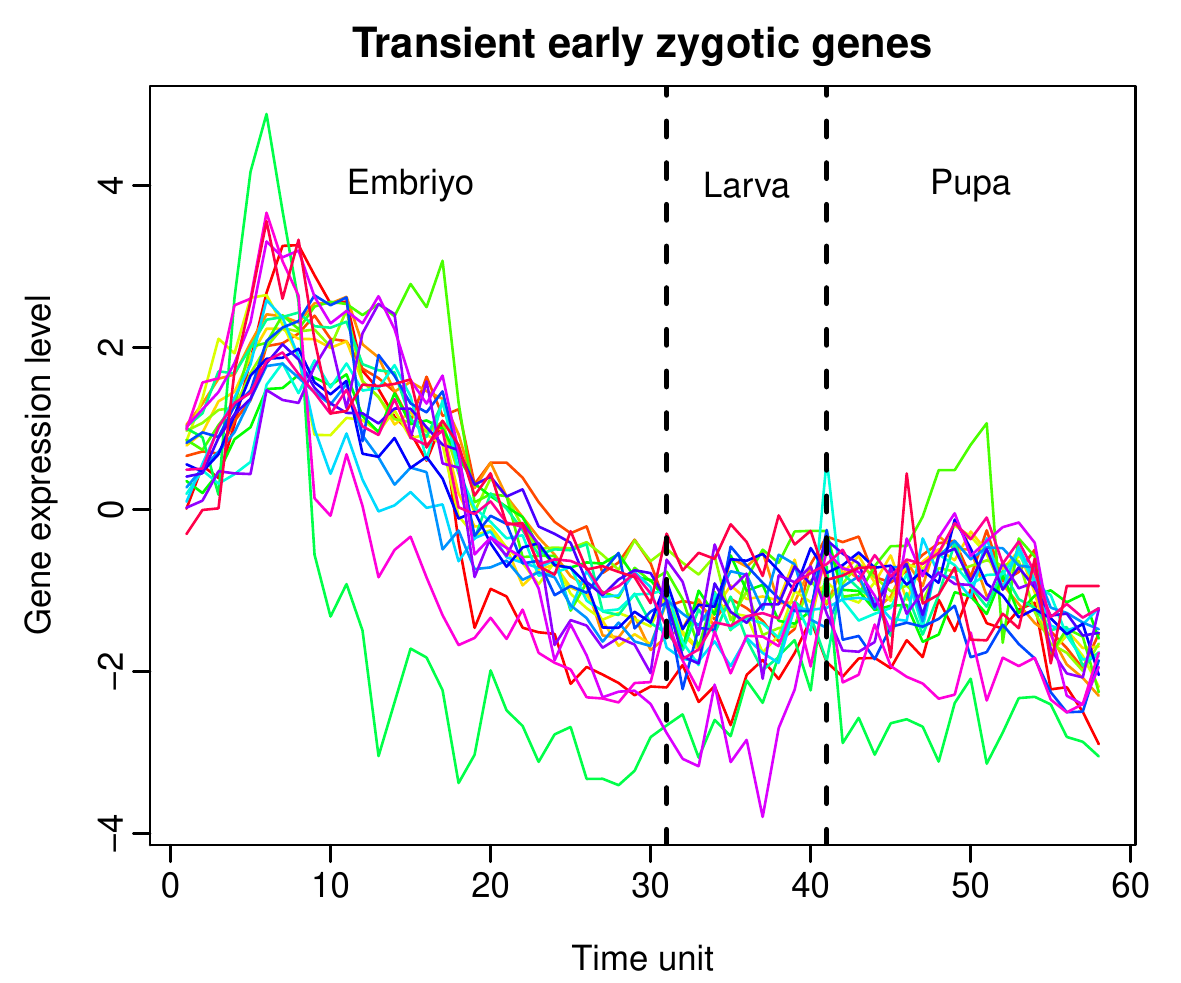}
  \includegraphics[width=5.7cm]{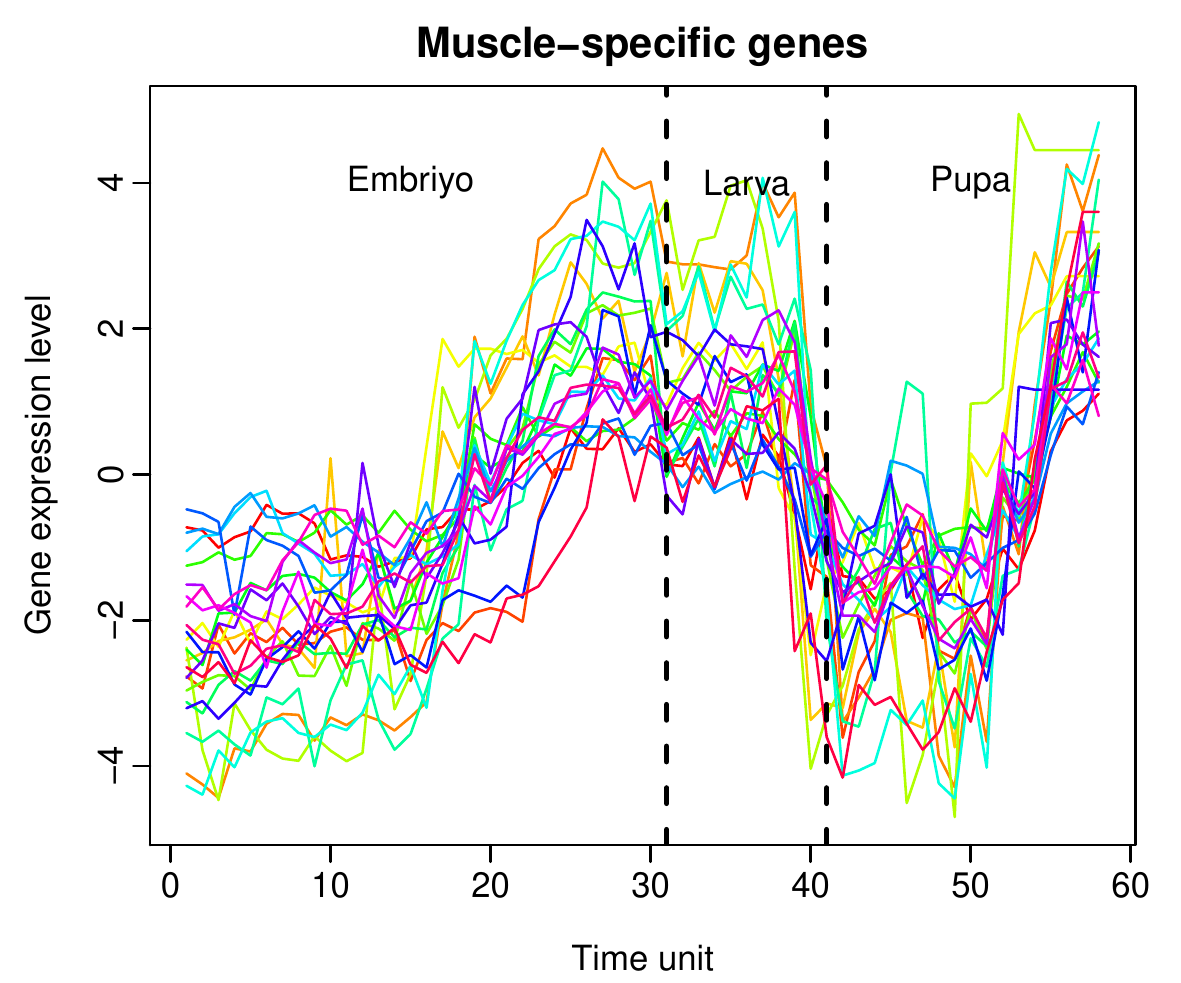}
  \includegraphics[width=5.7cm]{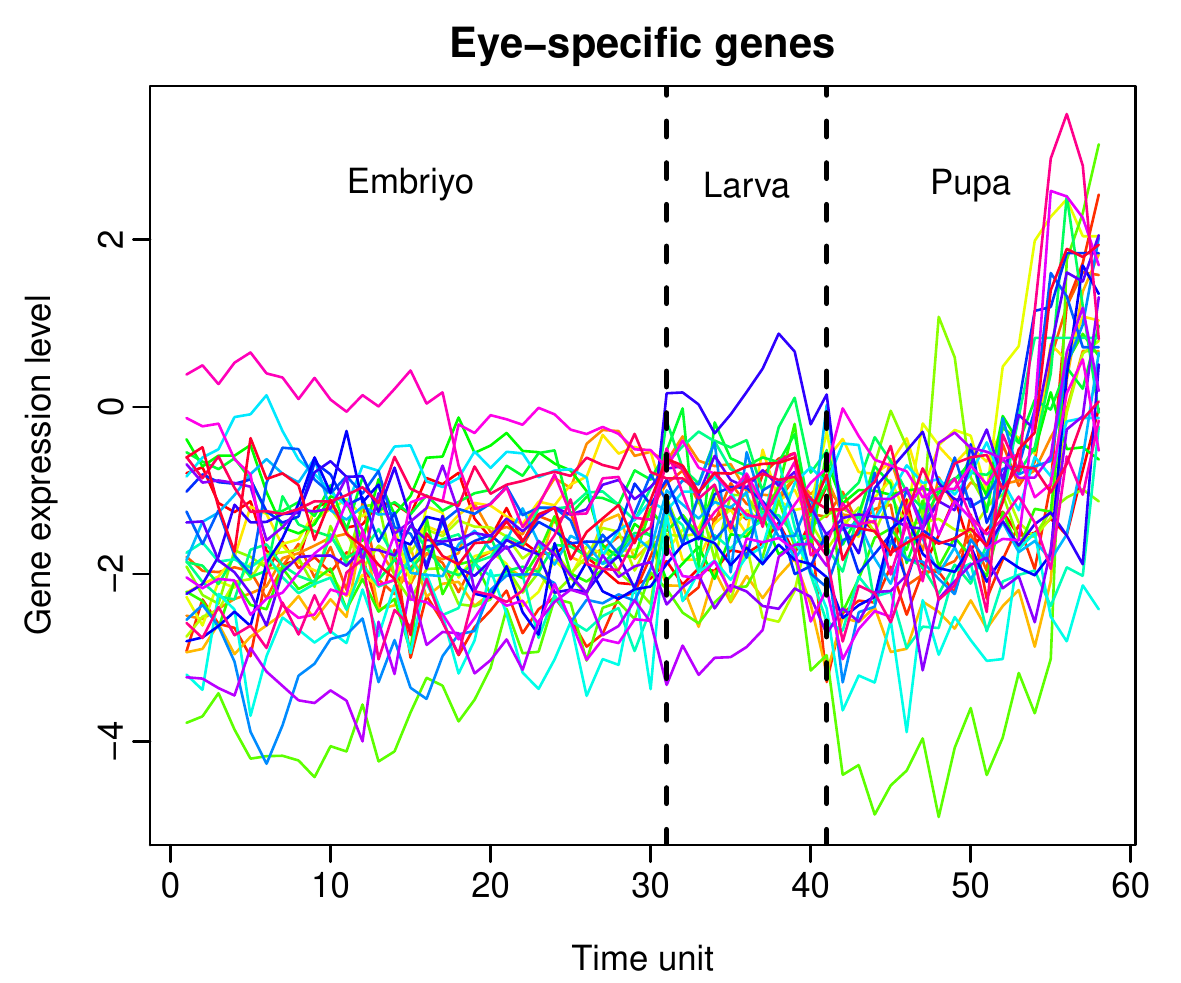}
  \caption{Graphical displays of the gene expression profiles; transient early zygotic genes (left panel), muscle-specific genes (middle panel), and eye-specific genes (right panel).}
  \label{fig:Fig_6}
\end{figure}

We consider the microarray data obtained by a cDNA microarray experiment, reported by \cite{Arbeitman}. The original microarray data consist of gene expression levels for 4028 genes involved in Drosophila melanogaster's life cycle during four stages of morphogenesis: embryo, larva, pupa, and adult. \cite{Arbeitman} reported gene expression patterns for nearly one-third of all Drosophila genes and identified the groups of co-expressed genes. In this study, we consider three specific groups of genes involved during the first three stages in the life cycle of Drosophila, e.g., embryo, larva, and pupa: 
\begin{inparaenum}
\item[1)] ``transient early zygotic'' consisting of 21 genes,
\item[2)] ``muscle-specific'' consisting of 23 genes, and
\item[3)] ``eye-specific'' consisting of 33 genes.
\end{inparaenum}
Transient early zygotic genes, which are the blastoderm-specific genes, are expressed at high levels only during the critical period of development when cellularization of the syncytial blastoderm embryo occurs \cite{Arbeitman}. These genes have a single-peak expression pattern in the embryo stage. Muscle-specific genes, which are tissue-specific genes, present a two-peak expression pattern that coincides with larval and pupa muscle development. Eye-specific genes, on the other hand, have a single-peak expression pattern in the pupa stage. The graphical displays of the gene expression profiles for all three groups of genes are presented in Figure~\ref{fig:Fig_6}.

Our aim with the gene expression levels is to explore the relationship between temporal patterns of gene expression levels measured at different development stages of Drosophila melanogaster. More precisely, our goal is to predict the later life gene expression profiles (pupa stage) of the three groups above of genes based on the gene expression patterns measured at embryo and larval stages. Using a mean regression-based FFR model, \cite{muller2008} studied the 23 muscle-specific gene expression levels to explore the dependence of adult gene expression patterns on larval patterns. In addition, they studied 27 strictly maternal gene expression levels and assessed gene expression pattern dependencies between the early embryo and the female pattern adult germline. The results produced by the analyses of \cite{muller2008} demonstrated that the gene expression profiles for pupa and adult phases are strongly related to the profiles of the same genes obtained during the embryo phase. Furthermore, their results showed a positive relationship in expression for muscle development-related genes and a negative relationship for Drosophila's strictly maternal genes.

In this study, we consider the following FFR model to explore the dependence of pupa gene expression patterns on embryo and larval patterns:
\begin{equation*}
\Y(t) = \beta_0(t) + \int_{s_1=1}^{31} \X_1(s_1) \beta_1(s_1,t) ds_1 + \int_{s_2=1}^{10} \X_2(s_2) \beta_2(s_2, t) ds_2, \; t \in [1, 17],
\end{equation*}
where $\Y(t)$, $\X_1(s_1)$, and $\X_2(s_2)$ denote the gene expression levels measured at pupa, embryo, and larval stages, respectively. From Figure~\ref{fig:Fig_6}, the muscle-specific genes are generally expressed at similar levels. On the other hand, some of the transient early zygotic and eye-specific genes are observed at higher or lower levels (potential outliers). Therefore, our proposed FFLQR model may robustly predict these groups' later life gene expression profiles compared with the existing models. To compare the prediction performance of our proposed method with the LS, FPLS, and FPC-based FFR models, we repeat the following procedure 100 times. For each group of genes, the entire datasets are randomly divided into two parts: 
\begin{inparaenum}
\item[1)] roughly, half of the datasets are used to construct models and
\item[2)] the remaining observations are used for validation.
\end{inparaenum}
For each replication, the MSPE is computed for each method. In addition, We apply the case-sampling-based bootstrap to construct pointwise prediction intervals for the later life gene expression profiles. We calculate performance metrics based on five FPCs and B-spline basis expansion functions. Note that, for the gene expression levels, we calculate the performance metrics under two models: the full model, which includes both $\X_1$ and $\X_2$ in the model, and the selected model, which includes only the significant predictor variable(s) determined by the proposed variable selection procedure.

\begin{figure}[!htb]
  \centering
  \includegraphics[width=7.7cm]{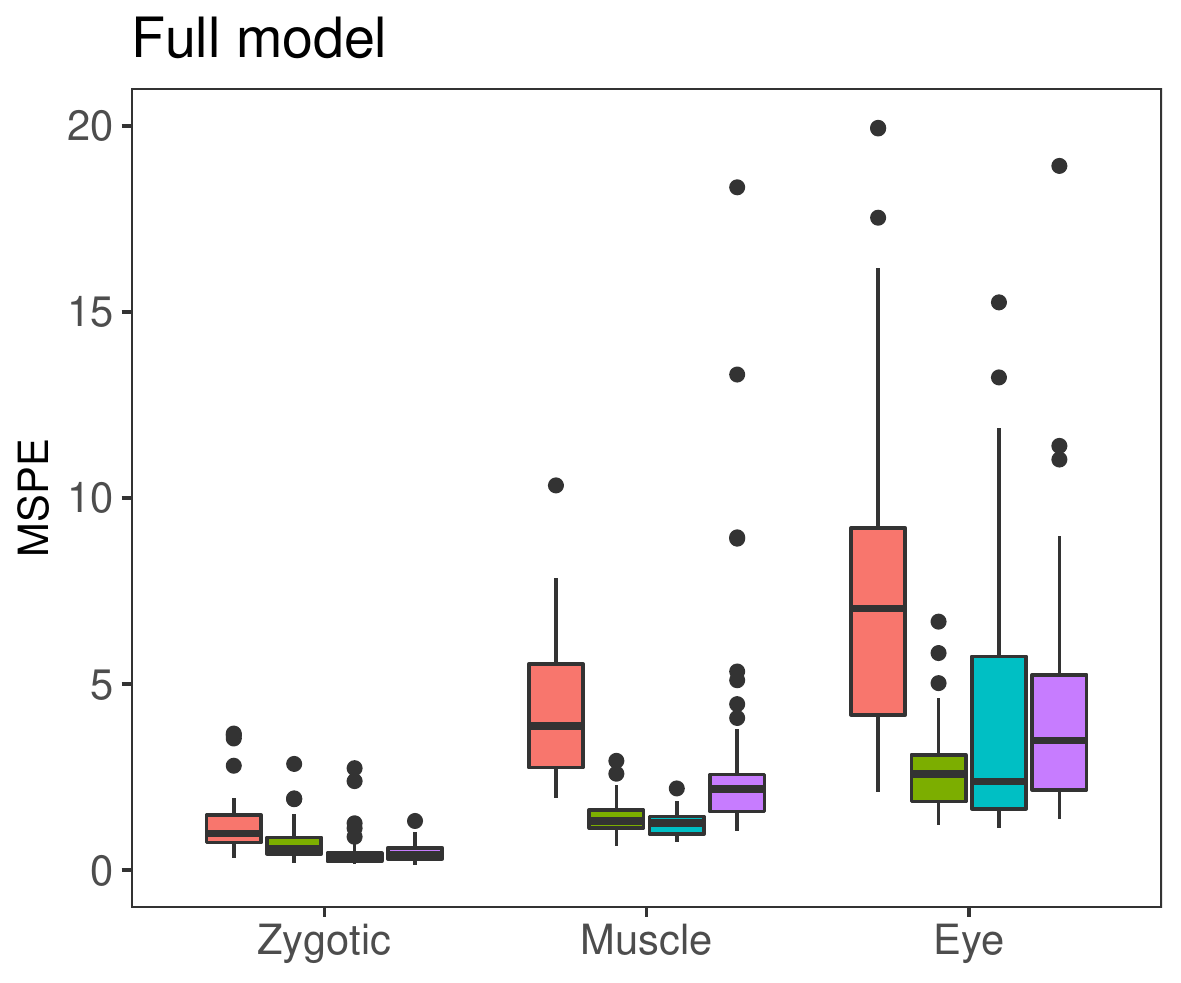}\quad
  \includegraphics[width=7.7cm]{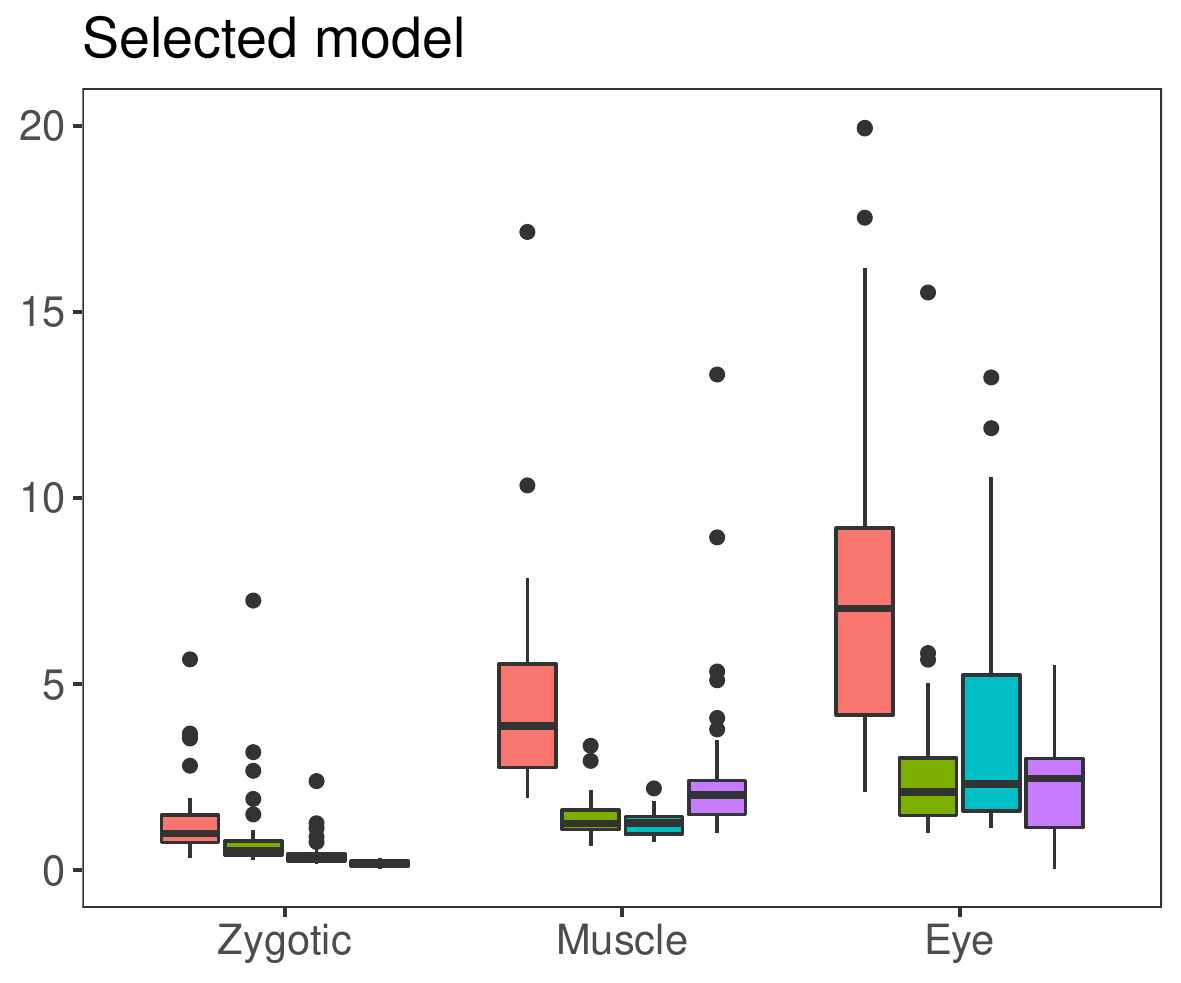}
  \\
  \includegraphics[width=7.7cm]{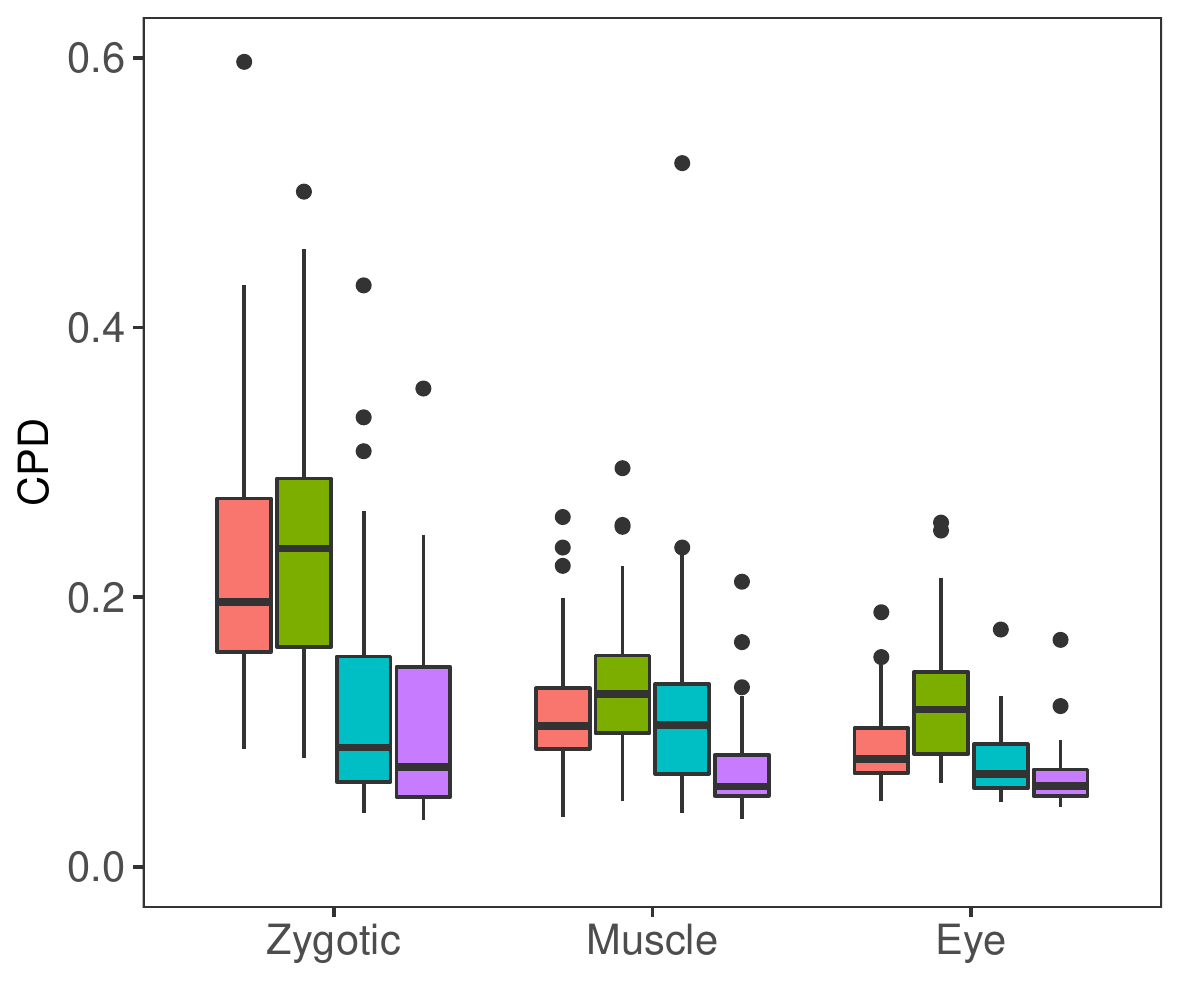}\quad
  \includegraphics[width=7.7cm]{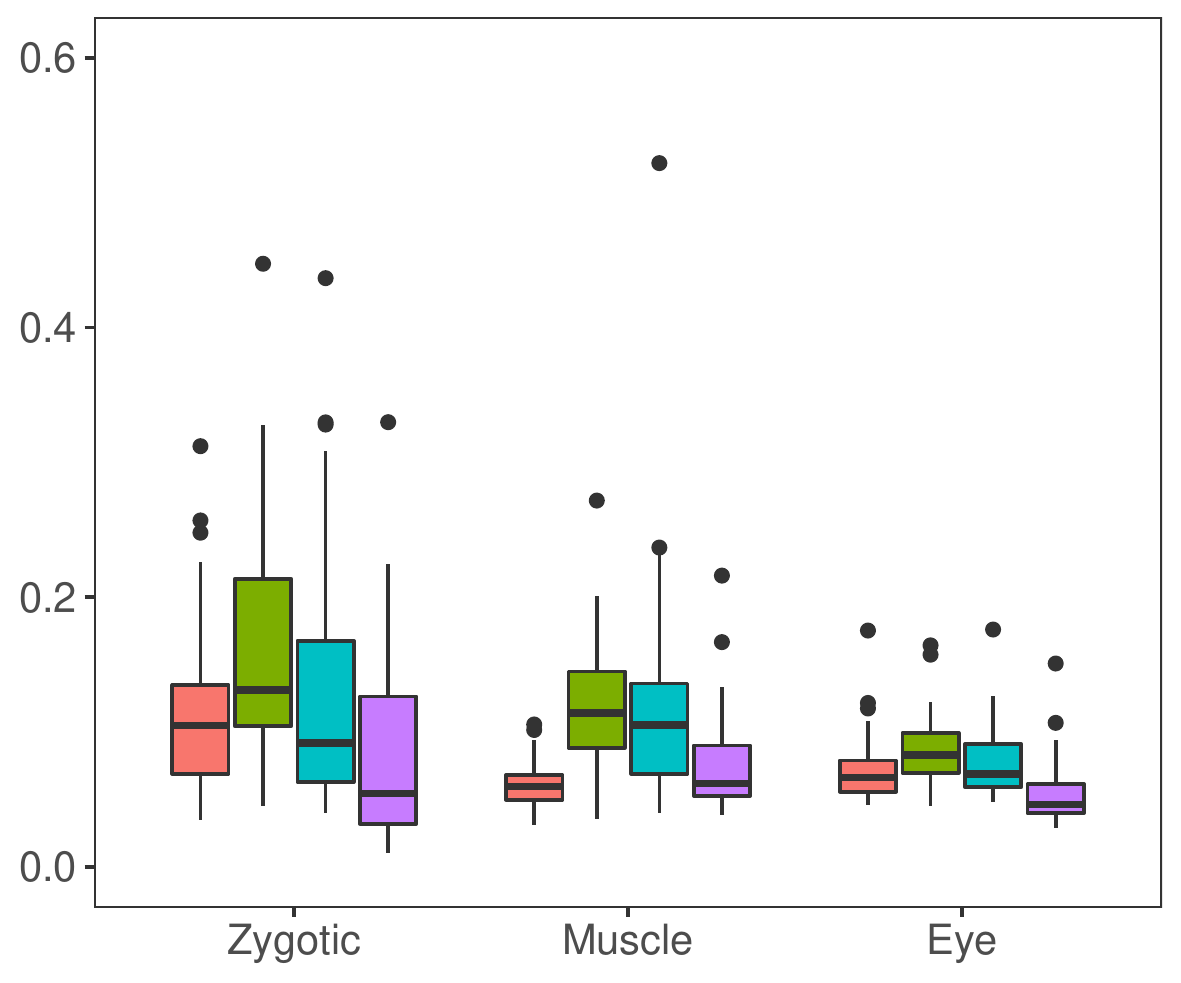}
  \\
  \includegraphics[width=7.7cm]{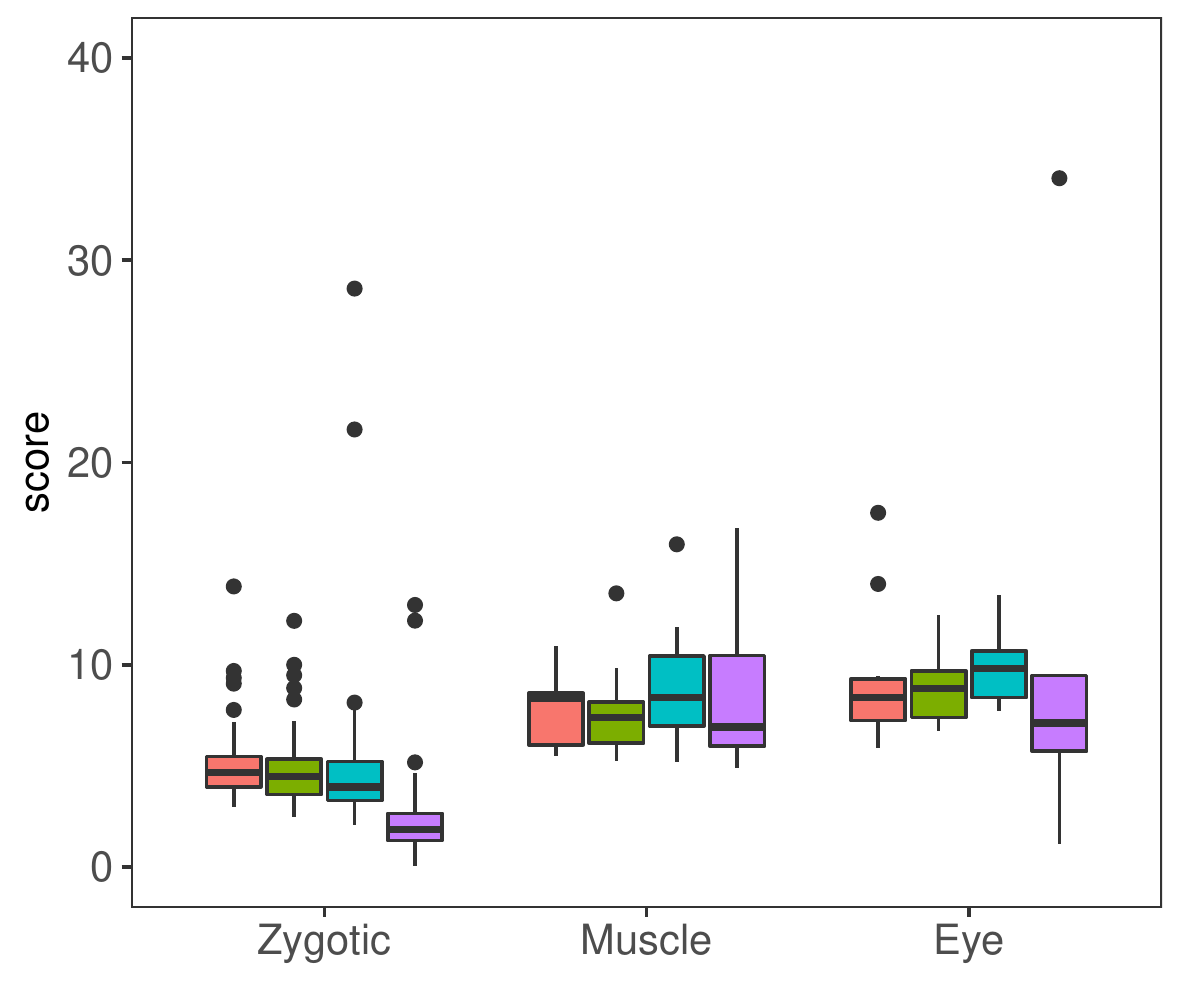}\quad
  \includegraphics[width=7.7cm]{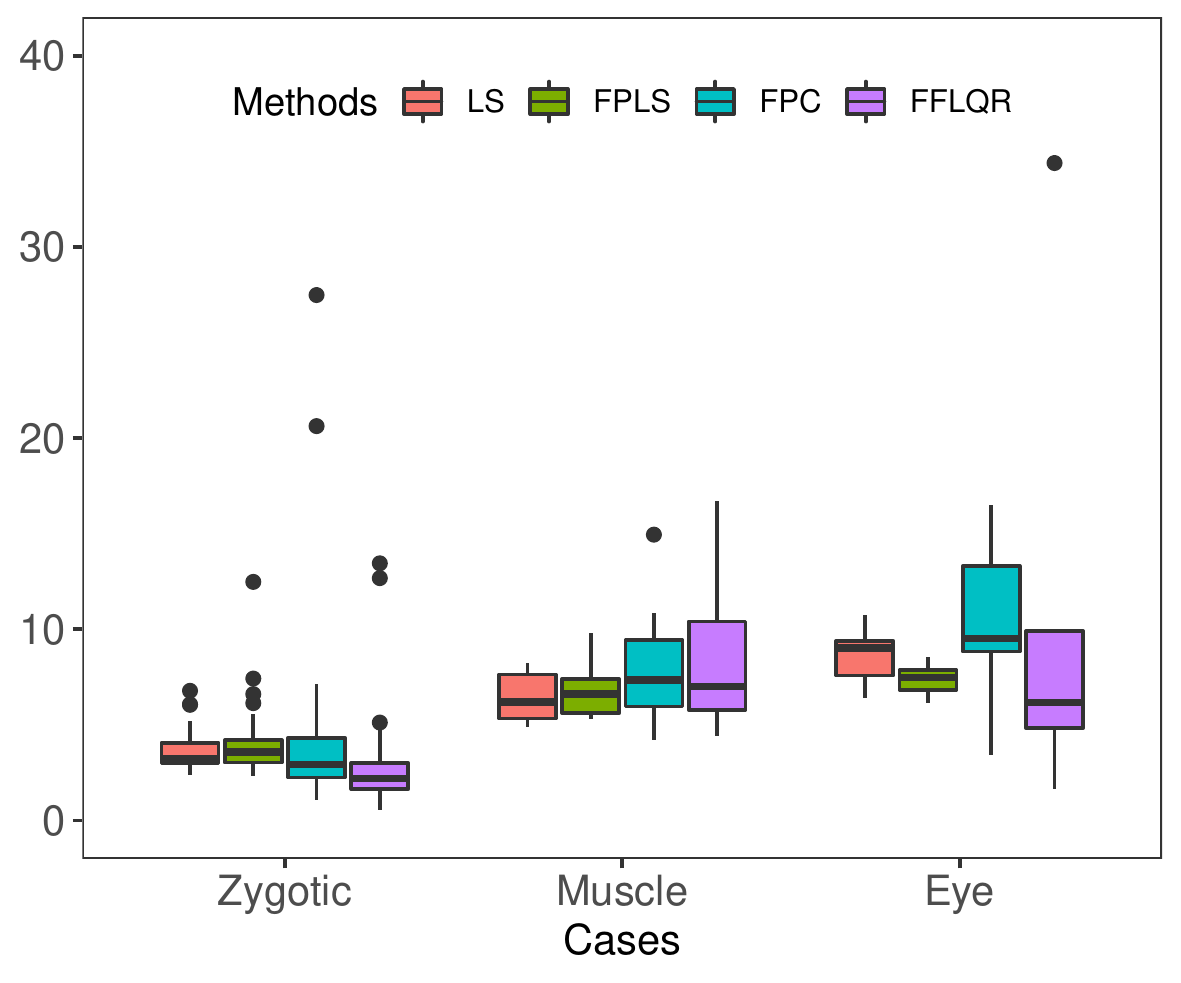}
  \caption{\textbf{Predictive model performances}: Calculated MSPE (first row), CPD (second row), and interval score (third row) values of the LS, FPLS, FPC, and FFLQR methods for the Drosophila life cycle gene expression time-series data; full model (first column) and selected model (second row).}
  \label{fig:Fig_7}
\end{figure}

The results produced by our analyses are presented in Figure~\ref{fig:Fig_7}. This figure shows that the selected model produces smaller prediction performance metrics than the full model for all the methods. When considering the performance of selected models, compared with the LS FPLS, and FPC methods, FFLQR produces smaller MSPE values for transient early zygotic and eye-specific genes. In contrast, it produces slightly larger error values than the FPLS and FPC for muscle-specific genes. This finding is because some of the transient early zygotic and eye-specific genes are observed at higher or lower levels (potential outliers). At the same time, all the muscle-specific genes are generally expressed at similar levels. Therefore, our proposed method produces robust results for the transient early zygotic and eye-specific genes. In addition, the proposed method produces generally smaller CPD and interval score values than the LS, FPLS, and FPC methods.

\vspace{.2in}

\begin{figure}[!htbp]
  \centering
  \includegraphics[width=5.7cm]{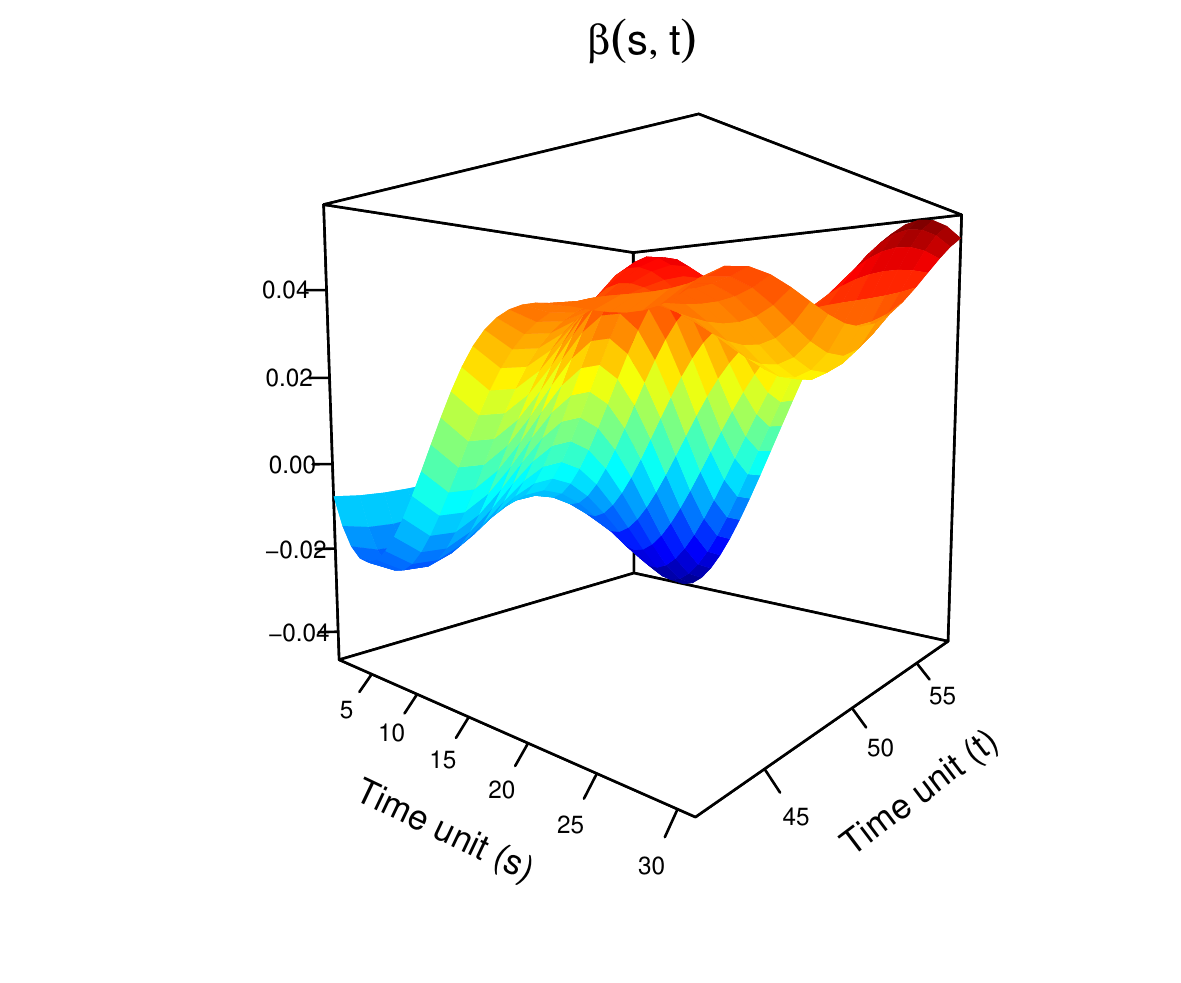}
  \includegraphics[width=5.7cm]{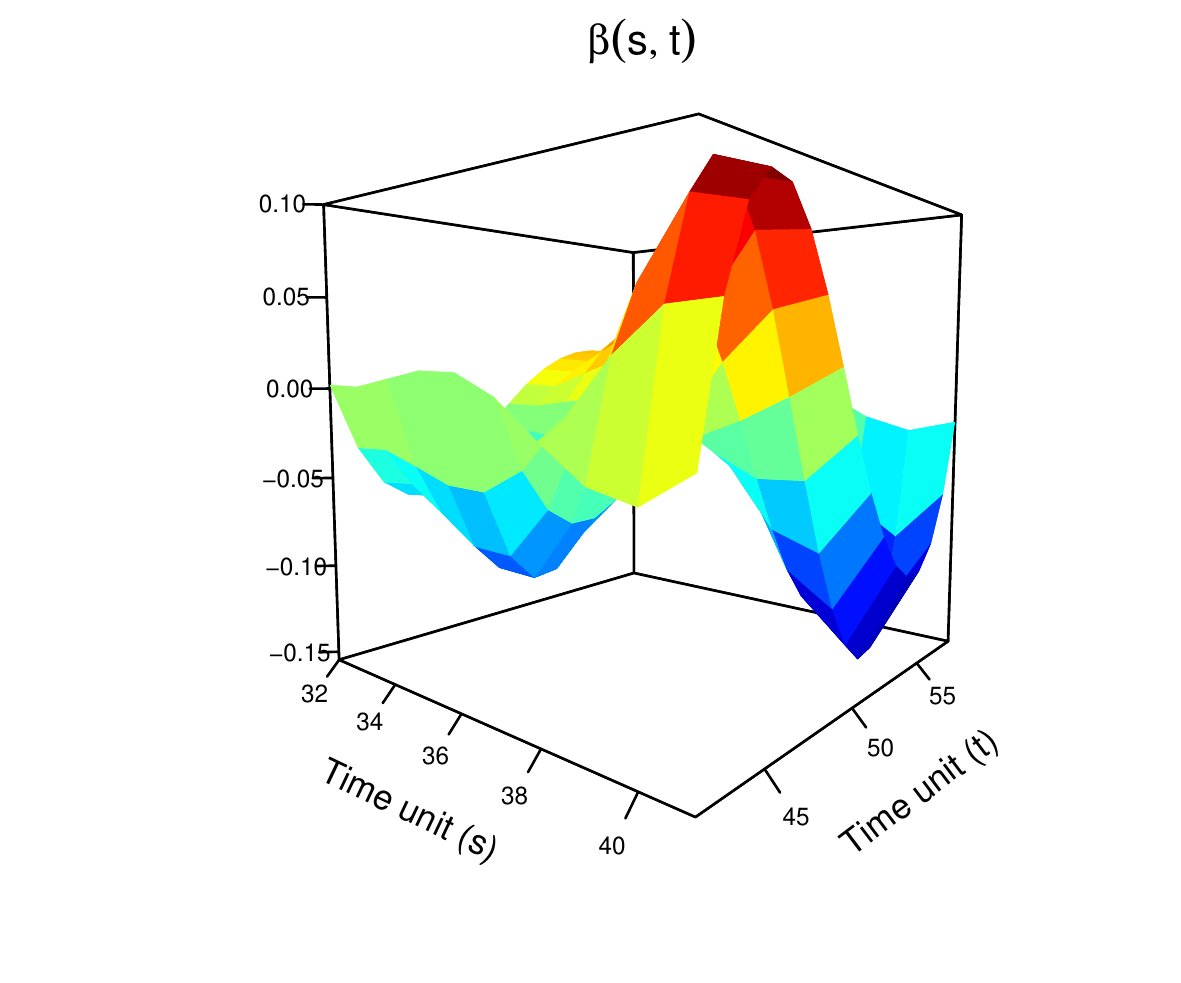}
  \includegraphics[width=5.7cm]{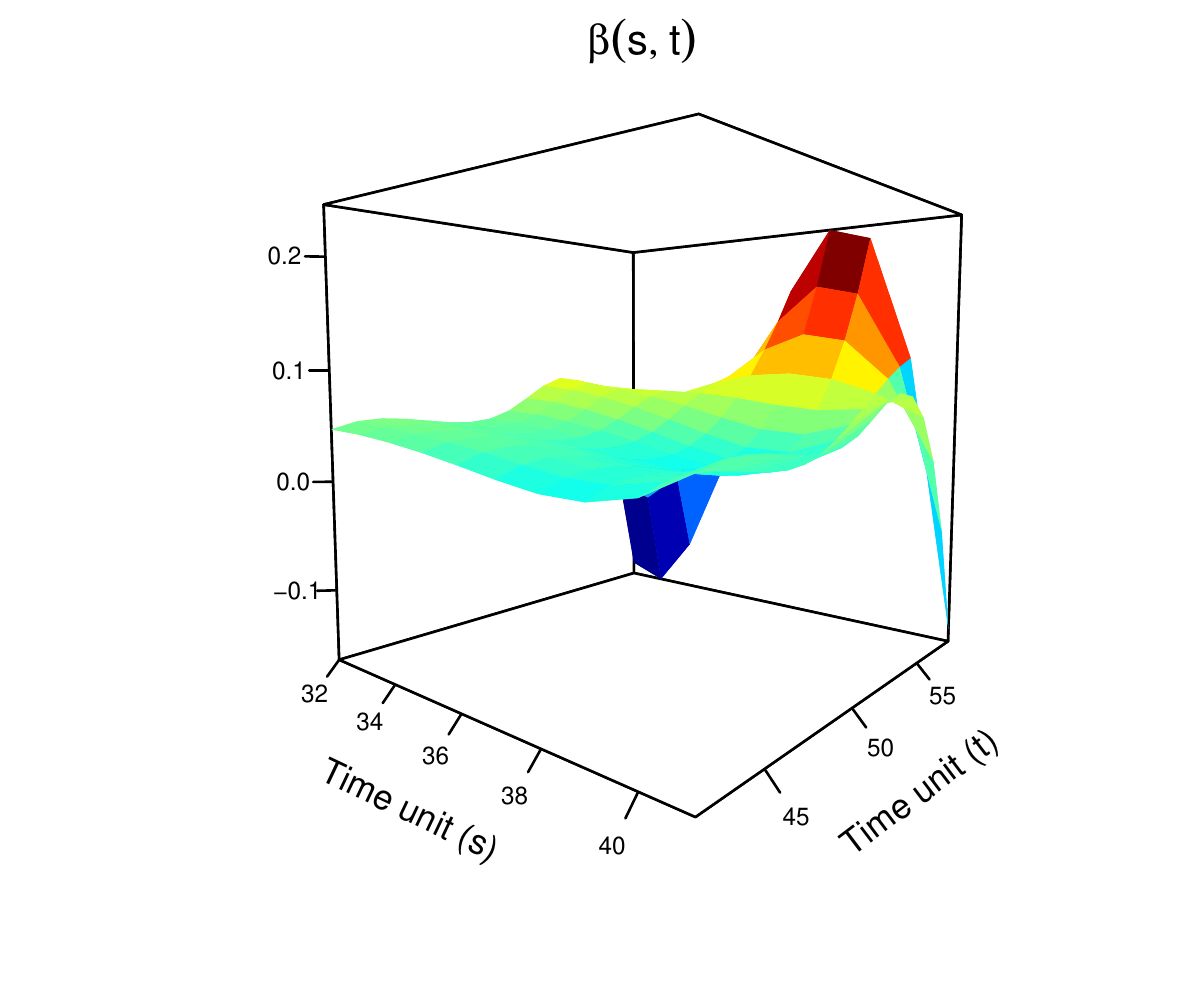}
  \caption{Estimated regression coefficient functions $\widehat{\beta}_{\tau}(s,t)$ when $\tau = 0.5$ for the Drosophila life cycle gene expression time-series data; transient early zygotic genes (left panel), muscle-specific genes (middle panel), and eye-specific genes (right panel).}
  \label{fig:Fig_8}
\end{figure}

We construct a model for all genes using the observed gene expression profiles to determine the significant variables. For transient early zygotic genes, the only larval stage ($\X_2$) is selected into the final model by the LS, FPLS, and FPC methods. In contrast, only the proposed method selected the embryo ($\X_1$) as significant. For muscle-specific and eye-specific genes, on the other hand, the only larval stage ($\X_2$) is selected as significant by the LS, FPC, and FFLQR methods, while both embryo ($\X_1$) and larval ($\X_2$) stages are selected as significant by the FPLS method. The estimated regression coefficient functions by the proposed method for all genes are presented in Figure~\ref{fig:Fig_8}.

\section{Conclusion} \label{sec:conc}

We propose a QR approach for the FFR models, an extension of the traditional FFR model to the QR framework. In our proposed method, all the functional objects are first transformed into a finite-dimensional space using the FPC method to overcome the ill-posed problem caused by the infinite-dimensional nature of the functional variables. It approximates the regression coefficient functions using the FPC decomposition of the functional response and predictor variables. A forward variable selection procedure and the proposed FFLQR method are introduced to determine the significant functional predictors for the model. Moreover, we employ a nonparametric bootstrap approach to investigate further the uncertainty of predictions produced by the proposed FFLQR method. Our method's finite-sample performance is investigated via several Monte Carlo experiments and empirical data analysis, and they are compared favorably with existing methods.

All the numerical analyses performed in our study have demonstrated that the proposed FFLQR method produces superior performance over the existing function-on-function mean regression models when the error terms follow a non-normal distribution or in the presence of outliers. 




\end{document}